\documentclass[11pt,a4paper]{article}
\pdfoutput=1
\usepackage{jheppub}
\usepackage{graphicx,psfrag,color}% Include figure files
\usepackage{bm}% bold math
\usepackage{mathbbol,verbatim}
\usepackage{slashed}
\usepackage{graphics}
\usepackage{color,ulem}
\usepackage{multirow}
\usepackage{subcaption}
\definecolor{greeen}{rgb}{0.03,0.84,0.13}
\definecolor{test}{rgb}{0.03,0.74,0.33}
\definecolor{viol}{rgb}{0.44,0,0.94}
\definecolor{or}{rgb}{0.95,0.65,0}

\title{CP Violating Effects in Heavy Neutrino Oscillations: Implications for Colliders and Leptogenesis}

\author[a]{P. S. Bhupal Dev,}
\author[b]{Rabindra N. Mohapatra,}
\author[a]{Yongchao Zhang}
\affiliation[a]{Department of Physics and McDonnell Center for the Space Sciences,  Washington University, St. Louis, MO 63130, USA}
\affiliation[b]{Maryland Center for Fundamental Physics, Department of Physics, University of Maryland, College Park, MD 20742, USA}

\date{\today}

\begin{document}

\abstract{Two of the important implications of the seesaw mechanism are: (i) a simple way to understand the small neutrino masses, and (ii) the origin of matter-anti-matter asymmetry in the universe via the leptogenesis mechanism. For TeV-scale seesaw models, successful leptogenesis requires that the right-handed neutrinos (RHNs) must be quasi-degenerate and if they have CP violating phases, they also contribute to the CP asymmetry. We investigate this in the TeV-scale left-right models for seesaw and point out a way to probe the quasi-degeneracy possibility with CP violating mixings for RHNs in hadron colliders using simple observables constructed out of same-sign dilepton charge asymmetry (SSCA). In particular, we isolate the parameter regions of the model, where the viability of leptogenesis can be tested using the SSCA at the Large Hadron Collider, as well as future 27 TeV and 100 TeV hadron
colliders. We also independently confirm an earlier result that there is a generic lower bound on the $W_R$ mass of about 10 TeV for leptogenesis to work.}

\maketitle

\section{Introduction}

Neutrino oscillation experiments have established that neutrinos have very small but nonzero masses. This begs for some new physics beyond the Standard Model (SM), since the SM predicts the neutrinos to be exactly massless to all orders in perturbation theory. While the nature of the underlying new physics is far from clear, a simple paradigm that provides a natural way to understand tiny neutrino masses is the seesaw mechanism~\cite{seesaw1, seesaw2, seesaw3, seesaw4, seesaw5}. The two key ingredients of the so-called type-I seesaw mechanism are the existence of right-handed neutrinos (RHNs) and their Majorana masses. In order to calculate the light neutrino masses using the seesaw formula, we need the Dirac Yukawa coupling matrix and the Majorana mass matrix of the RHNs. Understanding both the Dirac mass matrix and the RHN Majorana mass matrix is therefore key to understanding the origin of neutrino masses.

Another important implication of seesaw paradigm is that it provides a natural framework for understanding the origin of matter-antimatter asymmetry via the mechanism of leptogenesis~\cite{Fukugita:1986hr}. Typical scenarios considered are those with high-scale seesaw~\cite{Buchmuller:2004nz}. However if the seesaw scale is in the TeV range, successful leptogenesis requires that there must be at least two RHNs which are quasi-degenerate~\cite{RL4}. We investigate the collider signatures and leptogenesis in a TeV-scale left-right embedding of seesaw that includes quasi-degenerate RHNs and show that observations of Majorana signatures in pp-colliders for this model can provide a way to not only probe the mixings and CP violation in the RHN sector but also to test  leptogenesis.

To see the advantage of left-right models in this discussion, note that in the case of minimal seesaw models, where one extends the SM by including the RHNs `by hand', one can diagonalize the RHN mass matrix and absorb the required rotation matrices into the Dirac Yukawa couplings. The specific rotation angles and phases then lose their separate identity since they can be absorbed by redefining the Dirac Yukawa coupling (see Section~\ref{sec:minimal}). %\footnote{\blue{The resulting rotation angles and phases that arise after seesaw diagonalization with the new Dirac mass matrix appear in the weak interaction Lagrangian as RHN admixture with the light neutrino states and can be measured in principle~\cite{zhou}, although they appear with a coefficient suppressed by {the heavy-light neutrino mixings}. This makes it in general  difficult to observe except in special cases~\cite{Kersten} which postulate special texture for Dirac mass matrix.}}
Similar thing happens when the SM is extended by a flavor universal $U(1)$ gauge symmetry. %(see for example ~\cite{ Deshpande:1979df, Galison:1983pa, Buchmuller:1991ce, FileviezPerez:2010gw, Basso:2011hn}).
On the other hand the situation is different if the minimal seesaw is embedded into the left-right symmetric model (LRSM) based on the gauge group $SU(2)_L\times SU(2)_R\times U(1)_{B-L}$~\cite{Pati:1974yy, Mohapatra:1974gc, Senjanovic:1975rk} or a flavor-dependent $U(1)$ acting on the RHNs. Here we focus on the LRSM where any rotation $V_R$ of the RHN mass matrix (except for the overall phase) will manifest in the $W^\pm_R$ interaction and become in principle measurable if an on-shell $W_R$ is produced at the Large Hadron Collider (LHC) or future colliders. In this paper, we point out that observables which can help us to measure the mixing angles and phases in $V_R$ at colliders using on-shell $W_R$ production are the same-sign charge asymmetry (SSCA) ${\cal A}_{\alpha\beta}$ defined in Eq.~\eqref{eqn:Aab} and the ratio of SSCAs ${\cal R}_{\rm CP}^{(\ell)}$ defined in Eq.~\eqref{eqn:Rab}. Similar observables were previously considered in the context of minimal seesaw~\cite{Bray:2007ru} and $U(1)_{B-L}$ models~\cite{Blanchet:2009bu}.

%\footnote{A similar observable was considered as probe of leptogenesis in  a $Z'$ model in ~\cite{Blanchet:2009bu}.}

In $pp$-collisions, the same-sign (SS) dilepton final states, which accompany the opposite-sign (OS) dilepton events, have been known to provide a `smoking gun' test of the Majorana nature of the RHNs~\cite{KS} and hence of the seesaw paradigm. In the hierarchical RHN case, the number of SS and OS dilepton states arising from RHN production and decay for each flavor combination are equal and the relative strengths of signal between different flavor modes depends on the right-handed (RH) lepton mixing angles~\cite{juan}. However, as has been pointed out in Refs.~\cite{GJ, GJ2, DDM}, if the RHNs are quasi-degenerate enough so that the coherence condition is met~\cite{Akhmedov:2007fk}, the RHNs can oscillate into each other.\footnote{The same kind of oscillation can happen also in the inverse seesaw case, where a singlet neutrino is degenerate with the RHN, forming a pseudo-Dirac pair, see Refs.~\cite{DM, Hirsch, Antusch}.} In this case, the ratio of SS and OS dilepton signals due to Majorana RHNs becomes flavor-dependent and can become unequal for each flavor combination if there is CP violation (CPV) in the RHN mass matrix~\cite{DDM}. It is encouraging to note that recent CMS~\cite{Sirunyan:2018pom} and ATLAS~\cite{Aaboud:2019wfg} searches for RHNs have included both SS and OS dilepton events in their event selection.

In this paper, we study the CPV effect in the RHN mass matrix by focusing only on the SS final states, which have relatively less SM background, as compared to the OS final states. %will not consider the opposite-sign signals since there can be many back ground processes which contribute to opposite sign dileptons.  We extend the discussion of Ref.~\cite{DDM} to
In particular, we consider the SSCA observables ${\cal A}_{\alpha\beta}$ defined in Eq.~\eqref{eqn:Aab} and ${\cal R}_{\rm CP}^{(\ell)}$ defined in Eq.~\eqref{eqn:Rab}
%defined as~\cite{Bray:2007ru}
%\blue{introduce in the introduction the ratio of ratio ${\cal R}_{\rm CP}$ as in the paper by Pilaftsis et al???} %and discuss their implications for leptogenesis.
%\begin{align}
%\label{eqn:ssca}
%{\cal A}_{\alpha\beta} \ \equiv \ \frac{{\cal N}(\ell^+_\alpha\ell^+_\beta)-{\cal N}(\ell^-_\alpha\ell^-_\beta)}{{\cal N}(\ell^+_\alpha\ell^+_\beta)+{\cal N}(\ell^-_\alpha\ell^-_\beta)}
%\end{align}
for specific charged-lepton flavor combinations $(\alpha, \beta)$ and point out how one can extract the RHN mass matrix mixing and phase information from collider observations. It is worth noting that the SSCA is also well-suited to study the CP violating effects in other situations in particle physics, such as meson mixing; see e.g.~\cite{Nir:1999mg}.

We focus on a simple but a realistic example of two nearly degenerate RHNs $N_{e,\,\mu}$ in LRSM with TeV-scale $W_R$. We assume that third generation RHN $N_\tau$ is heavier so that it has no effect on the light or heavy neutrino masses. Our model can be easily generalized if the third RHN is also degenerate with the other two. We quantify the dependence of SSCAs on the CP phase $\delta_R$ in the mass matrix of the two oscillating RHNs and discuss when this asymmetry is observable.  We explicitly show that the charge asymmetry of SS dilepton states for different flavors can provide useful information on the phase as well as the mixing angles of the RHN mass matrix.
It turns out that due to an intrinsic difference between the production rates for $W^\pm_R$ in $pp$ collisions due to parton distribution function (PDF) asymmetries between the up and down quarks in proton, a non-zero ${\cal A}_{\alpha\beta}$ is present even in the absence of CPV. However, the asymmetries can still be measured at the LHC for a sizable SS dilepton signal from $W_R$ boson decay and can provide a useful probe of the RHN mass matrix.
%This makes the new CP-induced SSCAs hard to observe at the Large Hadron Collider (LHC) energies. However
At future higher-energy colliders, such as the High-Energy LHC (HE-LHC)~\cite{Zimmermann:2018wdi} with center-of-mass energy $\sqrt{s}=27$ TeV and future $100$ TeV collider, such as Future Circular Collider (FCC-hh)~\cite{Benedikt:2018csr} or Super Proton-Proton Collider (SPPC)~\cite{CEPC-SPPCStudyGroup:2015csa}, the charge asymmetries can be measured for larger $W_R$ masses. On the other hand, we will see that in the ratios ${\cal R}_{\rm CP}^{(\ell)}$, the PDF uncertainties cancel, and therefore, these are cleaner observables for probing the CPV in the RHN sector.

It is well known that if the RHNs are quasi-degenerate, the mixing angle and CP phase in the RHN sector could play an important role in resonant leptogenesis~\cite{RL1, RL2, RL3,RL4}. We show how the SSCA observations in colliders can provide key insight into TeV-scale leptogenesis. In fact, in combination with other collider observations such as the $W_R$ mass, it can even rule out TeV leptogenesis for certain parameter ranges (i.e. CP phases and mixing angles) of the model. Our work is largely complementary to the falsification scheme for high-scale leptogenesis~\cite{Deppisch:2013jxa}, as well as to other probes of CPV and low-scale leptogenesis at colliders~\cite{Vasquez:2014mxa, Caputo:2016ojx, Antusch:2017pkq}.

To carry out our leptogenesis calculation, we choose a generic form of the Dirac Yukawa coupling matrix using the Casas-Ibarra parameterization~\cite{Casas:2001sr}.\footnote{Our model has parity broken at higher scale leaving $SU(2)_R$ unbroken till the TeV scale and therefore we cannot use the Dirac mass matrix formula derived in Ref.~\cite{Nemevsek:2012iq} which is valid for $C$-symmetric LRSM.} It is known in the LRSM that for resonant leptogenesis to provide adequate lepton asymmetry, the dilution and washout effect from RH gauge interactions must be small~\cite{Frere:2008ct, Dev:2014iva, Dhuria:2015cfa}. This puts an {\it absolute} lower bound on the $W_R$ mass, at 9.4 TeV for normal hierarchy (NH) ordering of active neutrino masses and 8.9 TeV for inverted hierarchy (IH), as we will find in the scenario of this paper. These limits are close to an earlier result in Ref.~\cite{DLM} which uses a different form of the Yukawa texture. We also find that to make leptogenesis work in the LRSM the Yukawa couplings $y_{\alpha i}$ of RHNs can not be either too small or too large, turning out to be in the range of $1.0 \times 10^{-6} \lesssim |y|_{\rm max} \lesssim 8.6 \times 10^{-4}$. As a result, the leptogenesis constraints turn out to be stronger than the low-energy high-precision measurements such as neutrino-less double beta ($0\nu\beta\beta$) decays, the lepton flavor violating (LFV) decays such as $\mu \to e \gamma$ and the electric dipole moment (EDM) of electron.

This paper is organized as follows: In Section~\ref{sec:minimal}, we discuss the physical significance of the RHN CP phase in the minimal type-I versus left-right seesaw. In Section~\ref{sec:SSCA}, we provide the basic framework for the SSCAs, considering both three and two-body decays of RHNs. We also clarify the coherence conditions for RHN oscillation at high-energy colliders. In Section~\ref{sec:prospect}, we estimate the prospects of SSCAs at future high-energy hadron colliders, as shown in Figs.~\ref{fig:prospect} to \ref{fig:RCP3}. In Section~\ref{sec:leptogenesis}, we elaborate on the role of RHN mixing and CP phase in TeV-scale resonant leptogenesis. We also estimate the leptogenesis constraints on $W_R$ boson mass. In section~\ref{sec:constraints}, we discuss the constraints on the RHN sector from $0\nu\beta\beta$ decays, $\mu \to e\gamma$ and electron EDM. We conclude in Section~\ref{sec:conclusion}. The analytic formula for the square root of the RHN Majorana mass matrix ${\cal M}_N^{1/2}$ is given in Appendix~\ref{sec:analytic}, and more details about heavy-light neutrino mixing are presented in Appendix~\ref{sec:ST}.

%\section{\sout{Right-handed neutrino rotation matrix and associated CP violation}
\section{CP phase in the RHN sector} \label{sec:minimal}

In this section we introduce the problem we are addressing and our goal for this paper. The type-I seesaw~\cite{seesaw1, seesaw2, seesaw3, seesaw4, seesaw5} generally has the Dirac mass matrix ${\cal M}_D$, as well as the RHN Majorana mass matrix ${\cal M}_N$, both involving CP phases. Of course, as is well known, the rotation angles and the phases can be redefined by choice of basis depending on the full theory. If we consider a basis where the charged lepton mass matrix is diagonal, the phases and rotation angles of both ${\cal M}_D$ and ${\cal M}_N$ are physical parameters and their measurement would provide useful insight about theories of neutrino masses. The question we address in this paper is: is it possible to measure the phases and rotation angles of the unitary matrix $V$ that diagonalizes the ${\cal M}_N$ matrix in the basis where the charged lepton mass matrix is diagonal? This is important since those phases not only determine the final leptonic mixing angles and phases of the light neutrinos but also play a role in explaining the origin of matter via leptogenesis.

Consider the type-I seesaw extension of the SM (we call this the `SM seesaw'), where the leptonic sector of the Lagrangian can be written as:
\begin{eqnarray}
{\cal L}_I^{} ~=~
\frac{g_L}{{2}} \overrightarrow{W}^\mu \cdot \bar{\psi}_L\gamma_\mu  \vec{\tau}\psi_L +
h_e \bar{\psi}_L H \ell_R +
h_\nu\bar{\psi}_L\widetilde{H}N +
N^T{\cal M}_NN ~+~ {\rm H.c.} \,,
\end{eqnarray}
where $\overrightarrow{W}_\mu$ and $g_L$ are respectively the SM $SU(2)_L$ gauge fields and coupling constant, $\vec{\tau}$ is the vector of Pauli matrices, $\psi_L$, $H$ and $\ell_R$ are respectively the left-handed lepton doublets, Higgs doublet and right-handed lepton singlets in the SM, and $\widetilde{H}\equiv i\tau_2 H^\ast$. If we choose a basis such that the charged lepton Yukawa coupling matrix $h_e$ is diagonal, the RHN mass matrix ${\cal M}_N$ remains non-diagonal and this fixes the leptonic basis and all angles and phases in this basis are physical.
We can next diagonalize the mass matrix ${\cal M}_N$ by a unitary rotation $V^T{\cal M}_NV$. The question we ask now is whether we can measure the angles and the phases that parameterize $V$. The answer  in the SM type-I seesaw is that we can rewrite $h_\nu$ as $h^\prime_\nu = h_\nu V^\dagger$ and the matrix $V$ simply redefines the Dirac mass term ${\cal M}_D$ in the type I seesaw and does not have a separate identity and therefore cannot be measured separately. In other words, it is as if we had chosen the neutrino Yukawa matrix as $h^\prime_\nu$ to start with. The resulting rotation angles and phases that arise after seesaw diagonalization with the new Dirac mass matrix appear in the weak interaction Lagrangian as RHN admixtures with the light neutrino states and can be measured in principle~\cite{Bray:2007ru, zhou}. However, these effects always appear with a coefficient suppressed by the heavy-light neutrino mixing, which makes it in general  difficult to observe, except in special cases~\cite{Kersten, Ibarra:2010xw} with special textures for Dirac mass matrix to realize large light-heavy neutrino mixing.

On the other hand, in the LRSM, there is an extra gauge boson interaction and we can write the leptonic part of the Lagrangian  as:
\begin{eqnarray}
{\cal L}^{\rm LRSM}_I & \ = \ &
\frac{g_L}{2} \overrightarrow{W}^{\mu} \cdot \bar{\psi}_L\gamma_\mu \vec{\tau} \psi_L +
\frac{g_R}{2} \overrightarrow{W}_R^\mu \cdot\bar{\psi}_R\gamma_\mu \vec{\tau} \psi_R
\\ \nonumber
 && + h_\nu\bar{\psi}_L \Phi \psi_R
 + h_e \bar{\psi}_L \widetilde{\Phi} \psi_R
 + f_R N^T  \Delta_R^{0}N ~+~ {\rm H.c.} \,,
\label{eqn:lagLRSM}
\end{eqnarray}
where $\overrightarrow{W}_{R\mu}$ and $g_R$ are respectively the $SU(2)_R$ gauge fields and coupling constant, $\Phi$ and $\Delta_R$ are the bidoublet and $SU(2)_R$ triplet Higgs fields of the LRSM respectively, and $\widetilde{\Phi} = \tau_2 \Phi^\ast \tau_2$. For definiteness, let us choose $\langle \Phi \rangle = {\rm diag} (\kappa, 0)$, so that $h_e$ connects only charged leptons and $h_\nu$ connects $\nu_L$ with the RHN $N$. Again as in the type-I seesaw case, let us choose a basis where $h_e$ is diagonal. In this basis in general, the RHN mass matrix given by ${\cal M}_N=f_R v_R$ is non-diagonal, with $v_R$ the vacuum expectation value (VEV) of the $\Delta^0_R$ field. We can now ask if $V$ diagonalizes the RHN mass matrix, are the rotation angles and phases in this matrix $V$ observable? We claim that in the LRSM, due to the presence of the $W_R$ interaction, the $V$ matrix is unambiguously observable since it rotates the leptonic fields in the $W_R$ interaction. This also contributes to the CP phase in the leptonic sector as well as to leptogenesis.  Our goal in this paper is to show how to measure the rotation angles and CP phases in $V$ for the RHN sector at colliders. Since we also expect this phase to contribute to leptogenesis, we want to display the connection between collider information and leptogenesis requirements in the hope that we can test leptogenesis for this particular model at colliders.

\section{Same-sign charge asymmetries}
\label{sec:SSCA}

We first show how SS dilepton signals arise from the seesaw Lagrangian~\eqref{eqn:lagLRSM} in the LRSM. For this purpose, we start with the RH charged currents in the leptonic sector of LRSM, which originate from the second term in Eq.~\eqref{eqn:lagLRSM} and are explicitly given by
\begin{eqnarray}
\label{eqn:LRHN}
{\cal L}_I^{\rm LRSM} & \ \supset \ & \frac{g_R}{\sqrt{2}}\left[\bar{\ell}_{R,\,\alpha} \gamma_\mu N_{\alpha} W^{-,\mu}_R ~+~ \overline{N}_\alpha \gamma_\mu  \ell_{R,\alpha}W^{+,\mu}_R\right] \nonumber \\
& \ = \ & \frac{g_R}{\sqrt{2}}\left[\bar{\ell}_{R,\,\alpha} \gamma_\mu N_{\alpha} W^{-,\mu}_R ~+~ {N^T_\alpha} C^{-1}\gamma_\mu  \ell_{R,\alpha}W^{+,\mu}_R\right] \,,
\end{eqnarray}
where %$g_R$ is the gauge coupling for the $SU(2)_R$ group in the LRSM, and
$C$ is the charge conjugation operator. Note that the first term on the RHS is responsible for OS dileptons whereas the second term  uses the Majorana condition for RHNs and is the reason why SS dilepton states appear. To present our discussion involving heavy RHN oscillation, we assume that there are two quasi-degenerate RHNs\footnote{There is often a misconception that having two quasi-degenerate RHNs means that they are a pseudo-Dirac pair but this need not be so, depending on the relative CP phase between them. For example, consider a mass matrix of RHNs that have positive eigenvalues after diagonalization; in this case the RHN pair is not pseudo-Dirac. Only if they have opposite CP phases (i.e. one eigenvalue positive and one negative with similar magnitude), that means a pseudo-Dirac pair.} carrying the lepton flavors $\alpha = e,\,\mu$, and the third one $N_\tau$ to be much heavier so that it plays no role in our discussion. Thus, the LRSM scenario we are considering can be viewed as an effective theory with only two RHNs. The lighter flavor states $N_{e,\,\mu}$ are related to the two mass eigenstates $N_{1,2}$ via
\begin{eqnarray}
\label{eqn:UR}
\left( \begin{matrix} N_e \\ N_\mu \end{matrix} \right) \ = \
U_R \left( \begin{matrix} N_1 \\ N_2 \end{matrix} \right) \ = \
\left( \begin{matrix}
\cos\theta_R & \sin\theta_R e^{-i\delta_R} \\
-\sin\theta_R e^{i\delta_R} & \cos\theta_R
\end{matrix} \right)
\left( \begin{matrix} N_1 \\ N_2 \end{matrix} \right) \,.
\end{eqnarray}
Then we can write down explicitly all the terms in the Lagrangian~(\ref{eqn:LRHN}), which dictate the production of RHNs  from $W_R$ boson decay:
\begin{eqnarray}
\mathcal{L} & \ = \ &
\frac{g_R}{\sqrt{2}}\bigg[
\bar{e}_R\gamma_\mu \Big(c_\theta N_1+s_\theta e^{-i\delta_R}N_2 \Big) W^{-,\mu}_R
+ \bar{\mu}_R\gamma_\mu \Big( -s_\theta e^{i\delta_R} N_1+c_\theta N_2 \Big) W^{-,\mu}_R \bigg] ~+~ {\rm H.c.} \nonumber \\ &&
\end{eqnarray}
We assume that the two RHNs $N_{e,\,\mu}$ are lighter than the $W_R$ boson, i.e. $M_{N_{1,2}} < M_{W_R}$,\footnote{This choice is also preferred by vacuum stability arguments~\cite{Mohapatra:1986pj, Maiezza:2016bzp, Dev:2018foq}.} with $M_{N_{1,2}}$ the mass eigenvalues for the two RHNs and $M_{W_R}$ the $W_R$ mass. so that they can be produced on-shell from $W_R$ decay.  As a result of the Majorana nature, the heavy RHNs $N_{\alpha}$ decay into both positively and negatively charged leptons, when they are produced at colliders. This can happen either through an off-shell $W_R$ boson, i.e. the three-body decays $N_\alpha \to \ell_\alpha^\pm q\bar{q}'$ (with $q,q'$ being SM quark jets), or through the heavy-light neutrino mixing, i.e. the two-body decays $N_\alpha \to \ell_\beta^\pm W^\mp$. Note that in the latter case the flavor index $\beta = e,\,\mu,\,\tau$ and it might differ from the flavor of decaying RHN, i.e. $\alpha \neq \beta$. The SSCA, which is a CPV effect, arises from the propagation of RHN mass eigenstates, and can {\it in principle} come from either the three-body or two-body decays of RHNs. However, we will see in the following subsections that in the parameter space of interest, the contribution from the Yukawa coupling mediated two-body decays can be neglected, and we can only see the CP-induced SSCAs from $W_R$-mediated three-body decays.

\subsection{Effect of three-body decays $N_\alpha \to \ell_\alpha^\pm q \bar{q}'$}

In this section, we consider only the contributions from the $W_R$-boson mediated three-body decays $N_\alpha \to \ell_\alpha^\pm q\bar{q}'$. Following the notation of Ref.~\cite{DDM}, we denote by $A (\ell_\alpha^\pm\ell_\beta^\pm,t)$ the time evolution of the amplitudes for the SS dilepton events $W_R \to \ell_\alpha^\pm (N_\beta \to \ell_\beta^\pm q\bar{q}')$ at high-energy colliders. Then the flavor dependence is as follows:
\begin{eqnarray}
\label{eqn:SSee}
A (e^\pm e^\pm,t) & \ = \ &
\cos^2\theta_R \exp \left\{-iE_{N_1} t -\frac{1}{2} \Gamma_{N_{1}}t \right\} \nonumber \\
&& + \sin^2\theta_R \exp \left\{ \pm 2i\delta_R -iE_{N_2} t -\frac{1}{2} \Gamma_{N_2}t \right\} \,,  \\
\label{eqn:SSmm}
A (\mu^\pm \mu^\pm,t) & \ = \ &
\sin^2\theta_R \exp \left\{ \mp 2i\delta_R -iE_{N_1} t -\frac{1}{2} \Gamma_{N_1} t \right\} \nonumber \\
&& + \cos^2\theta_R  \exp \left\{ -iE_{N_2} t -\frac{1}{2} \Gamma_{N_2} t \right\} \,, \\
A (e^\pm \mu^\pm,t) \ = \ A (\mu^\pm e^\pm,t) & \ = \ & -\sin\theta_R \cos\theta_R
\left[ \exp \left\{ \mp i\delta_R -iE_{N_1} t -\frac{1}{2} \Gamma_{N_1} t \right\} \right.
\nonumber \\
&& \left. - \exp \left\{ \pm i\delta_R -i E_{N_2} t -\frac{1}{2} \Gamma_{N_1} t \right\} \right] \,,
\label{eqn:SSem}
\end{eqnarray}
where $\Gamma_{N_{1,2}}$ are the total decay widths of the two mass eigenstates $N_{1,2}$, and $E_{N_{1,2}}$ the energies of $N_{1,2}$ at colliders.  Note that the dependence of the amplitudes in Eqs.~(\ref{eqn:SSee})--(\ref{eqn:SSem}) for the positively-charged and negatively-charged leptons on the CP phase $\delta_R$ is different, which is the origin of the SSCAs defined below in Eqs.~(\ref{eqn:Aab}) and (\ref{eqn:RCP}). Integrating over time $\int_0^\infty {\rm d} t |A(t)|^2$, the SS dilepton event numbers are proportional to the factors below:
\begin{eqnarray}
\label{eqn:Nee}
{\cal N} (e^\pm e^\pm) & \ \propto \ & \Gamma_{\rm avg}
\left[\frac{\cos^4\theta_R}{\Gamma_{N_1}} + \frac{\sin^4\theta_R}{\Gamma_{N_2}} +
\frac{\sin^22\theta_R (\Gamma_{\rm avg} \cos 2\delta_R \pm \Delta E_N \sin 2\delta_R)}{2 \left( \Gamma_{\rm avg}^2 + (\Delta E_N)^2 \right)} \right] \,, \\
\label{eqn:Nmm}
{\cal N} (\mu^\pm \mu^\pm) & \ \propto \ & \Gamma_{\rm avg}
\left[\frac{\sin^4\theta_R}{\Gamma_{N_1}} + \frac{\cos^4\theta_R}{\Gamma_{N_2}} +
\frac{\sin^22\theta_R (\Gamma_{\rm avg} \cos 2\delta_R \pm \Delta E_N \sin 2\delta_R)}{2 \left( \Gamma_{\rm avg}^2 + (\Delta E_N)^2 \right)} \right] \,, \\
\label{eqn:Nem}
{\cal N} (e^\pm \mu^\pm) & \ = \ & {\cal N} (\mu^\pm e^\pm) \ \propto \
\frac14 \Gamma_{\rm avg}  \sin^22\theta_R
\left[\frac{1}{\Gamma_{N_1}} + \frac{1}{\Gamma_{N_2}} -
\frac{2 (\Gamma_{\rm avg} \cos 2\delta_R \pm \Delta E_N \sin 2\delta_R)}{\Gamma_{\rm avg}^2 + (\Delta E_N)^2} \right] \,, \nonumber \\
\end{eqnarray}
where $\Gamma_{\rm avg}\equiv (\Gamma_{N_1}+\Gamma_{N_2})/2$ is the average total width of RHNs, and $\Delta E_N \equiv E_{N_2} - E_{N_1}$ the energy difference of the two RHN mass eigenstates at high-energy colliders. Note that the SSCAs do not depend on the RHN energies $E_{N_{1,\,2}}$ but only on the energy difference $\Delta E_N$, which can be estimated depending on whether the RHNs $N_{1,\,2}$ are non-relativistic or relativistic (see Sec.~\ref{sec:coh}). Eqs.~(\ref{eqn:Nee})--(\ref{eqn:Nem}) can be simplified in the limit of $\Gamma_{N_1} = \Gamma_{N_2}$, which is a good approximation for a pair of quasi-degenerate RHNs in the parameter space of interest. Then the flavor-dependent SS dilepton event numbers are proportional to the factors
\begin{eqnarray}
\label{eqn:Ree}
&& R (e^\pm e^\pm) \ \simeq \ R(\mu^\pm \mu^\pm) \ \simeq \
\frac12 - R (e^\pm \mu^\pm) \, , \\
\label{eqn:Rem}
&& R (e^\pm \mu^\pm) \ = \ R (\mu^\pm e^\pm) \ \simeq \
\frac{1}{4} \sin^22\theta_R \left( 1 - \frac{\cos2\delta_R \pm x \sin2\delta_R}{1+x^2} \right) ,
\end{eqnarray}
with $x \equiv \Delta E_N / \Gamma_{\rm avg}$. The $R$ factors are normalized to follow the sum rule
\begin{eqnarray}
\sum_{\alpha,\beta \, = \, e,\,\mu} R (\ell_\alpha^\pm \ell_\beta^\pm) \ = \ 1 \,.
\end{eqnarray}

\subsection{Effect of two-body decays $N_\alpha \to \ell_\beta^\pm W^\mp$}
\label{sec:two-body-decay}

In the generic LRSM, the Dirac neutrino mass matrix ${\cal M}_D$ matrix is correlated with the charged lepton masses. However, the neutrino sector and charged lepton sector can be decoupled from each other if we choose the VEV configuration for the bidoublet $\Phi$ fields to be $\langle \Phi \rangle = {\rm diag} (\kappa, 0)$
%set one of the vacuum expectation values (VEVs) $\kappa'$ of the bidoublet field $\Phi$ to be zero in the scalar sector
(see e.g. Refs.~\cite{Deshpande:1990ip, Zhang:2007da, Dev:2016dja} for more details on the scalar sector of LRSM). This choice has the advantage that we do not have to worry about simultaneously fitting the  neutrino and charged lepton masses. In the specific type-I seesaw case with only two RHNs, one of the active neutrinos is massless, i.e. $m_1 = 0$ for normal hierarchy (NH) and $m_3 = 0$ for inverted hierarchy (IH), and ${\cal M}_D$ can be parameterized in the Casas-Ibarra form~\cite{Casas:2001sr}:  %which is equivalent to the Nemev\v{s}ek-Senjanovi\'{c}-Tello form in the LRSM~\cite{Nemevsek:2012iq}
\begin{eqnarray}
\label{eqn:MD}
{\cal M}_D = i U \widehat{m}_{\nu}^{1/2} {\cal O} {\cal M}_{N}^{1/2} \,,
\label{eqn:CI}
\end{eqnarray}
where $U$ is the PMNS mixing matrix for light neutrinos, and $\widehat{m}_{\nu} = {\rm diag} \{ m_1,\, m_2,\, m_3 \}$ is the diagonal mass matrix for the active neutrinos. The analytic formula of the square root of the $2\times 2$ RHN matrix ${\cal M}_N^{1/2}$ can be found in Appendix~\ref{sec:analytic2},\footnote{For the sake of completeness, we have also included the formula for ${\cal M}_N^{1/2}$ in the three RHN case in Section~\ref{sec:analytic3}, with the third RHN $N_\tau$ not mixing with the first two $N_{e,\,\mu}$.} and ${\cal O}$ is an arbitrary complex matrix in the form of
\begin{eqnarray}
\label{eqn:Ocal}
{\cal O} \ = \ %\left\{
\begin{cases}
\begin{pmatrix}
0 & 0 \\
\cos\zeta & \sin\zeta  \\
-\sin\zeta & \cos\zeta
\end{pmatrix}  & \text{ for NH} \,,  \\
\begin{pmatrix}
\cos\zeta & \sin\zeta  \\
-\sin\zeta & \cos\zeta \\
0 & 0
\end{pmatrix} & \text{ for IH} \,,
\end{cases} \;\;\; \text{with} \;\;\;
{\cal O} {\cal O}^{\sf T}  \ = \
\begin{cases}
\begin{pmatrix}
0 & 0 \\ 0 & {\bf 1}_{2\times2}
\end{pmatrix} &  \text{ for NH} \,,  \\
%&& {\cal O}  {\cal O}^{\sf T} =
\begin{pmatrix}
{\bf 1}_{2\times2} & 0 \\ 0 & 0
\end{pmatrix} &  \text{ for IH} \,,
\end{cases}
\end{eqnarray}
where $\zeta$ is a free parameter, either real or complex, and $v_{\rm EW}$ is  the electroweak VEV. If $\zeta$ is complex, having large Im$\zeta$ will significantly enhance the Yukawa couplings $y = {\cal M}_D/v_{\rm EW}$ (see e.g. Refs.~\cite{Dev:2014xea, Bambhaniya:2016rbb}), as required by leptogenesis in the TeV-scale LRSM (see Section~\ref{sec:leptogenesis}). For simplicity, we assume $\zeta$ is purely imaginary, with $\sin\zeta = i \sinh ({\rm Im}\zeta)$ and $\cos\zeta = \cosh ({\rm Im} \zeta)$. If ${\rm Im}\zeta>2$, which is preferred by leptogenesis (see Fig.~\ref{fig:asymmetry1}), we have
\begin{eqnarray}
\label{eqn:zeta}
|\sin\zeta| \simeq \cos\zeta \gg 1 \,.
\end{eqnarray}
Then the ${\cal M}_D$ matrix elements in Eq.~\eqref{eqn:CI} for the NH case can be explicitly written as follows:
\begin{eqnarray}
({\cal M}_D)_{\alpha 1} & \ = \ &
i \cos\zeta \left[ \left( {\cal M}_N^{1/2} \right)_{11} \sqrt{m_2} U_{\alpha 2} +
\left( {\cal M}_N^{1/2} \right)_{12} \sqrt{m_3} U_{\alpha 3}  \right] \nonumber \\
&& + i \sin\zeta \left[ \left( {\cal M}_N^{1/2} \right)_{12} \sqrt{m_2} U_{\alpha 2} -
\left( {\cal M}_N^{1/2} \right)_{11} \sqrt{m_3} U_{\alpha 3}  \right] \,, \\
({\cal M}_D)_{\alpha 2} & \ = \ &
i \cos\zeta \left[ \left( {\cal M}_N^{1/2} \right)_{12} \sqrt{m_2} U_{\alpha 2} +
\left( {\cal M}_N^{1/2} \right)_{22} \sqrt{m_3} U_{\alpha 3}  \right] \nonumber \\
&& + i \sin\zeta \left[ \left( {\cal M}_N^{1/2} \right)_{22} \sqrt{m_2} U_{\alpha 2} -
\left( {\cal M}_N^{1/2} \right)_{12} \sqrt{m_3} U_{\alpha 3}  \right] \,,
\end{eqnarray}
where the flavor index $\alpha = e,\mu,\tau$. Using the relation in Eq.~(\ref{eqn:MNhalf22}), it is straightforward to prove that
\begin{eqnarray}
\frac{|y_{\mu 1}|^2}{|y_{e 1}|^2} \ \simeq \
\frac{|y_{\mu 2}|^2}{|y_{e 2}|^2} \,, \quad
\frac{|y_{\tau 1}|^2}{|y_{e 1}|^2} \ \simeq \
\frac{|y_{\tau 2}|^2}{|y_{e 2}|^2} \,.
\end{eqnarray}
These relations are also true for the IH case, and imply that
\begin{eqnarray}
{\rm BR} (N_e \to \ell_\alpha^\pm W^\mp ) \ = \
{\rm BR} (N_\mu \to \ell_\alpha^\pm W^\mp )\, , \quad
\text{with } \alpha = e,\, \mu,\, \tau \,.
\end{eqnarray}

Defining the rescaled flavor-dependent BR for the two-body decays
\begin{eqnarray}
B_{\alpha\beta} \ \equiv \ \frac{\Gamma (N_\alpha \to \ell_\beta^\pm W^\mp)}{\Gamma (N_\alpha \to \sum_\beta \ell_\beta^\pm W^\mp)} \,,
\end{eqnarray}
if the two-body decays of RHNs dominate over the three-body decays, then the SS dilepton event numbers  for the flavor $\ell_\alpha^\pm \ell_\beta^\pm$ combinations are respectively proportional to the following factors:
\begin{eqnarray}
e^\pm e^\pm: &&
R (e^\pm e^\pm) B_{ee} + R (e^\pm \mu^\pm) B_{\mu e} \ = \
\frac12 B_{ee} \,. \\
\mu^\pm e^\pm: &&
R (\mu^\pm \mu^\pm) B_{\mu\mu} + R (\mu^\pm e^\pm) B_{e\mu} \ = \
\frac12 B_{e\mu} \,. \\
e^\pm \mu^\pm: &&
R (e^\pm \mu^\pm) B_{\mu\mu} + R (e^\pm e^\pm) B_{e\mu} \ = \
\frac12 B_{e\mu} \,. \\
\mu^\pm \mu^\pm: &&
R (\mu^\pm \mu^\pm) B_{\mu e} + R (\mu^\pm e^\pm) B_{ee} \ = \
\frac12 B_{ee} \,. \\
e^\pm \tau^\pm: &&
R (e^\pm e^\pm) B_{e\tau} + R (e^\pm \mu^\pm) B_{\mu\tau} \ = \
\frac12 B_{e\tau} \,. \\
\mu^\pm \tau^\pm: &&
R (\mu^\pm e^\pm) B_{e\tau} + R (\mu^\pm \mu^\pm) B_{\mu\tau} \ = \
\frac12 B_{e\tau} \,.
\end{eqnarray}
From this result, we can see that the CPV effects are cancelled out in the two-body decays of both $N_e$ and $N_\mu$, at least in the parameter space of interest in this paper.

\subsection{Combining three- and two-body decays}

After taking into account both three- and two-body decays of $N_{\alpha}$ of the RHNs, we define the SSCA ${\cal A}_{\alpha\beta}$ at proton-proton colliders as
\begin{align}
\label{eqn:Aab}
{\cal A}_{\alpha\beta}  \ \equiv \
\frac{\sigma (pp \to W_R^+) {\cal R} (\ell_\alpha^+ \ell_\beta^+) -
\sigma (pp \to W_R^-) {\cal R} (\ell_\alpha^- \ell_\beta^-)}
{\sigma (pp \to W_R^+) {\cal R} (\ell_\alpha^+ \ell_\beta^+) +
\sigma (pp \to W_R^-) {\cal R} (\ell_\alpha^- \ell_\beta^-)} \,,
\end{align}
which depends on the factor
\begin{eqnarray}
\label{eqn:Rab}
{\cal R} (\ell_\alpha^\pm \ell_\beta^\pm) \ \simeq \ {\rm BR}_g
R(\ell_\alpha^\pm \ell_\beta^\pm) + \frac{1}{4}
{\rm BR}_y B_{\alpha\beta} \,,
\end{eqnarray}
where the first and second terms are respectively the three-body (gauge-mediated) and two-body (Yukawa-mediated) decay contributions, with the respective branching ratios (BRs):
\begin{eqnarray}
\label{eqn:BRy}
{\rm BR}_g \ \equiv \ \frac{\Gamma (N_\alpha \to \ell q\bar{q}')}{\Gamma (N_\alpha \to \ell q\bar{q}') + 2\Gamma (N_\alpha \to \sum_\beta \ell_\beta^\pm W^\mp)} \,, \\
\label{eqn:BRy}
{\rm BR}_y \ \equiv \ \frac{2\Gamma (N_\alpha \to \sum_\beta \ell_\beta^\pm W^\mp)}{\Gamma (N_\alpha \to \ell q\bar{q}') + 2\Gamma (N_\alpha \to \sum_\beta \ell_\beta^\pm W^\mp)} \,.
\end{eqnarray}
%The prefactor $(1-{\rm BR}_y)$ in the first term of Eq.~(\ref{eqn:Rab}) is the BR for the three-body decays.
The extra factor of $1/2$ in the second term of Eq.~(\ref{eqn:Rab}) and the factor of $2$ in Eq.~(\ref{eqn:BRy}) account for the fact that RHNs decay both into charged leptons and active neutrinos through the Yukawa couplings and
\begin{eqnarray}
\Gamma (N \to \ell W) \ \simeq \ \Gamma (N \to \nu Z) + \Gamma (N \to \nu h)
\end{eqnarray}
in the large RHN mass limit due to the Goldstone equivalence theorem.  Although the two-body decays do not contribute to the CP-induced SSCA, they will affect the measurement of SSCAs at high-energy colliders, when the corresponding branching ratio ${\rm BR}_y$ is sizable. In particular, in the limit of ${\rm BR}_y \gg {\rm BR}_g$, the CP-induced SSCA will be highly suppressed.

As long as the CP phase $\delta_R \neq 0, \pi/2, \pi$ and three-body decay BRs are sizable, the SSCAs ${\cal A}_{\alpha\beta}$ can be induced by CPV in the heavy neutrino sector. Furthermore, as long as the RHN mixing angle $\theta_R \neq 0$, we can have the $e^\pm \mu^\pm$ events from RHN decay, and the asymmetry ${\cal A}_{e\mu}^{}$ does not depend on the RHN mixing angle $\theta_R$, as the factor $\sin^22\theta_R$ cancels out in the ratio (\ref{eqn:Aab}).\footnote{The RHN mixing angle $\theta_R$ affects the factors $B_{\alpha\beta}$ in Eq.~(\ref{eqn:Rab}) via the Casas-Ibarra formula in Eq.~(\ref{eqn:MD}); however, in the parameter space of interest the resulting effects on the SSCAs and the dependence of ${\cal A}_{e\mu}$ on $\theta_R$ can be safely neglected.} In the limit of ${\rm BR}_g \gg {\rm BR}_y$, i.e. when the three-body decay BR is much larger than the two-body decay BR, we have the relation
\begin{eqnarray}
\label{eqn:relation}
{\cal A}_{e\mu}^{} (\delta_R) & \ = \ & {\cal A}_{ee,\,\mu\mu}^{} \left( \theta_R = \frac{\pi}{4}, \delta_R + \frac{\pi}{2} \right) \,.
\label{eqn:Aemu}
\end{eqnarray}
If the two-body decays are sizable, the SSCAs ${\cal A}_{ee}$ and ${\cal A}_{\mu\mu}$ might be different, as in general the ratios $B_{ee} \neq B_{e\mu}$, depending on the Yukawa coupling structure. Any significant violation of the relation~\eqref{eqn:Aemu} and ${\cal A}_{ee} \neq {\cal A}_{\mu\mu}$ would imply sizable mixing of a third RHN $N_\tau$ with $N_{e,\,\mu}$  or imply that the two-body decays of RHNs are important. In the limits of $\delta_R \to 0,\pi/2,\pi$ or in the two-body decay dominated regime, the factors $R (\ell_\alpha^+ \ell_\beta^+) = R (\ell_\alpha^- \ell_\beta^-)$ and Eq.~\eqref{eqn:Aab} reduces to
\begin{align}
{\cal A}^{(0)} \ = \ \frac{\sigma (pp \to W_R^+) - \sigma (pp \to W_R^-)}{\sigma (pp \to W_R^+)+\sigma (pp \to W_R^-)} \, ,
\label{eqn:Aab0}
\end{align}
which is non-zero purely due to the proton PDF effects, and to some extent, similar to the pure PDF-induced charge asymmetry in the SM $W^\pm$ production at $pp$ colliders:
\begin{eqnarray}
{\cal A}^{(W,\,0)} \ = \
\frac{\sigma (pp \to W^+ \to \ell^+ \nu) - \sigma (pp \to W^- \to \ell^- \bar\nu)}{\sigma (pp \to W^+ \to \ell^+ \nu) + \sigma (pp \to W^- \to \ell^- \bar\nu)} \, ,
\end{eqnarray}
which has been measured by both ATLAS~\cite{Aaboud:2016btc} and CMS~\cite{CMS:2015ois} collaborations in the $pp$ collisions, as well as in Pb-Pb collisions~\cite{Aad:2014bha, Chatrchyan:2012nt} at the LHC.

The flavor-dependent SSCA ${\cal A}_{\alpha\beta}^{}$ defined in Eq.~\eqref{eqn:Aab} can be used to probe directly the mixing angles and CP phases in the RHN sector. For instance, if a significant deviation of ${\cal A}_{\alpha\beta}^{}$ from the pure PDF-induced ${\cal A}^{(0)}$ is observed at LHC or future higher energy colliders, then it is expected that there are (at least) two quasi-degenerate RHNs and there is a  new CP phase $\delta_R \neq 0$, $\pi/2$ and $\pi$ in the RHN sector. The CP phase $\delta_R$ can be directly determined by the asymmetry ${\cal A}_{e\mu}^{}$, up to a two-fold ambiguity.
%%\footnote{To make sure there are $e^\pm \mu^\pm$ events from RHN production and decay, it is required that the mixing angle $\theta_R \neq 0$.}
Then we can determine the RHN mixing $\theta_R$ and remove the ambiguity of $\delta_R$ from the measurement of ${\cal A}_{ee}$ or ${\cal A}_{\mu\mu}$,  or at least narrow down its value to a limited range, as shown in Section~\ref{sec:Aab}.

\subsection{Same-sign ratio ${\cal R}_{\rm CP}$}

As discussed above, the CP-induced SSCA effect can be potentially smeared by the PDF effects. Therefore, it is useful to define the following ratio of ratios~\cite{Bray:2007ru}, which depends only on the CPV effects:
%\begin{eqnarray}
%{\cal R}_{\rm CP}^{(e)} & \ \equiv \ &
%\frac{\frac{\sigma (pp \to W_R \to e^+ e^+ jj)}{\sigma (pp \to W_R \to e^+ \mu^+ jj)} - \frac{\sigma (pp \to W_R \to e^- e^- jj)}{\sigma (pp \to W_R \to e^- \mu^- jj)}}
%{\frac{\sigma (pp \to W_R \to e^+ e^+ jj)}{\sigma (pp \to W_R \to e^+ \mu^+ jj)} + \frac{\sigma (pp \to W_R \to e^- e^- jj)}{\sigma (pp \to W_R \to e^- \mu^- jj)}} \nonumber \\ & \ = \ &
%\frac{\frac{{\cal R} (e^+ e^+)}{{\cal R} (e^+ \mu^+) + {\cal R} (\mu^+ e^+)} - \frac{{\cal R} (e^- e^-)}{{\cal R} (e^- \mu^-) + {\cal R} (\mu^- e^-)}}{\frac{{\cal R} (e^+ e^+)}{{\cal R} (e^+ \mu^+) + {\cal R} (\mu^+ e^+)} + \frac{{\cal R} (e^- e^-)}{{\cal R} (e^- \mu^-) + {\cal R} (\mu^- e^-)}} \,, \\
%{\cal R}_{\rm CP}^{(\mu)} & \ \equiv \ &
%\frac{\frac{\sigma (pp \to W_R \to \mu^+ \mu^+ jj)}{\sigma (pp \to W_R \to e^+ \mu^+ jj)} - \frac{\sigma (pp \to W_R \to \mu^- \mu^- jj)}{\sigma (pp \to W_R \to e^- \mu^- jj)}}
%{\frac{\sigma (pp \to W_R \to \mu^+ \mu^+ jj)}{\sigma (pp \to W_R \to e^+ \mu^+ jj)} + \frac{\sigma (pp \to W_R \to \mu^- \mu^- jj)}{\sigma (pp \to W_R \to e^- \mu^- jj)}} \nonumber \\ & \ = \ &
%\frac{\frac{{\cal R} (\mu^+ \mu^+)}{{\cal R} (e^+ \mu^+) + {\cal R} (\mu^+ e^+)} - \frac{{\cal R} (\mu^- \mu^-)}{{\cal R} (e^- \mu^-) + {\cal R} (\mu^- e^-)}}{\frac{{\cal R} (\mu^+ \mu^+)}{{\cal R} (e^+ \mu^+) + {\cal R} (\mu^+ e^+)} + \frac{{\cal R} (\mu^- \mu^-)}{{\cal R} (e^- \mu^-) + {\cal R} (\mu^- e^-)}} \,.
%\end{eqnarray}
\begin{eqnarray}
\label{eqn:RCP}
{\cal R}_{\rm CP}^{(\ell)} & \ \equiv \ &
\left[ \frac{\sigma (pp \to W_R^+ \to \ell^+ \ell^+ jj)}{\sigma (pp \to W_R^+ \to e^+ \mu^+ jj)} - \frac{\sigma (pp \to W_R^- \to \ell^- \ell^- jj)}{\sigma (pp \to W_R^- \to e^- \mu^- jj)}\right] \nonumber \\
&& \times \left[\frac{\sigma (pp \to W_R^+ \to \ell^+ \ell^+ jj)}{\sigma (pp \to W_R^+ \to e^+ \mu^+ jj)} + \frac{\sigma (pp \to W_R^- \to \ell^- \ell^- jj)}{\sigma (pp \to W_R^- \to e^- \mu^- jj)}\right]^{-1} \label{eq:Rab} \\
& \ = \ & \left[\frac{{\cal R} (\ell^+ \ell^+)}{{\cal R} (e^+ \mu^+) + {\cal R} (\mu^+ e^+)} - \frac{{\cal R} (\ell^- \ell^-)}{{\cal R} (e^- \mu^-) + {\cal R} (\mu^- e^-)} \right] \nonumber \\
&& \times \left[ \frac{{\cal R} (\ell^+ \ell^+)}{{\cal R} (e^+ \mu^+) + {\cal R} (\mu^+ e^+)} + \frac{{\cal R} (\ell^- \ell^-)}{{\cal R} (e^- \mu^-) + {\cal R} (\mu^- e^-)} \right]^{-1} \,,
\end{eqnarray}
with $\ell = e,\, \mu$ and ${\cal R}(\ell\ell')$ defined in Eq.~\eqref{eqn:Rab}.
In the definitions above, the dependence of ratios of production cross sections on the proton PDFs are cancelled out,
%e.g.
%\begin{eqnarray}
%\frac{\sigma (pp \to e^\pm e^\pm jj)}{\sigma (pp \to e^\pm \mu^\pm jj)}
%& \ = \ &
%\frac{\sigma (pp \to W_R^\pm) {\rm BR} (W_R \to e^\pm e^\pm jj)}{\sigma (pp \to W_R^\pm) {\rm BR} (W_R \to e^\pm \mu^\pm jj)} \nonumber \\
%& \ = \ & \frac{{\cal R} (e^\pm e^\pm)}{{\cal R} (e^\pm \mu^\pm) + {\cal R} (\mu^\pm e^\pm)} \,,
%\end{eqnarray}
and we are left with only the CP-induced asymmetries in the ratios ${\cal R}_{\rm CP}^{(e,\,\mu)}$, encoded in the ${\cal R} (\ell_\alpha^\pm \ell_\beta^\pm)$ factors. In the absence of CP violation, i.e. for $\delta_R = 0$, $\pi/2$ or $\pi$, the factors ${\cal R} (\ell_\alpha^+ \ell_\beta^+) = {\cal R} (\ell_\alpha^- \ell_\beta^-)$ and the ratios ${\cal R}_{\rm CP}^{(e,\,\mu)} = 0$. Furthermore, in the limit of vanishing two-body decay contributions, $R (e^\pm e^\pm) = R(\mu^\pm \mu^\pm)$ (cf. Eq.~(\ref{eqn:Ree})) and ${\cal R}_{\rm CP}^{(e)} = {\cal R}_{\rm CP}^{(\mu)}$.

As long as we can collect enough SS events at the LHC and future high-energy colliders to have a sizable signal-to-background ratio, the ratios ${\cal R}_{\rm CP}^{(e,\,\mu)}$ can be used to directly probe the mixing angles and CP phases in the RHN sector. As  ${\cal R}_{\rm CP}^{(e,\,\mu)}$ do not depend on the proton PDFs, unlike the asymmetries ${\cal A}_{\alpha\beta}$, any deviation of ${\cal R}_{\rm CP}^{(e,\, \mu)}$ from zero will indicate the existence of CP violation in the RHN sector.

\subsection{Coherence conditions} \label{sec:coh}

If the heavy RHNs are non-relativistic at colliders, i.e. their energies $E_{N_{1,\,2}} \simeq M_{N_{1,\,2}}$, we need only to replace $x = \Delta E_N / \Gamma_{\rm avg}$ by $R = \Delta M_N / \Gamma_{\rm avg}$ in Eq.~(\ref{eqn:Rem}), with $\Delta M_N \equiv M_{N_2} - M_{N_1}$ the RHN mass splitting. If the RHN masses are much smaller than the $W_R$ boson mass, the RHNs are relativistic at high-energy colliders and the energy difference $\Delta E_N \simeq \Delta p_N$ with $\Delta p_N$ the momentum difference for the two RHNs. To observe RHN oscillation from the SSCA signals at colliders, we have to make sure the coherence conditions are satisfied~\cite{DDM, Akhmedov:2007fk}: (i) the two RHNs are coherent when they are produced, i.e. the uncertainty in their mass square $\sigma_{M_N^2}$ is greater than their actual mass square difference $\Delta M_N^2$, and (ii) the coherence is maintained until they decay into lighter particles, i.e. $\sigma_{x} / \delta v_g > 1/ \Gamma_N$, with $\sigma_x$ the RHN wave-packet size and $\delta v_g$ the group velocity difference of the two RHNs. These coherence conditions impose upper bounds on the RHN mass splitting~\cite{Akhmedov:2007fk}, depending on how the RHNs are produced and decay at colliders. For the production of RHNs from $W_R$ decay, the first coherence condition leads to
\begin{eqnarray}
\label{eqn:coherence}
\Delta M_N^2 \ \lesssim \ 2\sqrt2 E_N \Gamma_{W_R} \, ,
\end{eqnarray}
with $\Gamma_{W_R}$ being the total $W_R$ decay width. For TeV-scale $W_R$, the condition~\eqref{eqn:coherence} requires that $\Delta M_N \lesssim {\cal O} (100 \, {\rm GeV})$ at LHC. The second condition provides a more stringent limit
\begin{eqnarray}
\label{eqn:coherence2}
\frac{(\Delta M_N^2)^2}{4\sqrt2 E_N^3} \ < \ \Gamma_{\rm avg} \, ,
\end{eqnarray}
which leads to the upper bound $\Delta M_N \lesssim {\cal O} ({\rm GeV})$. Throughout this paper we assume the two RHNs are quasi-degenerate with a mass splitting $\Delta M_N \ll {\rm GeV}$ and the coherence conditions are satisfied.

\section{Prospects at future high-energy colliders}
\label{sec:prospect}

The smoking-gun signals of a heavy $W_R$ boson are the SS dileptons plus jets without missing energy, i.e. $pp \to W_R^\pm \to \ell^\pm \ell^\pm jj$. The dominant SM backgrounds are mainly from diboson, $Z + {\rm jets}$, $t\bar{t}$ processes, or ``fake'' leptons, i.e. jets misidentified as leptons~\cite{Aaboud:2018spl, Mitra:2016kov} or lepton charge misidentified, with a fake rate of ${\cal O}(10^{-4})$ or smaller depending on the transverse momentum $p_T$ of jets and leptons~\cite{Alvarez:2016nrz}. The dominant uncertainty for measuring SSCAs comes from the PDFs, which is more significant than the reducible backgrounds from fake leptons and other processes.

\begin{figure}[!t]
  \centering
  \includegraphics[height=0.335\textwidth]{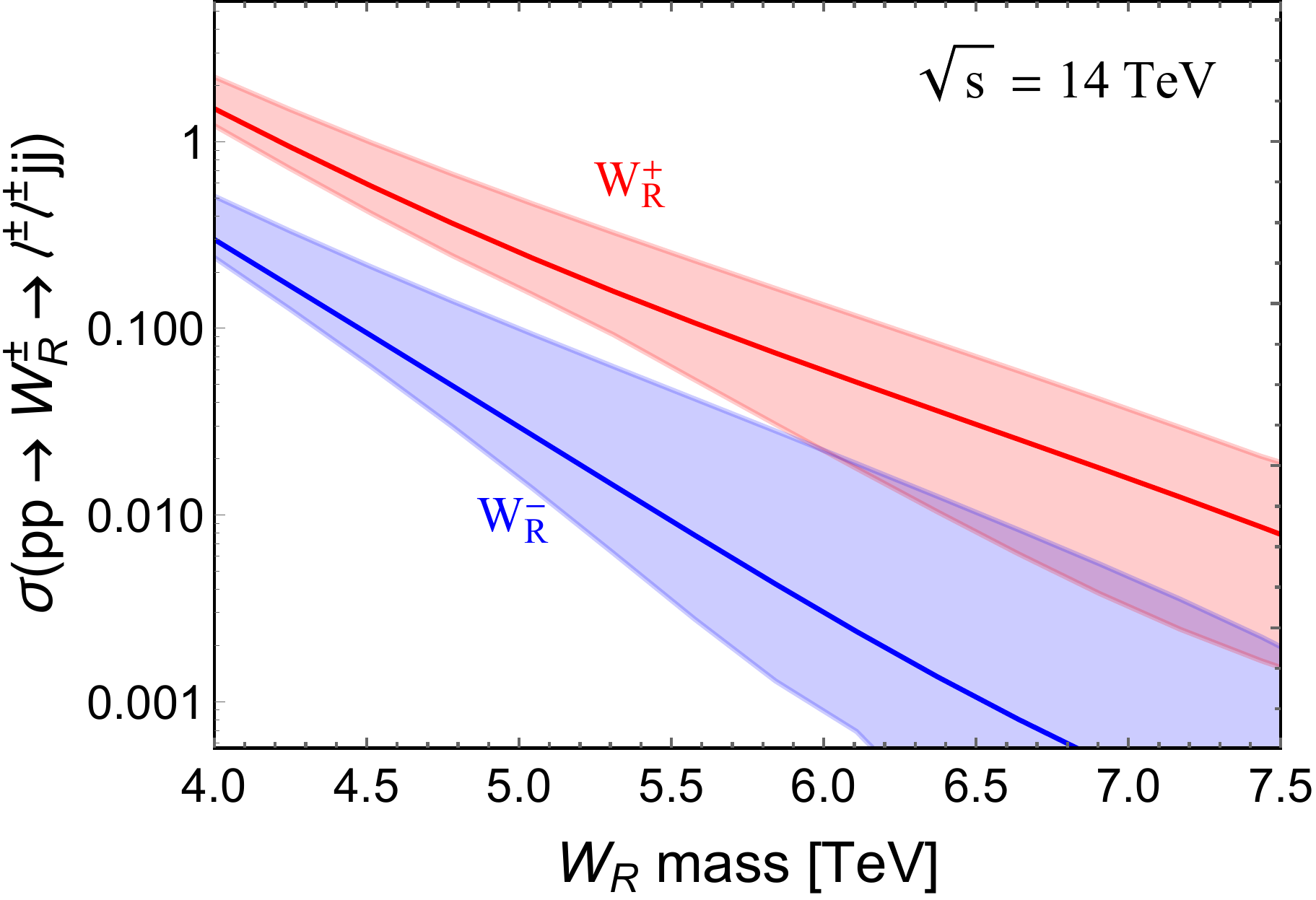} \vspace{5pt} \\
  \includegraphics[height=0.335\textwidth]{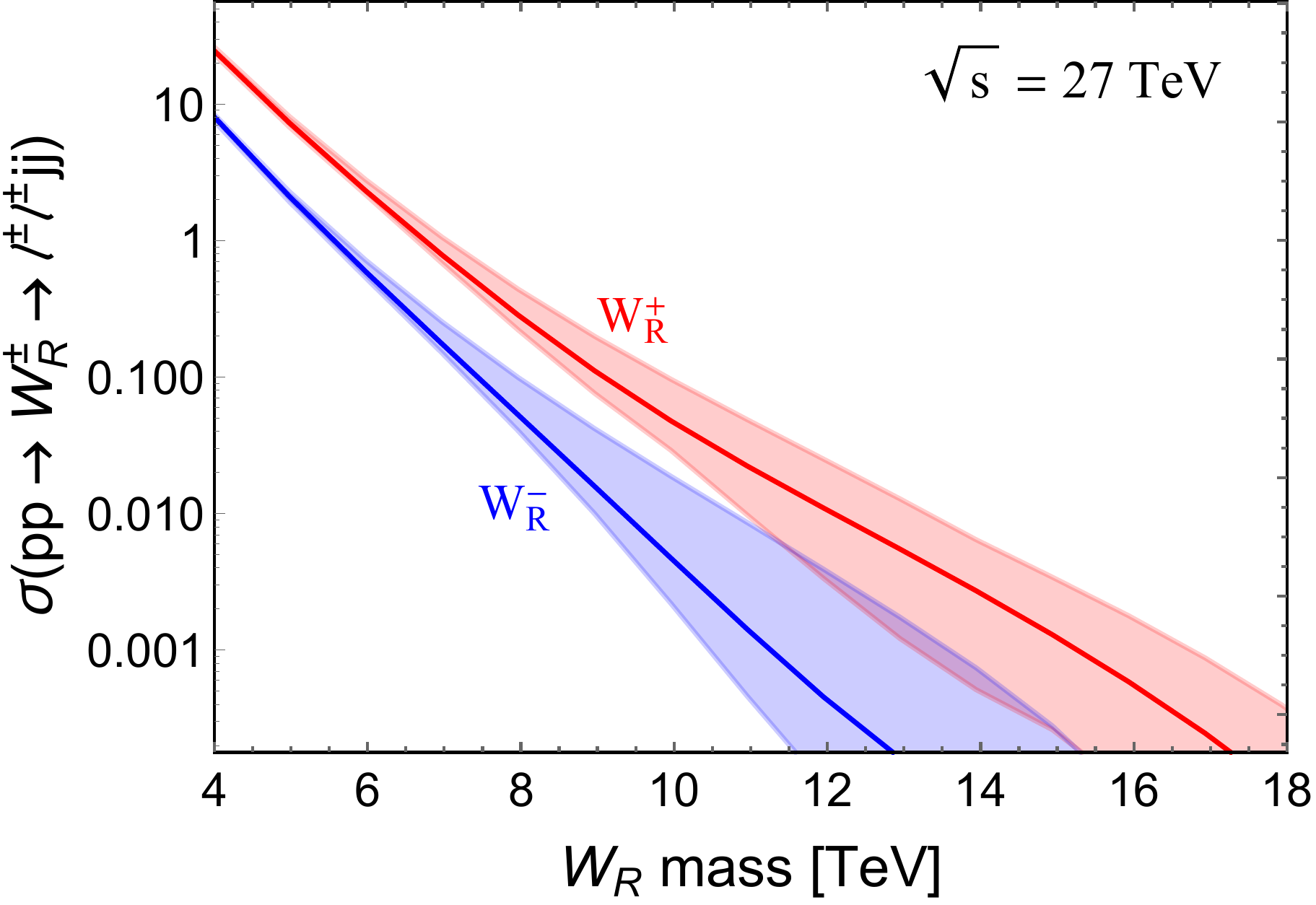}
  \includegraphics[height=0.335\textwidth]{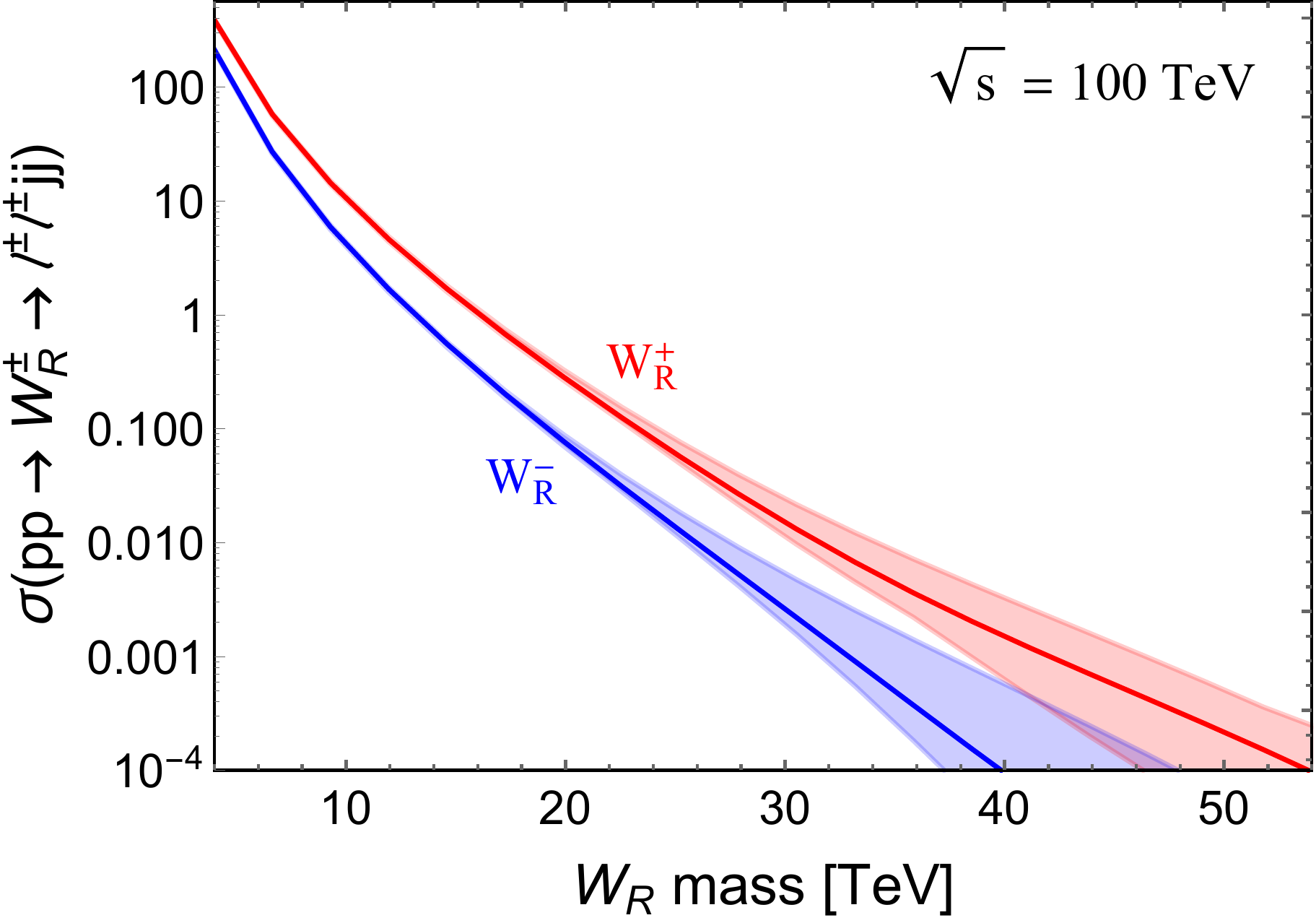}
  \caption{Parton-level, leading-order SS dilepton production cross section $\sigma (pp \to W_R^\pm) \times {\rm BR}(W_R^\pm \to \ell^\pm \ell^\pm jj)$ (with $\ell = e,\,\mu$) at the 14 TeV LHC (upper), 27 TeV HE-LHC (lower left) and 100 TeV FCC-hh (lower right), as a function of $W_R$ mass, with $g_R = g_L$. The solid red (blue) lines correspond to the central values for $W_R^+$ ($W_R^-$) production and the shaded bands are due to PDF uncertainties. }
  \label{fig:production}
\end{figure}

At $pp$ colliders, the production cross section for $W^+$ is larger than that for $W^-$, i.e. $\sigma(pp \to W^+X) > \sigma(pp \to W^-X)$, due to different PDFs for the $u$ and $d$ quarks in proton. Based on the same logic, we can produce more $W_R^+$ than $W_R^-$ at the LHC and future $pp$ colliders, which implies that even without any CPV in the RHN sector, we should expect
\begin{eqnarray}
\sigma (pp \to W_R^+ \to \ell^+ \ell^+ jj) \ > \
\sigma (pp \to W_R^- \to \ell^- \ell^- jj) \, .
\end{eqnarray}
As a result, a nonzero ${\cal A}^{(0)}_{\alpha\beta}$ [cf.~Eq.~\eqref{eqn:Aab0}] is induced purely by the proton PDF effects. The PDFs might suffer from large uncertainties, depending largely on the $W_R$ mass, or equivalently on the parton energy fraction $x_1 x_2 = \hat{s}/s \sim M^2_{W_R}/s$ (with $s$ the center-of-mass energy)~\cite{Mitra:2016kov}. Adopting the {\tt NNPDF3.1lo} PDF datasets with $\alpha_s (m_Z) = 0.118$~\cite{Ball:2017nwa} from {\tt LHAPDF6.1.6} library~\cite{Buckley:2014ana}, and using the LRSM model file in {\tt CalcHEP}~\cite{Belyaev:2012qa}, we estimated $\sigma (pp \to W_R^\pm) \times {\rm BR}(W_R^\pm \to \ell^+ \ell^+ jj)$ (with $\ell =e,\,\mu$) at the $\sqrt s=14$ TeV LHC, $\sqrt s=27$ TeV HE-LHC and $\sqrt s=100$ TeV FCC-hh. The results are presented in Fig.~\ref{fig:production}, with the central values for $W_R^+$ ($W_R^-$) shown as the solid red (blue) lines, and the shaded bands due to the PDF uncertainties.\footnote{{We repeated the simulations using {\tt CT14lo}~\cite{Dulat:2015mca} and {\tt MMHT2014lo}~\cite{Harland-Lang:2014zoa} PDF datasets, and found that although the production cross sections of $W_R^\pm$ vary to some extent, and there are uncertainties intrinsic to PDF extraction from data~\cite{AbdulKhalek:2019bux}, these issues do not affect qualitatively the main results of this paper.}} The higher-order QCD corrections depend on the center-of-mass energy $s$ and $W_R$ mass, and an average NLO $k-$factor of $1.1$ is included for the higher-order effects~\cite{Mitra:2016kov}, which is assumed to be the same for both $W_R^+$ and $W_R^-$.\footnote{The $k$-factors might be different for $W_R^+$ and $W_R^-$, and the difference might depend on the center-of-mass energy and $W_R$ mass~\cite{Richard}. In any case, we expect the difference to be small, just like the SM case, where $k_{\rm NLO}^{W^+}=1.17$ and $k_{\rm NLO}^{W^-}=1.21$~\cite{Catani:2010en}.} For the sake of concreteness, we have assumed the gauge couplings for $SU(2)_L$ and $SU(2)_R$ to be the same, i.e. $g_L = g_R$.
%, $M_{W_R} > M_{{1,2}}$ and the third RHN heavier than the $W_R$ boson.
It turns out that  the PDF uncertainties are more significant than the reducible backgrounds from the fake leptons and other processes; in other words, the cross section uncertainties are expected to be dominated by the proton PDFs, in particular when the ratio $M^2_{W_R}/s$ is large, as shown in Fig.~\ref{fig:production}.

\subsection{${\cal A}_{\alpha\beta}$} \label{sec:Aab}

\begin{figure}[!t]
  \centering
  \includegraphics[height=0.335\textwidth]{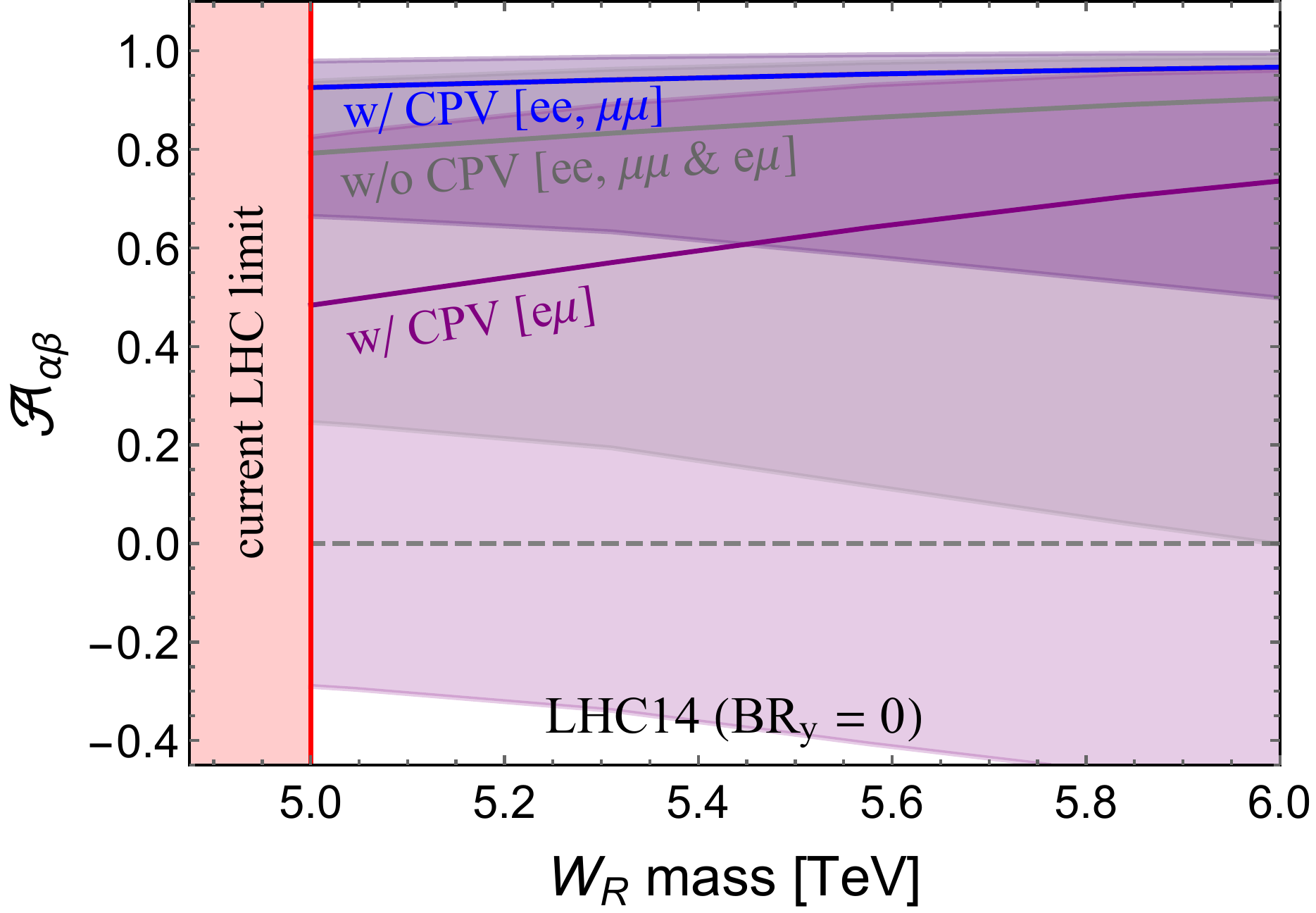}
  \includegraphics[height=0.335\textwidth]{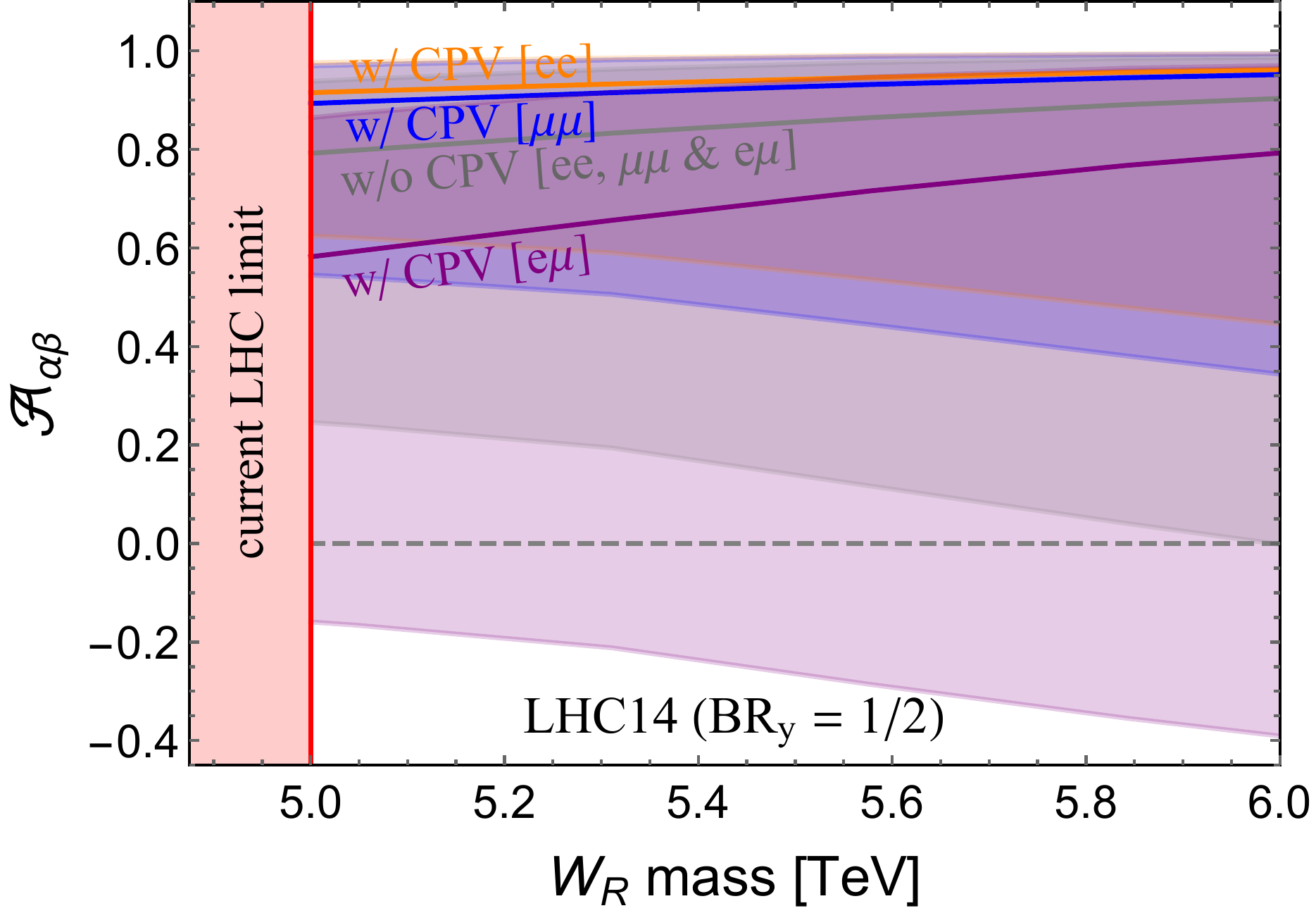}
  \includegraphics[height=0.335\textwidth]{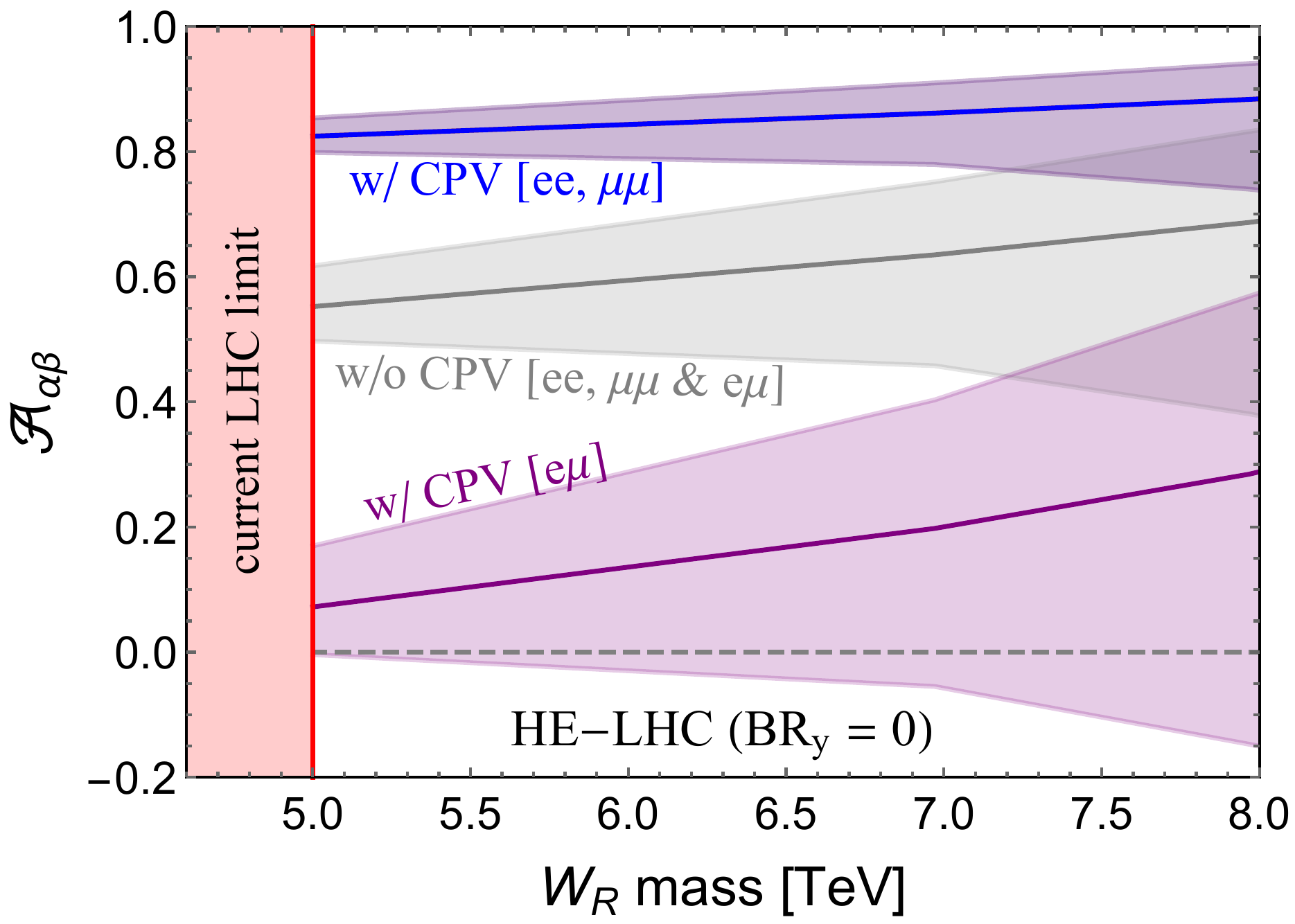}
  \includegraphics[height=0.335\textwidth]{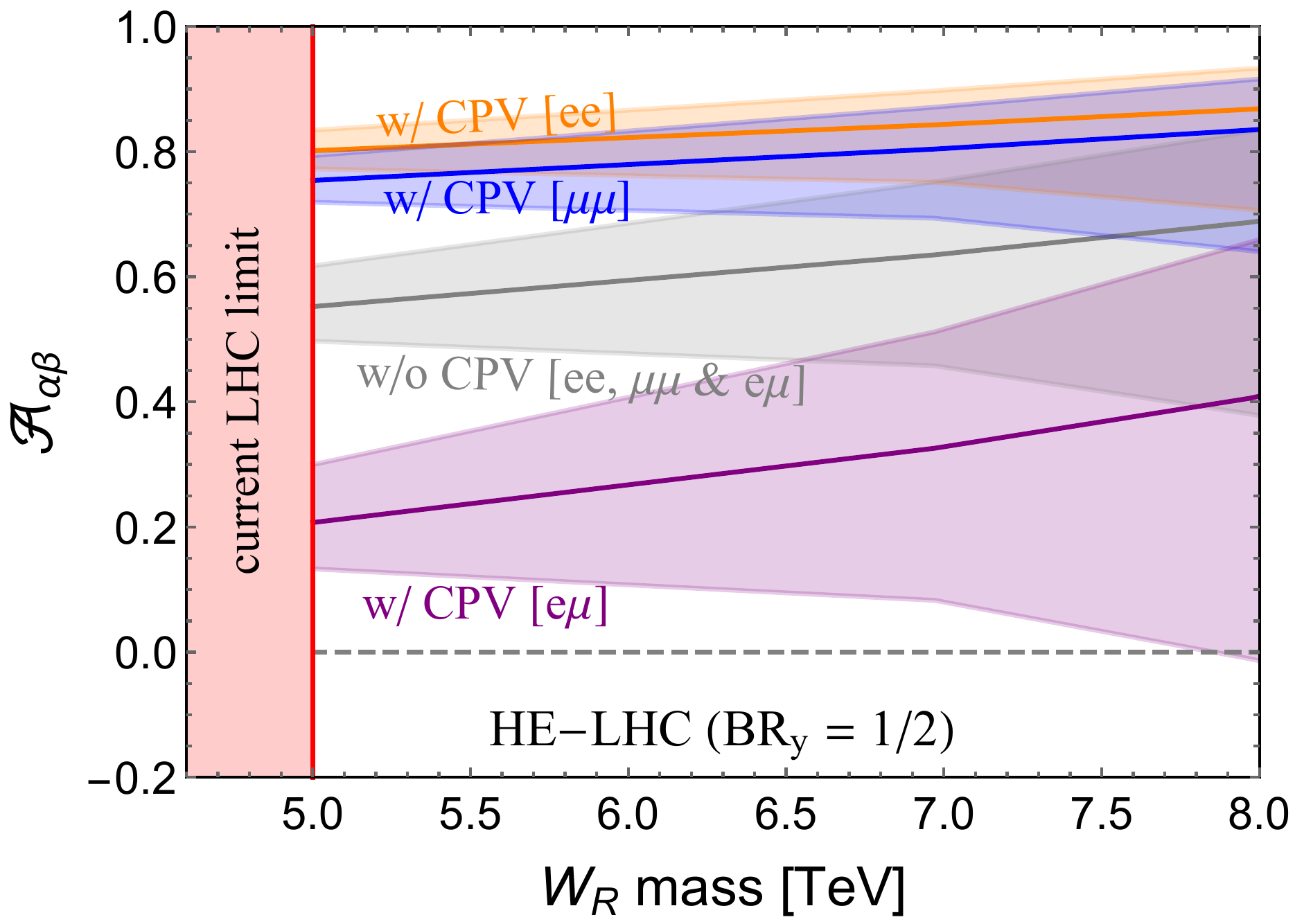}
  \includegraphics[height=0.335\textwidth]{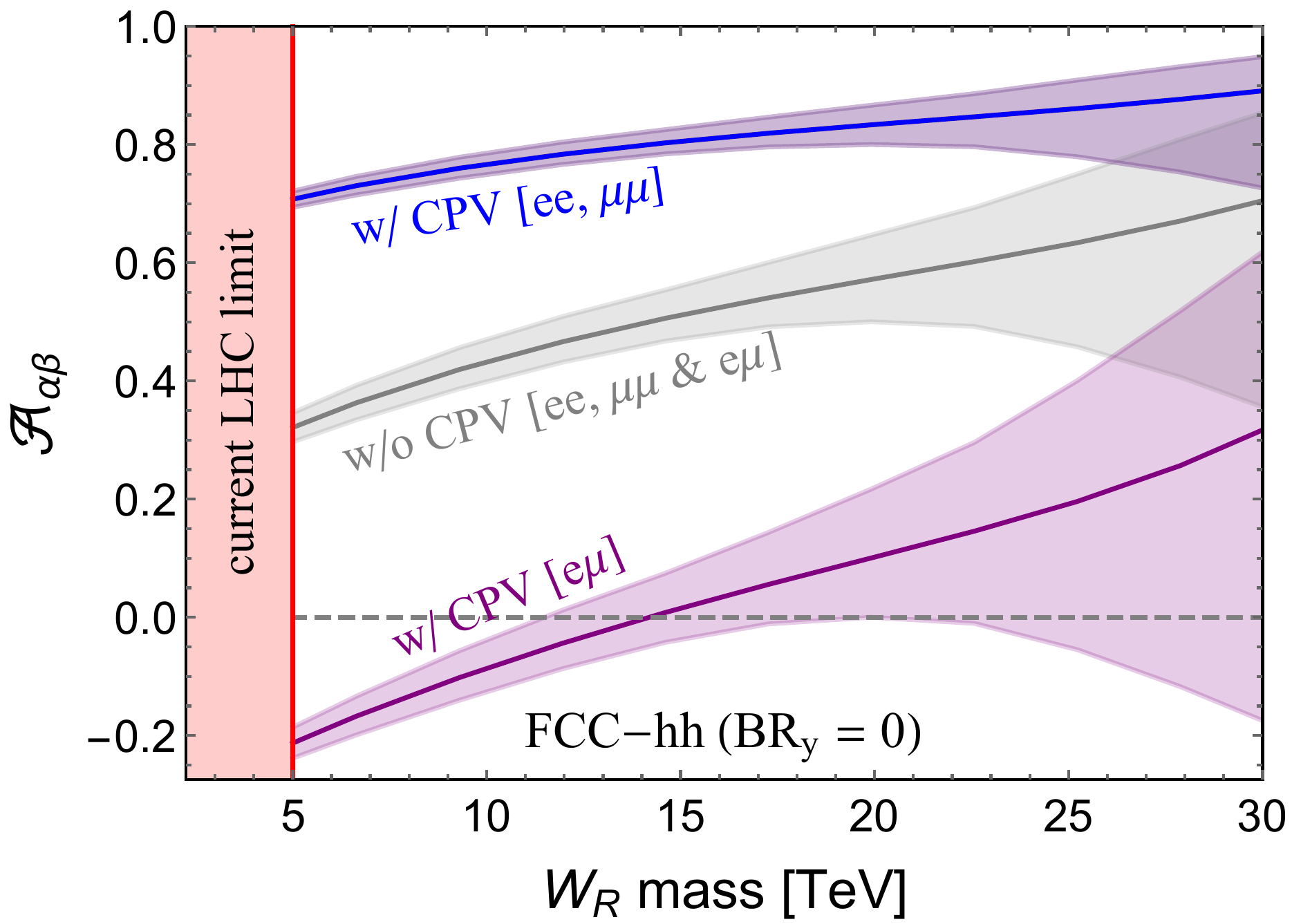}
  \includegraphics[height=0.335\textwidth]{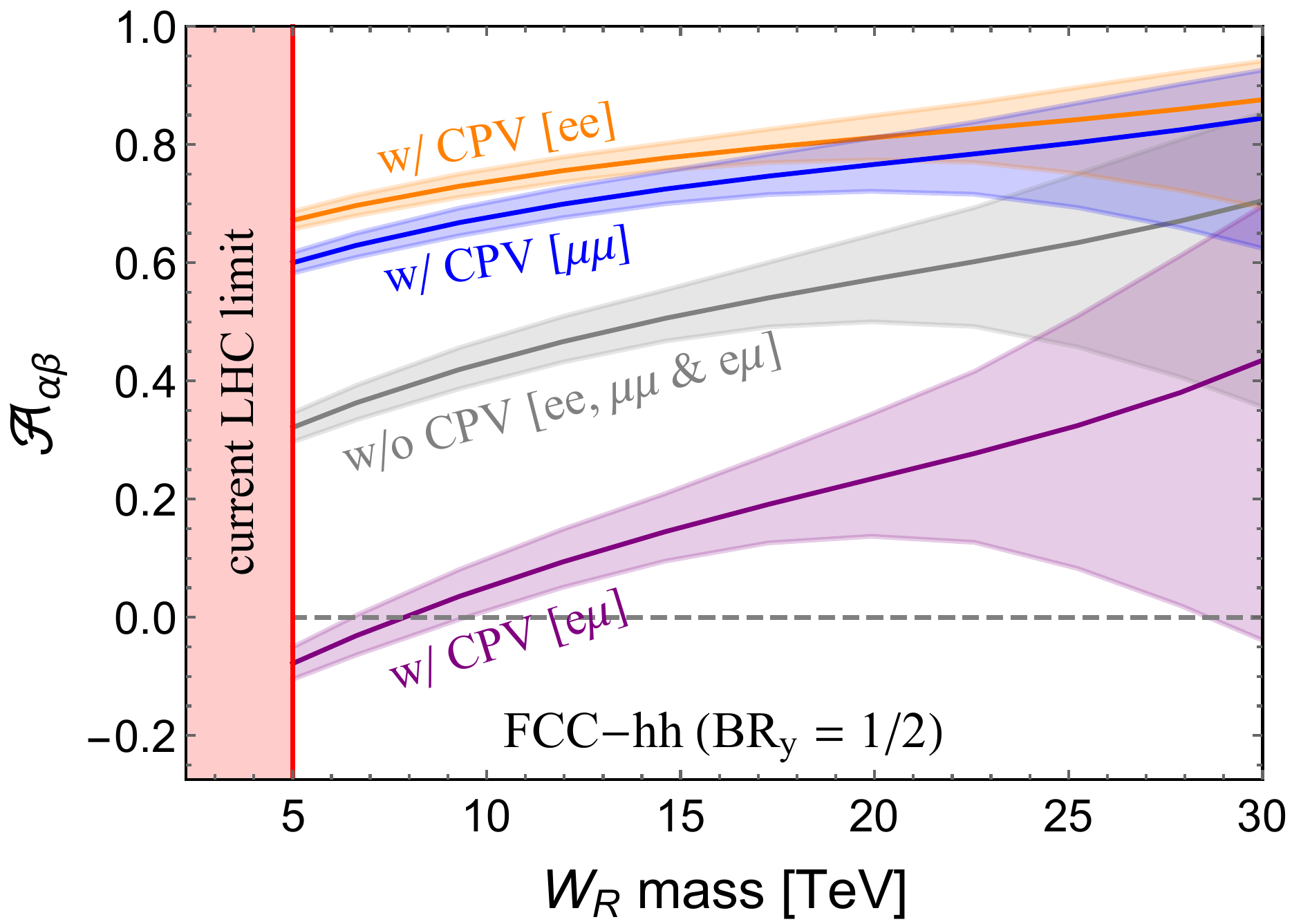}
  \caption{Prospects of ${\cal A}_{\alpha\beta}$ at the 14 TeV LHC (upper), 27 TeV HE-LHC (middle) and 100 TeV FCC-hh (lower), as functions of $W_R$ mass, with  ${\rm BR}_y = 0$ (left panels) and ${\rm BR}_y = 1/2$ (right panels). The solid gray line in each plot corresponds to the values of ${\cal A}_{ee,\,\mu\mu,\,e\mu}$ without CPV in the RHN sector, and the solid orange, blue and purple lines are respectively for ${\cal A}_{ee}$, $A_{\mu\mu}$ and ${\cal A}_{e\mu}$ with $\theta_R = \pi/4$ and $\delta_R = \pi/4$ (in the three left panels the $ee$ and $\mu\mu$ bands overlap with each other). The shaded bands are due to the PDF uncertainties, and the vertical (red) shaded region is excluded by the current LHC searches of $W_R$ boson with $g_R = g_L$~\cite{Sirunyan:2018pom, Aaboud:2018spl, Aaboud:2019wfg}. The horizontal dashed gray lines indicate ${\cal A}_{\alpha\beta} = 0$. }
  \label{fig:prospect}
\end{figure}

The expected charge asymmetries ${\cal A}_{\alpha\beta}$ (with $\alpha\beta = ee,\,\mu\mu,\,e\mu$) purely due to the PDFs and without any CPV in the RHN sector at the 14 TeV LHC, 27 TeV HE-LHC and 100 TeV FCC-hh are shown by the gray lines in Fig.~\ref{fig:prospect}, with the gray bands due to the PDF uncertainties. The CPV in the RHN sector induces extra SSCA, as given in Eq.~(\ref{eqn:Aab}). The special case with the largest CPV effect $\theta_R = \pi/4$ and $\delta_R = \pi/4$ is shown in Fig.~\ref{fig:prospect} by the orange, blue and purple bands, which are respectively for ${\cal A}_{ee}$, ${\cal A}_{\mu\mu}$, and ${\cal A}_{e\mu}$. As a benchmark scenario, we have taken ${\rm BR}_y = 0$ in the left panels of Fig.~\ref{fig:prospect},  where the three-body decays of $N_\alpha$ dominate over the two-body decays.
%In this case the asymmetries ${\cal A}_{ee,\,\mu\mu}$ are expected to be the same, without considering the different efficiencies of electrons and muons at hadron colliders, therefore the $ee$ and $\mu\mu$ bands overlap with each other.
In the right panels of Fig.~\ref{fig:prospect} we show the asymmetries for a second benchmark scenario with ${\rm BR}_y = 1/2$, in which case the three- and two-body decays are comparable to each other.
%One can clearly see that due to the different Yukawa couplings of $N_{e,\,\mu}$, the asymmetries ${\cal A}_{ee,\,\mu\mu}$ differ from each other, and can even be distinguished at future 100 TeV collider.
%Comparing the two scenarios in Fig.~\ref{fig:prospect}, one can see that the CP-induced SSCAs in the second case is to some extent ``diluted'' by the two-body decay of RHNs.
For the sake of concreteness, we have taken the Dirac CP phase in the PMNS matrix to be $-\pi/2$, which is favored by recent T2K~\cite{Abe:2018wpn} and NO$\nu$A~\cite{Acero:2019ksn} data, and the single Majorana phase is set to be zero, and $x = 1$ in Eq.~(\ref{eqn:Rem}).

When the current 13 TeV LHC constraints on $W_R$ mass of 5 TeV are taken into consideration~\cite{Sirunyan:2018pom, Aaboud:2018spl, Aaboud:2019wfg},\footnote{The LHC limits on the $W_R$ boson mass depend on the RHN mass involved. Here we have taken the most stringent bound of 5 TeV from Ref.~\cite{Aaboud:2019wfg}.} the energy fraction $x_1 x_2 = \hat{s}/s \sim 0.1$ and the PDF uncertainties are so large that the SSCAs ${\cal A}_{\alpha\beta}$ can not be measured at 14 TeV LHC, as can be seen in the two upper panels of Fig.~\ref{fig:prospect}. At 27 TeV HE-LHC and future 100 TeV colliders, the ratio $M_{W_R}^2/s$ could be significantly smaller, even for larger $W_R$ boson masses, and we can distinguish the CP-induced SSCAs from the PDF-induced effects. In the optimal case with ${\rm BR}_y = 0$, the CPV in the RHN sector can be observed at HE-LHC if $M_{W_R} < 7.2$ TeV, as shown in the middle left panel of Fig.~\ref{fig:prospect}. At future 100 TeV colliders like FCC-hh and SPPC, one could probe CPV through SSCA up to a $W_R$ mass of 26 TeV. In particular, if $M_{W_R} \lesssim 11.5$ TeV, the asymmetry ${\cal A}_{e\mu} <0$ at FCC-hh, which would otherwise be positive if there is no CPV in the RHN sector. If the BRs for two-body decays of RHNs are sizable, the CP-induced SSCAs are to some extent ``diluted'' by the second term in Eq.~(\ref{eqn:Rab}), as shown in the three right panels of Fig.~\ref{fig:prospect}. In the case of ${\rm BR}_y = 1/2$ with the three- and two-body decay widths being equal, the CP phase in the RHN sector can be probed up to a $W_R$ mass of 6.4 TeV at HL-LHC and improved up to 23 TeV at the future 100 TeV colliders.

Since the gauge interactions do not distinguish between electron and muon flavors, we expect the asymmetries ${\cal A}_{ee}$ and ${\cal A}_{\mu\mu}$ to be similar in the case of ${\rm BR}_y = 0$ (assuming similar efficiencies for electrons and muons at future hadron colliders), as shown in the three left panels of Fig.~\ref{fig:prospect}. However, for ${\rm BR}_y \neq 0$, if the Yukawa couplings of $N_{e,\,\mu}$ are different, then ${\cal A}_{ee}$ and ${\cal A}_{\mu\mu}$ values differ, as shown in the three right panels of Fig.~\ref{fig:prospect}. They can be distinguished at future 100 TeV colliders, even after taking into account the PDF uncertainties, as shown in the bottom right panel of Fig.~\ref{fig:prospect}. In general, any significant deviation from the gray bands in Fig.~\ref{fig:prospect} would be a clear signal of CPV in the RHN sector. Note that the SSCA measurements can be performed irrespective of the total integrated luminosity, as long as we can collect enough SS dilepton events to suppress the reducible SM backgrounds.  Actually we need only ${\cal O} (100 \, {\rm fb}^{-1})$ of data to have at least 100 events of both $\ell^+ \ell^+$ and $\ell^- \ell^-$ at FCC-hh (HE-LHC) for a $W_R$ mass of 10 (5) TeV. Therefore, we expect the SSCAs to be a `smoking gun' observable to reveal the existence of CPV in the RHN sector.

\begin{figure}[!t]
  \centering
  \includegraphics[height=0.33\textwidth]{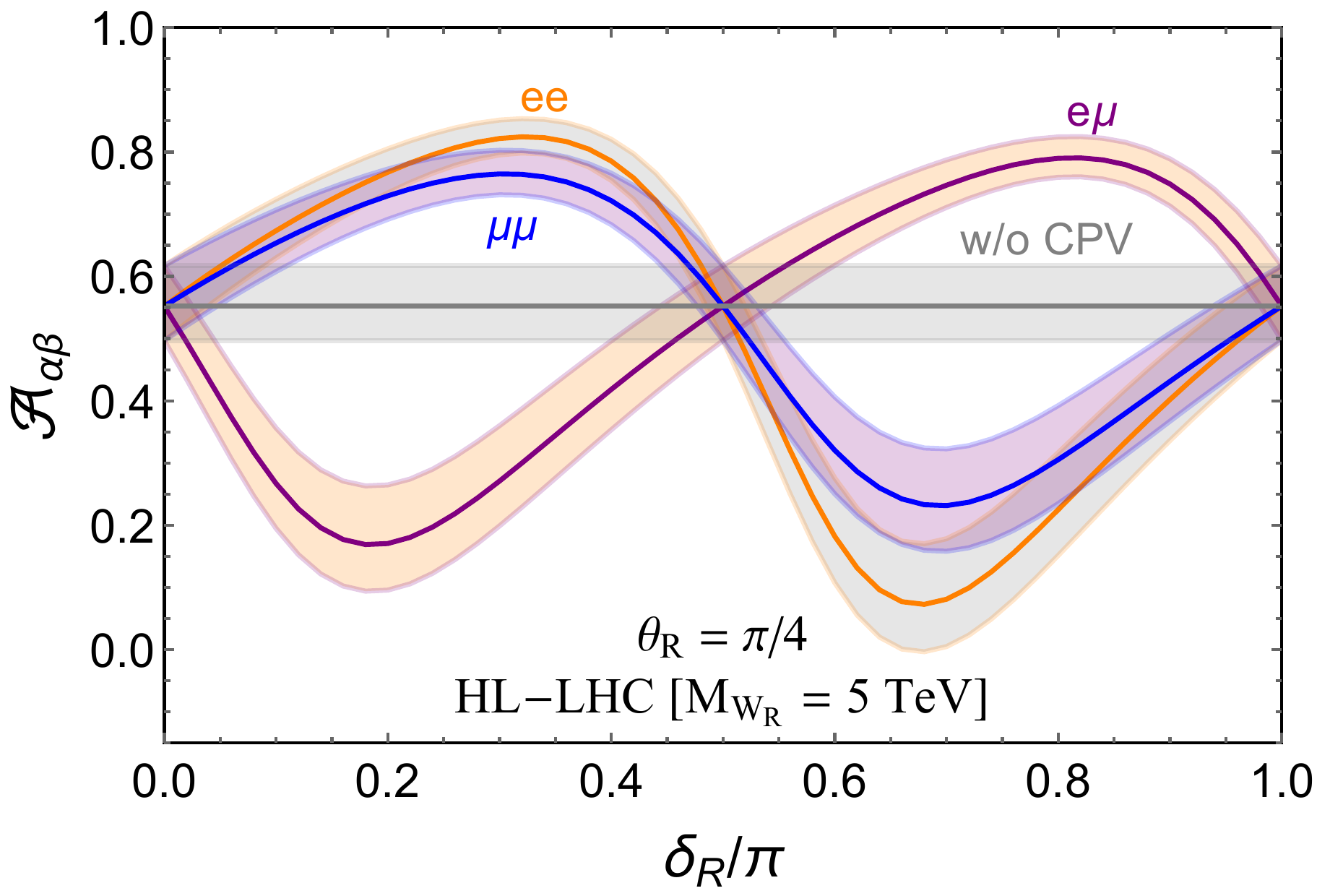}
  \includegraphics[height=0.33\textwidth]{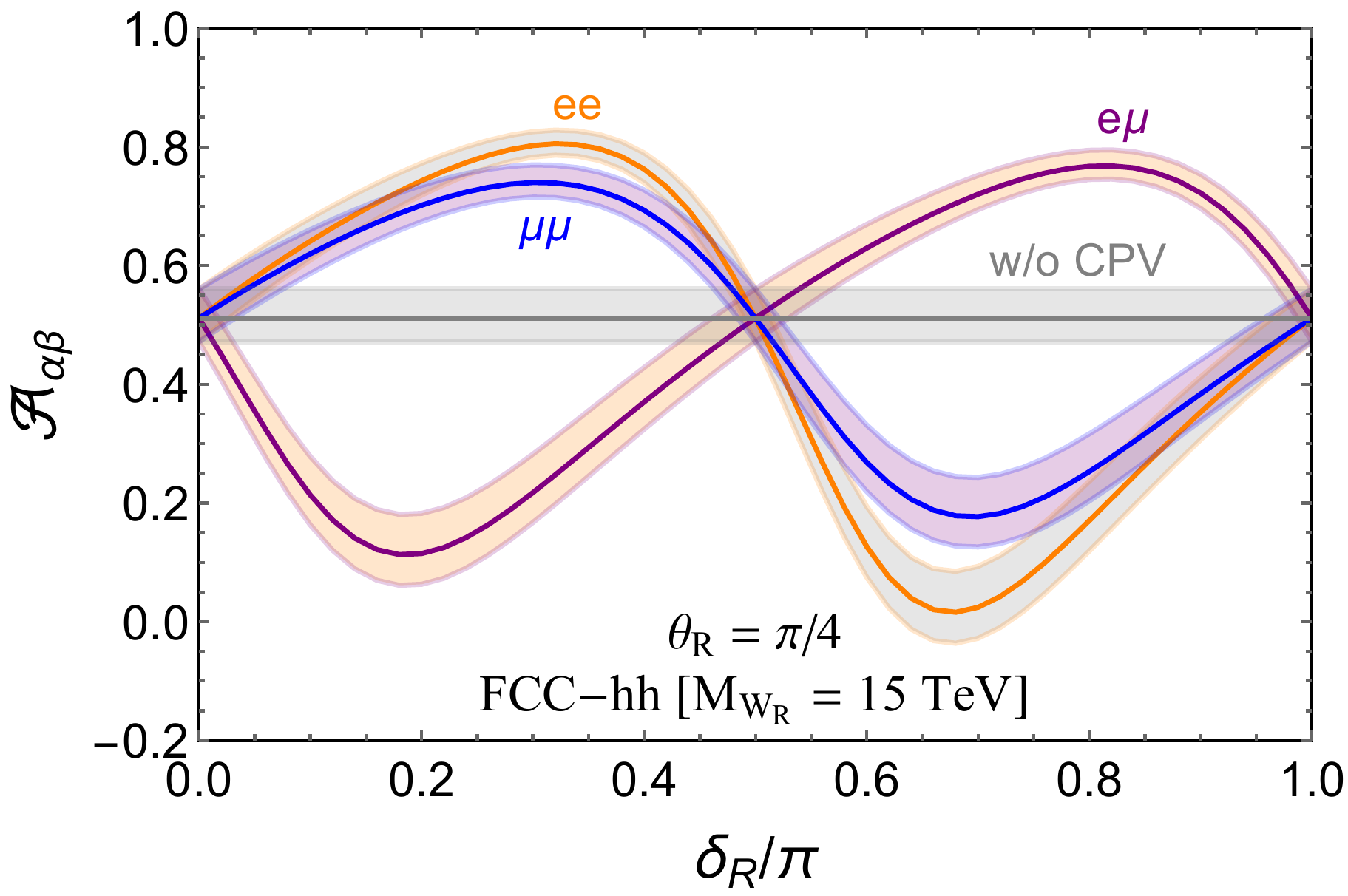}
  \caption{Prospects of charge asymmetry ${\cal A}_{\alpha\beta}$ at 27 TeV HE-LHC with $M_{W_R} = 5$ TeV (left) and future 100 TeV collider FCC-hh (right) with $M_{W_R} = 15$ TeV, as functions of the CP phase $\delta_R$, with $\theta_R = \pi/4$. The flavor combinations $ee$, $\mu\mu$ and $e\mu$  are depicted respectively in orange, blue and purple, and the horizontal gray bands correspond to the case without CPV. The solid lines denote the central values and the shaded bands are due to the PDF uncertainties. }
  \label{fig:CPV}
\end{figure}

The dependence of ${\cal A}_{\alpha\beta}$ on the CP phase $\delta_R$ at HE-LHC and FCC-hh is shown in Fig.~\ref{fig:CPV}, for the special case of $\theta_R = \pi/4$. The $ee$, $\mu\mu$ with $e\mu$ channels are shown respectively in orange, blue and purple. Note that the $e\mu$ channel does not depend on $\theta_R$ (the overall $\sin^22\theta_R$ term in Eq.~\eqref{eqn:Rem} cancels out in the ratio for ${\cal A}_{e\mu}$), so the result  for ${\cal A}_{e\mu}$ shown in Fig.~\ref{fig:CPV} is applicable for arbitrary $\theta_R$. As in Fig.~\ref{fig:prospect}, the solid lines correspond to the central values while the shaded bands are due to  the PDF uncertainties. The $W_R$ mass is set to be 5 TeV for the HL-LHC  case (left) and 15 TeV for the FCC-hh case (right) as a benchmark choice. The latter choice also respects the leptogenesis constraints on $W_R$ mass (see Section~\ref{sec:leptogenesis}). For comparison, the case without any CPV in the RHN sector is shown by the horizontal gray band in both panels. As expected from Fig.~\ref{fig:prospect}, the PDF uncertainties on the SSCA at 100 TeV collider are smaller than at HE-LHC, even though the $W_R$ mass is taken to be larger in the right panel of Fig.~\ref{fig:CPV} than in the left panel.

%If the asymmetries ${\cal A}_{\alpha\beta}$ could be found at future high-energy colliders, it is promising that we could determine the CPV in the RHN sector, or at least narrow it down to a limited range.

\begin{figure}[t]
  \centering
  \includegraphics[height=0.39\textwidth]{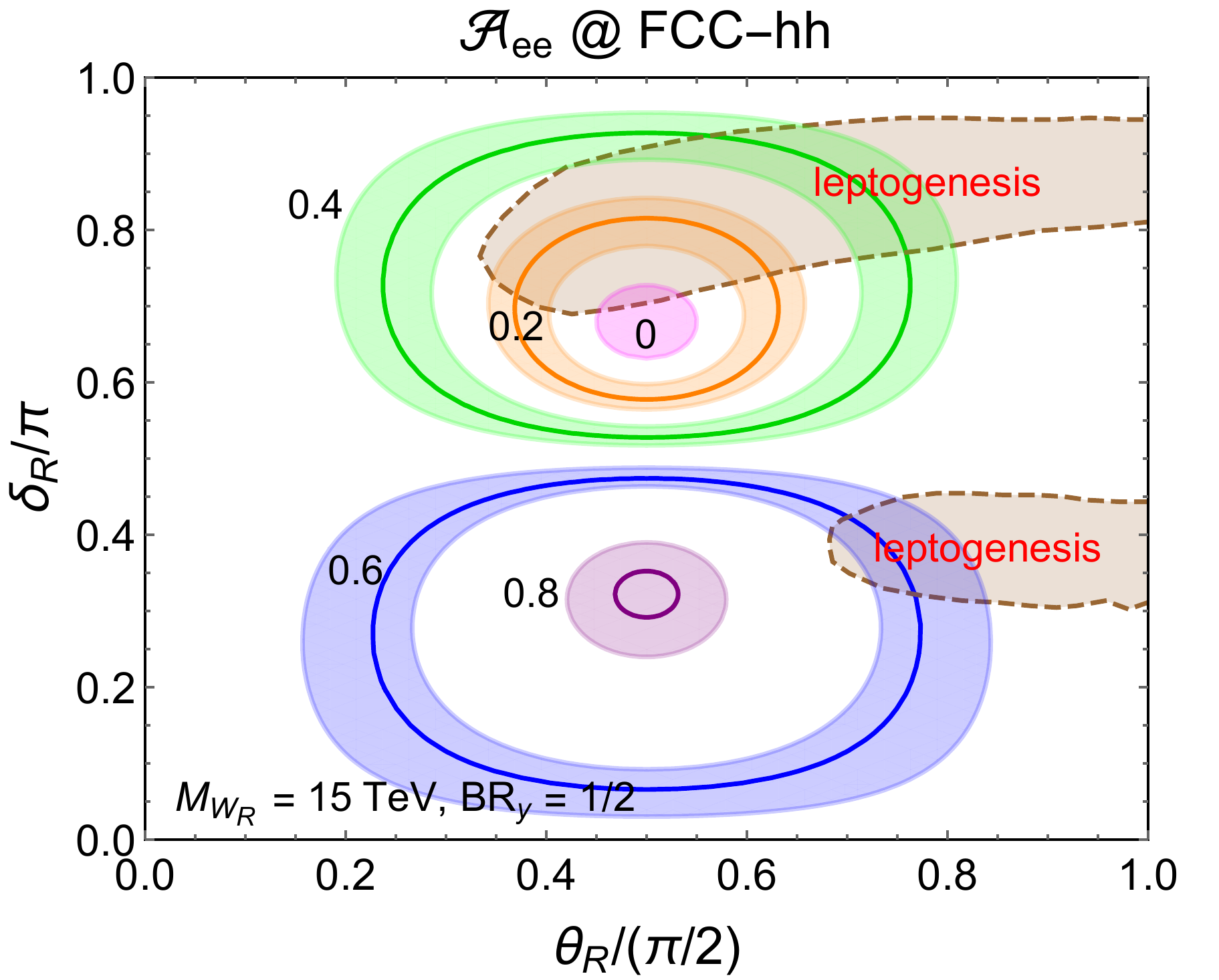}
  \includegraphics[height=0.39\textwidth]{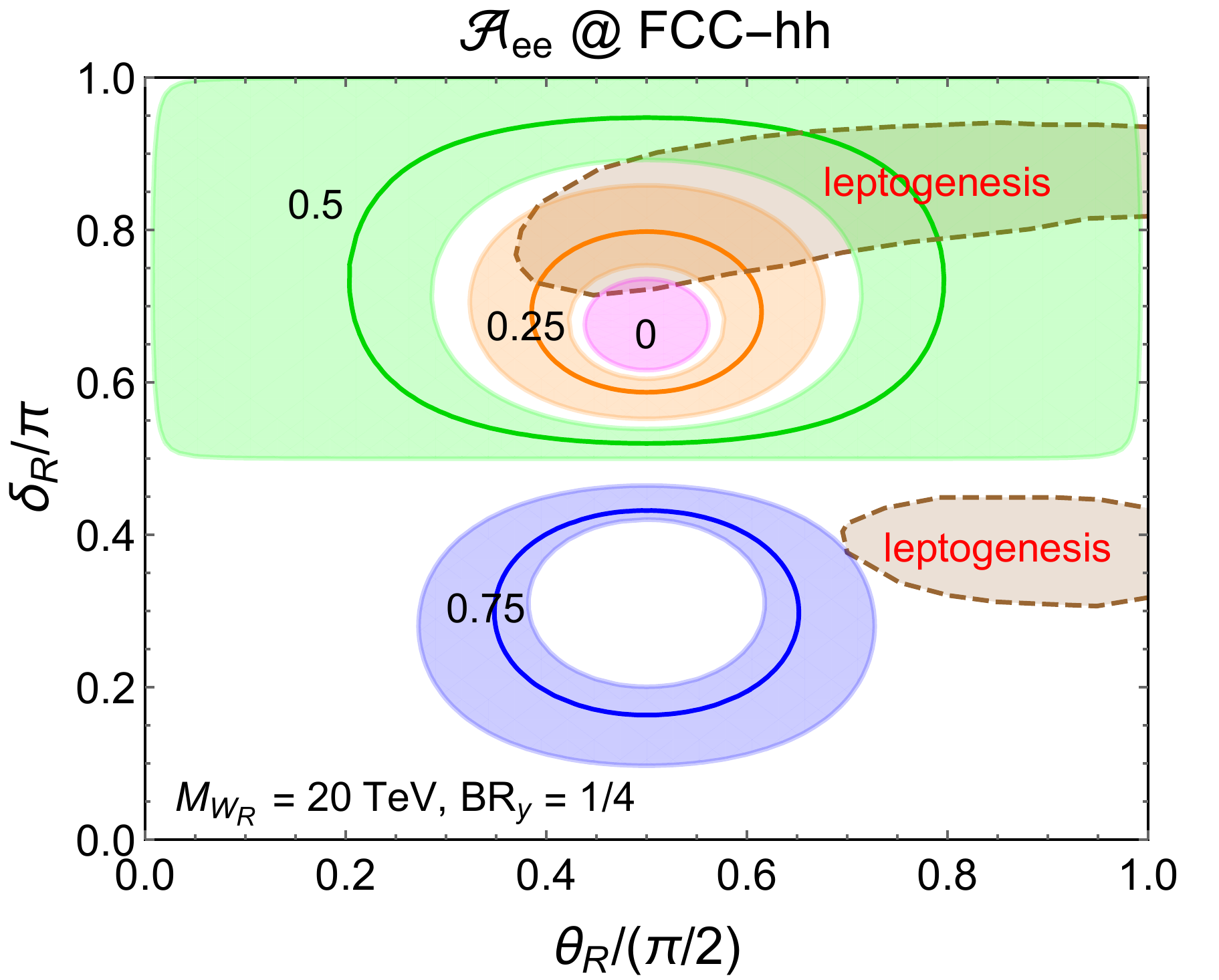}
  \caption{Contours of ${\cal A}_{ee}$ at future 100 TeV collider FCC-hh, as functions of the RHN mixing angle $\theta_R$ and CP phase $\delta_R$. Here we have fixed the RHN mass at 1 TeV. In the left panel, we show the lines with ${\cal A}_{ee} = 0.8$, 0.6, 0.4, 0.2 and 0, with $W_R$ mass at 15 TeV  and ${\rm BR}_y = 1/2$, while in the right panel we show the contours of ${\cal A}_{ee} = 0.75$, 0.5,  0.25, and $0$, with $W_R$ mass at 20 TeV  and ${\rm BR}_y = 1/4$. The shaded bands are due to the PDF uncertainties. The brown shaded regions bounded by the dashed lines in the left and right panels are respectively the regions covered by the blue and red points in Fig.~\ref{fig:example2}, which could generate successful leptogenesis;  see Section~\ref{sec:lepcol} for more details on the leptogenesis constraints. }
  \label{fig:contours}
\end{figure}

More generic dependence of ${\cal A}_{ee}$ at FCC-hh on the RHN mixing angle $\theta_R$ and CP phase $\delta_R$ is shown in Fig.~\ref{fig:contours}.  For the purpose of comparison and direct test of leptogenesis at future high-energy colliders (see Section~\ref{sec:leptogenesis}), two benchmark scenarios are considered, which both respect the leptogenesis constraints on the $W_R$ mass. For the sake of concreteness, we have taken $M_N = 1$ TeV and the ratio $x = 1$. In the left panel we have taken the $W_R$ mass to be 15 TeV and ${\rm BR}_y = 1/2$, which means that the three- and two-body decay widths are equal to each other. We show the values of ${\cal A}_{ee,\,\mu\mu} = 0.8$, $0.6$, $0.4$, $0.2$ and $0$, which are respectively depicted in purple, blue, green, orange and magenta. In the right panel, we have $M_{W_R} = 20$ TeV and ${\rm BR}_y = 1/4$, with the three-body decay width of RHNs being three times larger than that for the two-body decays. In this panel we show the contours of ${\cal A}_{ee,\,\mu\mu} = 0.75$, $0.5$, $0.25$ and $0$, respectively in blue, green, orange and magenta. In both panels, the solid lines correspond to the central values, while the shaded bands are due to the PDF uncertainties.  If the $W_R$ boson can be observed at high-energy colliders, its mass can be fixed by the invariant mass of the SS dileptons and jets. Then by measuring the asymmetry ${\cal A}_{ee}$ (and/or ${\cal A}_{\mu\mu}$), the values of $\theta_R$ and $\delta_R$ can be limited to a (narrow) band, as shown in Fig.~\ref{fig:contours}, which can be further narrowed down by the measurement of ${\cal A}_{e\mu}$, up to a twofold ambiguity, as exemplified in Fig.~\ref{fig:CPV}. As indicated by the brown shaded regions in Fig.~\ref{fig:contours}, only certain regions of $\theta_R$ and $\delta_R$ could generate successful leptogenesis in the LRSM, therefore the SSCAs at future high-energy hadron colliders can be used to test leptogenesis; see more details on the leptogenesis constraints in Section~\ref{sec:leptogenesis}.

\subsection{${\cal R}_{\rm CP}$}

\begin{figure}[!t]
  \centering
  \includegraphics[height=0.5\textwidth]{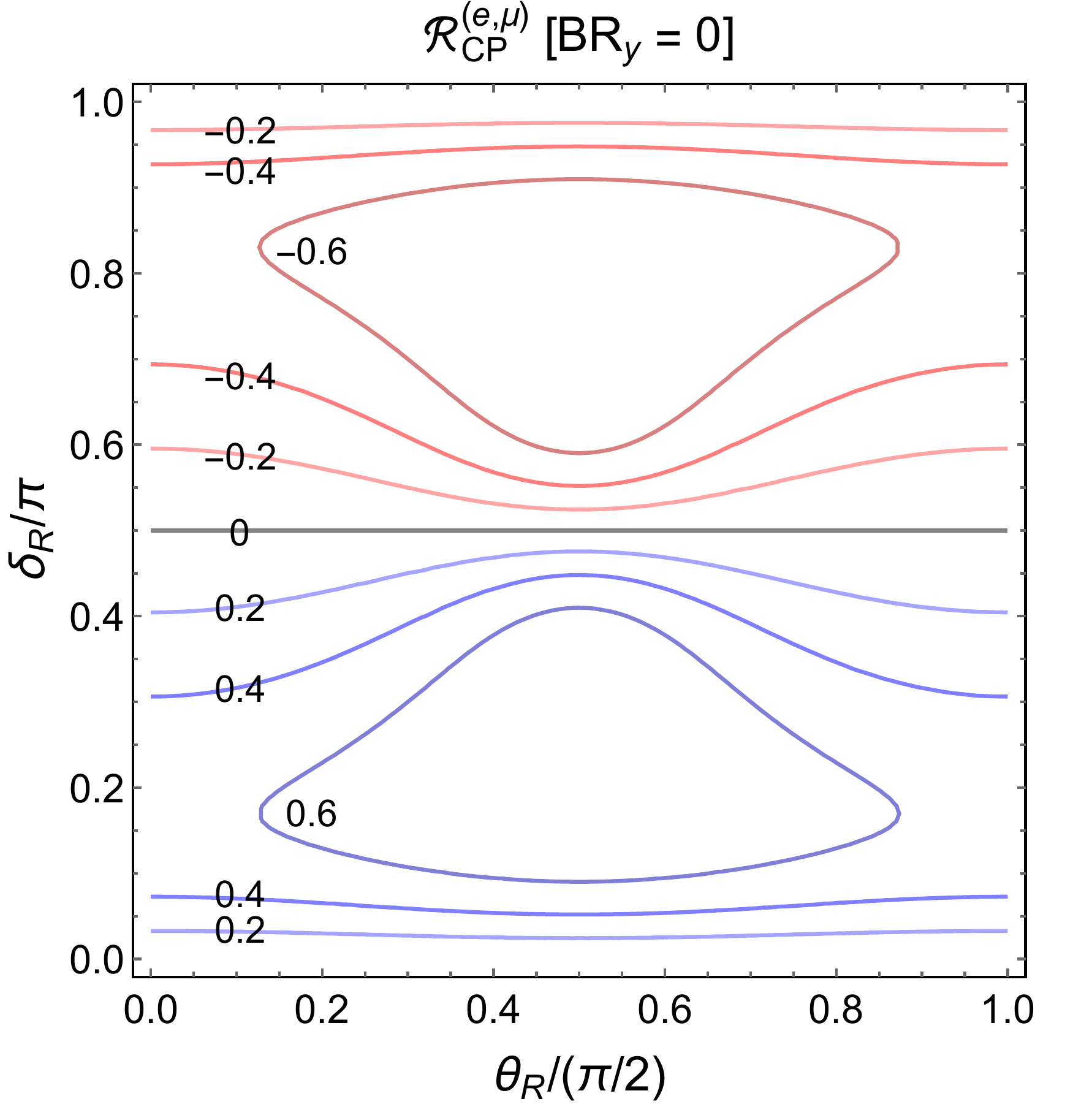}
  \caption{Ratios of charge asymmetry ratios ${\cal R}_{\rm CP}^{(e,\,\mu)}$ as functions of the RHN mixing angle $\theta_R$ and CP phase $\delta_R$ for the case ${\rm BR}_y = 0$. In this case, leptogenesis is not viable, since the RHNs do not have CPV decays to charged leptons. Here we have taken $M_{W_R}=10$ TeV for concreteness, but the ${\cal R}_{\rm CP}$ contours are independent of $M_{W_R}$ for ${\rm BR}_y = 0$. }
  \label{fig:RCP1}
\end{figure}

\begin{figure}[!t]
  \centering
  \includegraphics[height=0.5\textwidth]{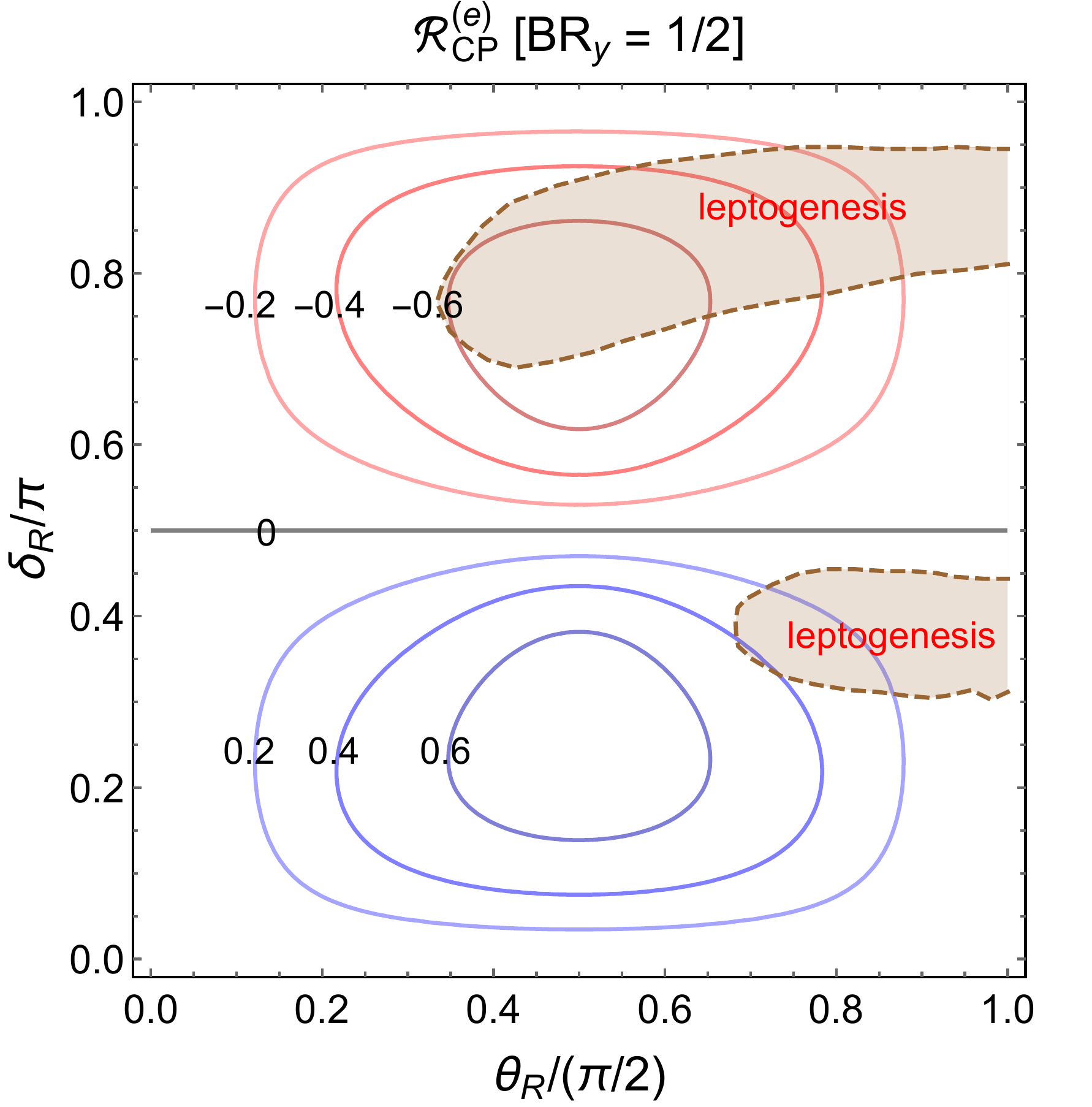}
  \includegraphics[height=0.5\textwidth]{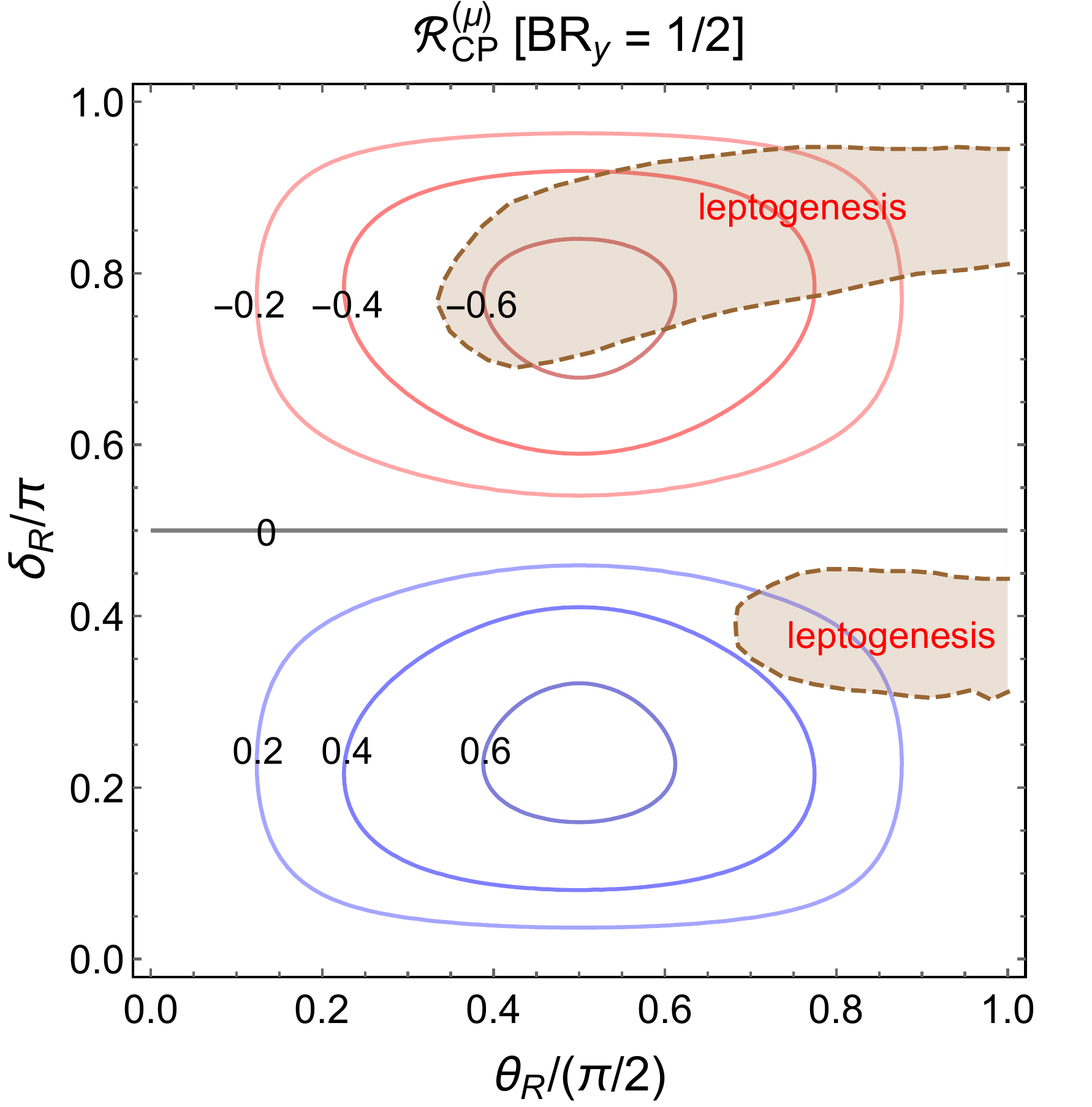}
  \caption{Ratios of charge asymmetry ratio charge asymmetry ratios ${\cal R}_{\rm CP}^{(e)}$ (left) and ${\cal R}_{\rm CP}^{(\mu)}$ (right) as functions of the RHN mixing angle $\theta_R$ and CP phase $\delta_R$ for the case ${\rm BR}_y = 1/2$ with  $M_{W_R} = 15$ TeV. The brown shaded regions bounded by the dashed lines are the regions covered by the blue points in Fig.~\ref{fig:example2}, which generates successful leptogenesis (see Section~\ref{sec:lepcol}).}
  \label{fig:RCP2}
\end{figure}

\begin{figure}[!t]
  \centering
  \includegraphics[height=0.5\textwidth]{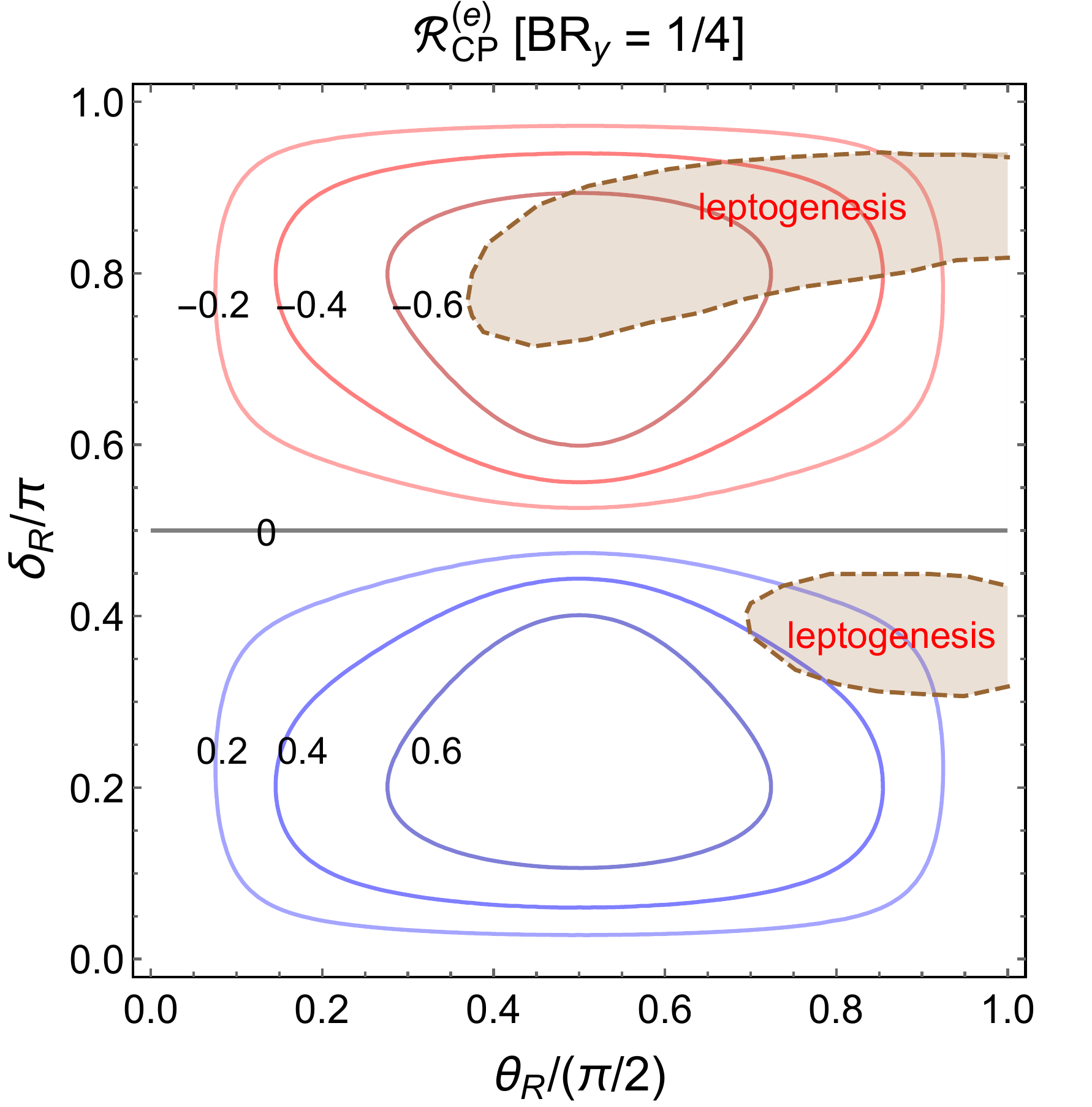}
  \includegraphics[height=0.5\textwidth]{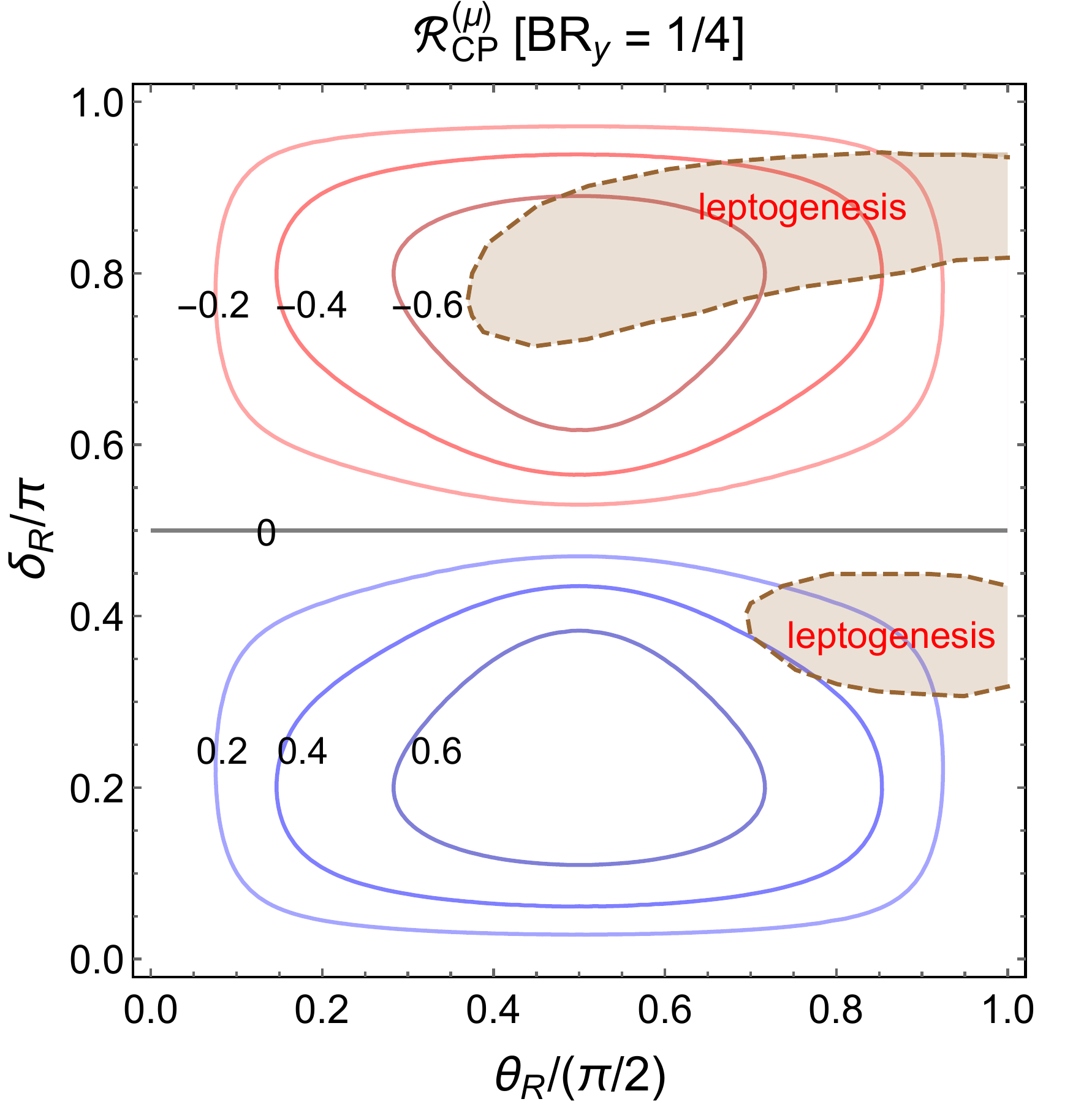}
  \caption{Charge asymmetry ratios ${\cal R}_{\rm CP}^{(e)}$ (left) and ${\cal R}_{\rm CP}^{(\mu)}$ (right) as functions of the RHN mixing angle $\theta_R$ and CP phase $\delta_R$ for the case ${\rm BR}_y = 1/4$ with  $M_{W_R} = 20$ TeV. The brown shaded regions bounded by the dashed lines are the regions covered by the red points in Fig.~\ref{fig:example2}, which generates successful leptogenesis (see Section~\ref{sec:lepcol}).}
  \label{fig:RCP3}
\end{figure}

As shown in Fig.~\ref{fig:prospect}, the prospects of the asymmetries ${\cal A}_{ee,\, \mu\mu,\, e\mu}$ are strongly affected by the PDF uncertainties. In contrast, the ratios ${\cal R}_{\rm CP}^{(e,\, \mu)}$ defined in Eq.~\eqref{eqn:Rab} are independent of the PDF uncertainties, as shown in Figs.~\ref{fig:RCP1}, \ref{fig:RCP2} and \ref{fig:RCP3} for the dependence of ${\cal R}_{\rm CP}^{(e,\, \mu)}$ on the RHN mixing angle $\theta_R$ and CP phase $\delta_R$. Fig.~\ref{fig:RCP1} is for the case ${\rm BR}_y = 0$ where we have only the gauge coupling mediated three-body decays in which case ${\cal R}_{\rm CP}^{(e)}$ and ${\cal R}_{\rm CP}^{(\mu)}$ are the same, irrespective of the $W_R$ mass. Here we have taken $M_{W_R}=10$ TeV for concreteness. The left and right panels of Fig.~\ref{fig:RCP2} are respectively for ${\cal R}_{\rm CP}^{(e)}$ and ${\cal R}_{\rm CP}^{(\mu)}$ with the three- and two-body decay contributions equal, i.e. ${\rm BR}_y = 1/2$, and the $W_R$ mass at 15 TeV. Fig.~\ref{fig:RCP3} is the same as Fig.~\ref{fig:RCP2} but with the benchmark values of ${\rm BR}_y = 1/4$ and $M_{W_R} = 20$ TeV.
%For the purpose of comparison and direct test of leptogenesis at future high-energy colliders (see Section~\ref{sec:leptogenesis}), we have set $W_R$ mass at 15 TeV.
Comparing the contours in Figs.~\ref{fig:RCP1}--\ref{fig:RCP3}, one can see that the ratios ${\cal R}_{\rm CP}^{(e,\, \mu)}$ depend on both three- and two-body decays, as expected. But the key point is that unlike the asymmetries ${\cal A}_{\alpha\beta}$, the ratios ${\cal R}_{\rm CP}^{(e,\, \mu)}$ are free from the PDF uncertainties, and thus provide a clean probe of CPV at future colliders and up to a higher $W_R$ mass. In particular, as long as the $W_R$ mass lies below the prospect of $\sim 6$ TeV at the HL-LHC, the ratios ${\cal R}_{\rm CP}^{(e,\, \mu)}$ can be measured the CPV in the RHN sector can be deciphered, whereas  we need a higher center-of-mass energy to unambiguously measure the asymmetries ${\cal A}_{\alpha\beta}$, as shown in Fig.~\ref{fig:prospect}. %One should note, however, that the asymmetries ${\cal A}_{\alpha\beta}$ have three degrees of freedom, i.e. $ee$, $\mu\mu$ and $e\mu$, while there are only two for ${\cal R}_{\rm CP}^{(e,\, \mu)}$, and ${\cal A}_{\alpha\beta}$ can in principle provide complementary information of the RHN sector to the ratios ${\cal R}_{\rm CP}^{(e,\, \mu)}$.
Furthermore, as in the case of ${\cal A}_{\alpha\beta}$, to probe ${\cal R}_{\rm CP}^{(e,\, \mu)}$ we need only ${\cal O} (100 \, {\rm fb}^{-1})$ of data to effectively suppress the SM backgrounds at FCC-hh (HE-LHC) for a $W_R$ mass of 10 (5) TeV. As indicated by the brown shaded regions in Fig.~\ref{fig:RCP2} and \ref{fig:RCP3}, only certain regions of $\theta_R$ and $\delta_R$ could generate successful leptogenesis in the LRSM, and therefore, the measurements of SSCAs at future high-energy hadron colliders can be effectively used to test leptogenesis; see more details in Section~\ref{sec:leptogenesis}.

\section{Testing leptogenesis at future hadron colliders}
\label{sec:leptogenesis}

In this section we study the implications of RHN mixing and associated CP phase for leptogenesis, where lepton asymmetry is generated from the CP violating decays of RHNs which is then transferred into the baryon asymmetry through electroweak sphaleron processes.
%The amount of baryon asymmetry therefore depends on the CP phases in the RHN sector.
In the type-I seesaw with hierarchical RHNs (the so-called `vanilla leptogenesis'), the RHN masses are required to be $\gtrsim 10^9$ GeV~\cite{Davidson:2002qv, Buchmuller:2003gz} for successful leptogenesis.
%In the type-I seesaw without left-right embedding, if we have only CP violation in the PMNS matrix but not in the RHN sector, the RHN masses are required to be $\gtrsim 10^9$ GeV~\cite{Davidson:2002qv, Buchmuller:2003gz};
With fine tuning and flavor effects taken into account, the hierarchical RHN masses can be lowered down to $10^6$ GeV~\cite{Moffat:2018smo}. In any case, these RHN masses are too heavy to be produced at foreseeable colliders. The situation  can be alleviated in the framework of resonant leptogenesis~\cite{RL1, RL2, RL3, RL4}, where (at least) two RHNs are quasi-degenerate, thereby resonantly enhancing the lepton asymmetry even for TeV-scale masses~\cite{Pilaftsis:2005rv, Dev:2014laa}; see Ref.~\cite{Dev:2017trv} for a review. Interestingly, this quasi-degeneracy is also a requirement for generating a non-zero SSCA~\cite{DDM}. Thus, measuring the SSCAs can not only provide important insight into the nature of the RHN mass matrix but also the origin of baryon asymmetry via resonant leptogenesis. In this section we will find that only in certain parameter space of $\theta_R$ and $\delta_R$ can the baryon asymmetry observed in the Universe be successfully produced with the right sign. The measurement of SSCAs ${\cal A}_{\alpha\beta}$ and ${\cal R}_{CP}^{(e,\,\mu)}$ would then help to test resonant leptogenesis directly at future high-energy hadron colliders.

\subsection{Dependence on the RHN sector}
%heavy RHN mixing and the associated CP phase will be the central ingredients to produce the observed baryon asymmetry, which is also related to the Yukawa couplings of RHNs via the seesaw mechanism.

%We now turn our attention to see how the mixing angle $\theta_R$ and CP phase $\delta_R$ in the RHN sector affect leptogenesis and if collider observations can enhance our understanding of leptogenesis.

%\subsection{Resonant leptogenesis}

%Then for the purpose of leptogenesis, we have
%\begin{eqnarray}
%(M_D^\dagger M_D) = (M_N^{1/2})^\dagger {\cal O}^\dagger \widehat{m}_\nu {\cal O} M_N^{1/2}
%\end{eqnarray}
%which does not depend on the PMNS matrix.

%Then the flavor-independent CP asymmetry~\cite{Xing:2006ms}
%\begin{eqnarray}
%\varepsilon_{{\rm CP},\,i} =
%\frac{{\rm Im} [(y^\dagger y)_{ij}^2]}{(y^\dagger y)_{11} (y^\dagger y)_{22}}
%\frac{(M_i^2 - M_j^2) M_i \Gamma_j}{(M_i^2 - M_j^2)^2 + M_i^2 \Gamma_j^2 } \times {\rm BR}_Y (N_i)
%\end{eqnarray}
%with $i \neq j$, $y = M_D/ v_{\rm EW}$, $\Gamma_i$ the partial decay width of $N_i$ through the  Yukawa couplings,
%\begin{eqnarray}
%{\rm BR}_Y (N_i) =
%\frac{\Gamma_i}{ \Gamma_i + \Gamma (N_i \to \ell qq')}
%\end{eqnarray}
%with $N_i \to \ell qq'$ mediated by the $W_R$ boson in the LR model.

In the minimal framework of resonant leptogenesis, the lepton asymmetry is generated from the on-shell decays of RHNs $N_{i} \to L\phi$ (with $i= 1,\,2$) via the Yukawa coupling $y = {\cal M}_D/v_{\rm EW}$, with $L$ and $\phi$ respectively the SM lepton and Higgs doublets. Given the analytic form of ${\cal M}_D$ in Eq.~(\ref{eqn:MD}), it is straightforward to calculate the flavor-dependent lepton asymmetry from the interference of tree- and loop-level diagrams, which can be approximated as~\cite{Dev:2014iva}
\begin{eqnarray}
\label{eqn:eta}
\eta^{\Delta L}_i \ \simeq \ \frac{3}{2z_c K_{\alpha}^{\rm eff}}
\sum_i \varepsilon_{i\alpha} d_i,
\end{eqnarray}
where $z_c = M_N/T_c$ with $M_N \equiv ( M_{N_1} + M_{N_2} )/2$ the averaged RHN mass, $T_c = 149.4$ GeV the critical temperature for electroweak phase transition, and
\begin{eqnarray}
\label{eqn:Keff}
K_\alpha^{\rm eff} \ = \ \kappa_\alpha
\left( 1 + \frac{\sum_i\gamma^{N_i}_{W_R}}{\sum_i B_{i\alpha} \gamma^{N_i}_{L\phi}} \right)
\end{eqnarray}
is the washout factor in presence of both Yukawa and $W_R$-mediated gauge interactions, with $B_{i\alpha}$ the BR for the decay $N_i \to L_\alpha + \phi$ and $\kappa_\alpha = \sum_i B_{i\alpha} (\Gamma_i/H_N)$, where $H_N  \simeq 17 M_N^2 / M_{\rm Pl}$ ($M_{\rm Pl}$ being the Planck mass) is the Hubble parameter at temperature $T=M_N$. Similarly,
\begin{eqnarray}
\label{eqn:dilution}
d_i \ = \ \frac{ \gamma^{N_i}_{L\phi} }{ \gamma^{N_i}_{L\phi} + \gamma^{N_i}_{Lqq} + \gamma^{N_i}_{W_R} }
\label{eqn:dilution}
\end{eqnarray}
is the dilution factor due to the $W_R$-mediated RH gauge interactions of RHNs. Here $\gamma^{N_i}_{L\phi}$, $\gamma^{N_i}_{Lqq}$ and $\gamma^{N_i}_{W_R}$ are the thermally-averaged rates due to the decays of $N_i$ through the Yukawa and gauge interactions and the $2 \leftrightarrow 2$ processes mediated by the $W_R$ boson, respectively. The main dependence on the RHN mixing and CP phase is contained in the flavor-dependent CP asymmetry\footnote{This is a simplified version of the fully flavor-dependent CP asymmetry, where we have neglected the intricacies related to the RHN mixing versus oscillation effects in the denominator of Eq.~(\ref{eqn:epsilon2})~\cite{Dev:2017trv}. }
\begin{eqnarray}
\label{eqn:epsilon2}
\varepsilon_{i\alpha} \ =
\ \frac{2(M_{N_i}^2 - M_{N_j}^2) {\rm Im} [y_{\alpha i}^\ast y_{\alpha j}]
{\rm Re} [(y^\dagger y)_{ij}]}{8\pi \left[4 (\Delta M_N)^2 + \Gamma_{N_j}^2 \right] (y^\dagger y)_{ii}} \times {\rm BR}_y (N_i) \,,
\end{eqnarray}
with $i,\,j = 1,\,2$ but $i \neq j$.

%then the expected flavor-denpendent lepton asymmetry
%\begin{eqnarray}
%\eta^{\Delta L}_\alpha \simeq
%\frac{3}{2z_c K_{\alpha}^{\rm eff}}
%\sum_i \varepsilon_{{\rm CP},\, i\alpha} d_i
%\end{eqnarray}

\subsection{Leptogenesis constraints on Yukawa couplings and $W_R$ mass}

Without any significant cancellation or fine-tuning in the ${\cal M}_D$ matrix, the Yukawa couplings $y$ are expected to much smaller than the gauge coupling $g_R$. In this scenario, the $W_R$-mediated gauge interactions in the LRSM, apart from giving an extra contribution to the total width of $N_i$, lead to $\Delta L = 1$ scattering processes $N \ell \leftrightarrow q \bar{q}\prime$, which cause significant dilution and/or washout of the lepton asymmetry produced by the Yukawa coupling induced RHN decays. This sets a lower bound on the $W_R$ mass, which is typically higher than the TeV scale, depending on the $M_N$ mass, the mass splitting $\Delta M_N$ and other parameters~\cite{Frere:2008ct, Dev:2014iva, Dhuria:2015cfa, DLM}. For TeV-scale RHNs, typically $|y| \sim \sqrt{|m_\nu M_N|} \sim 10^{-6}$, and the $W_R$ mass is required to be larger than roughly 50 TeV, to keep the dilution factor in Eq.~\eqref{eqn:dilution} not too small. This is even beyond the reach of future 100 TeV colliders~\cite{Mitra:2016kov, Dev:2015kca}.\footnote{With an integrated luminosity of 30 ab$^{-1}$, the $W_R$ boson can be probed at future 100 TeV colliders up to a mass of roughly 40 TeV if $m_N \lesssim 1$ TeV. A luminosity of $40 - 50 \, {\rm ab}^{-1}$ can probably reach the 50 TeV mark of $W_R$ mass~\cite{Ruiz:2017nip, CidVidal:2018eel}.} However, if the $\zeta$ parameter in the matrix ${\cal O}$ in Eq.~(\ref{eqn:MD}) is complex and $|\sin\zeta|, |\cos\zeta| \gg 1$, then the Yukawa couplings $y$ can be significantly enhanced (at the cost of fine-tuning in the seesaw formula), and the leptogenesis bound on $W_R$ mass can be relaxed accordingly. On the other hand, the couplings $|y|$ can not made too large, otherwise the $\Delta L = 0$ scattering process $L_\alpha \phi \leftrightarrow L_\beta \phi$, the $\Delta L = 2$ process $L\phi \leftrightarrow \bar{L}\phi^\dag$ and/or the inverse decay $L\phi\to N_i$ will induce significant dilution/washout effect and render leptogenesis ineffective.

%\begin{figure}[!t]
%  \centering
%  \includegraphics[width=0.48\textwidth]{example_NH.pdf}
%  \includegraphics[width=0.48\textwidth]{example_IH.pdf}
%  \caption{Example of lepton asymmetry $\eta^{\Delta L}_\alpha$ in resonant leptogenesis, as functions of the CP phase $\delta_R$ in the RHN sector, for the NH case with $m_1=0$ (left) and IH with $m_3=0$ (right). The central values of neutrino mass and mixing parameters are shown in Table~\ref{tab:neutriodata} and the other parameters are given in Table~\ref{tab:example}. The red, green and blue curves are respectively for the flavors $\alpha = e,\,\mu,\,\tau$, and the black curve for the total lepton asymmetry. The horizontal dashed gray line indicates the observed value of $\eta^{\Delta L}_\alpha$.}
%  \label{fig:examples}
%\end{figure}

\begin{table}[!t]
  \centering
  \caption[]{The $1\sigma$ values of the neutrino mass square differences and mixing angles for NH and IH used in our numerical analysis~\cite{Tanabashi:2018oca}.}
  \label{tab:neutriodata}
  %\begin{tabular}[t]{C{3.2cm}|C{2.8cm}|C{3.5cm}|C{3.5cm}}
  \begin{tabular}[t]{ccc}
  \hline\hline
  parameters & NH & IH \\ \hline
  $\Delta m^2_{21}$ [eV$^2$] & $(7.53 \pm 0.18) \times 10^{-5}$ & $(7.53 \pm 0.18) \times 10^{-5}$ \\
  $|\Delta m^2_{32}|$ [eV$^2$] & $(2.44 \pm 0.034) \times 10^{-3}$ & $(2.53 \pm 0.05) \times 10^{-3}$ \\ \hline

  $\sin^2\theta_{12}$ & $0.307^{+0.013}_{-0.012}$ & $0.307^{+0.013}_{-0.012}$ \\
  $\sin^2\theta_{23}$ &  $0.542^{+0.019}_{-0.022}$ &
  $0.536^{+0.023}_{-0.028}$ \\
  $\sin^2\theta_{13}$ & $0.0218 \pm 0.0007$ & $0.0218 \pm 0.0007$ \\ %\hline
  %$\delta_{\rm CP}$ & $[0,\, 2\pi]$ & $[0,\, 2\pi]$ \\
  %$\alpha$ & $[0,\, 2\pi]$ & $[0,\, 2\pi]$ \\
  \hline\hline
  \end{tabular}
\end{table}

\begin{figure}[!t]
  \centering
  \includegraphics[width=0.49\textwidth]{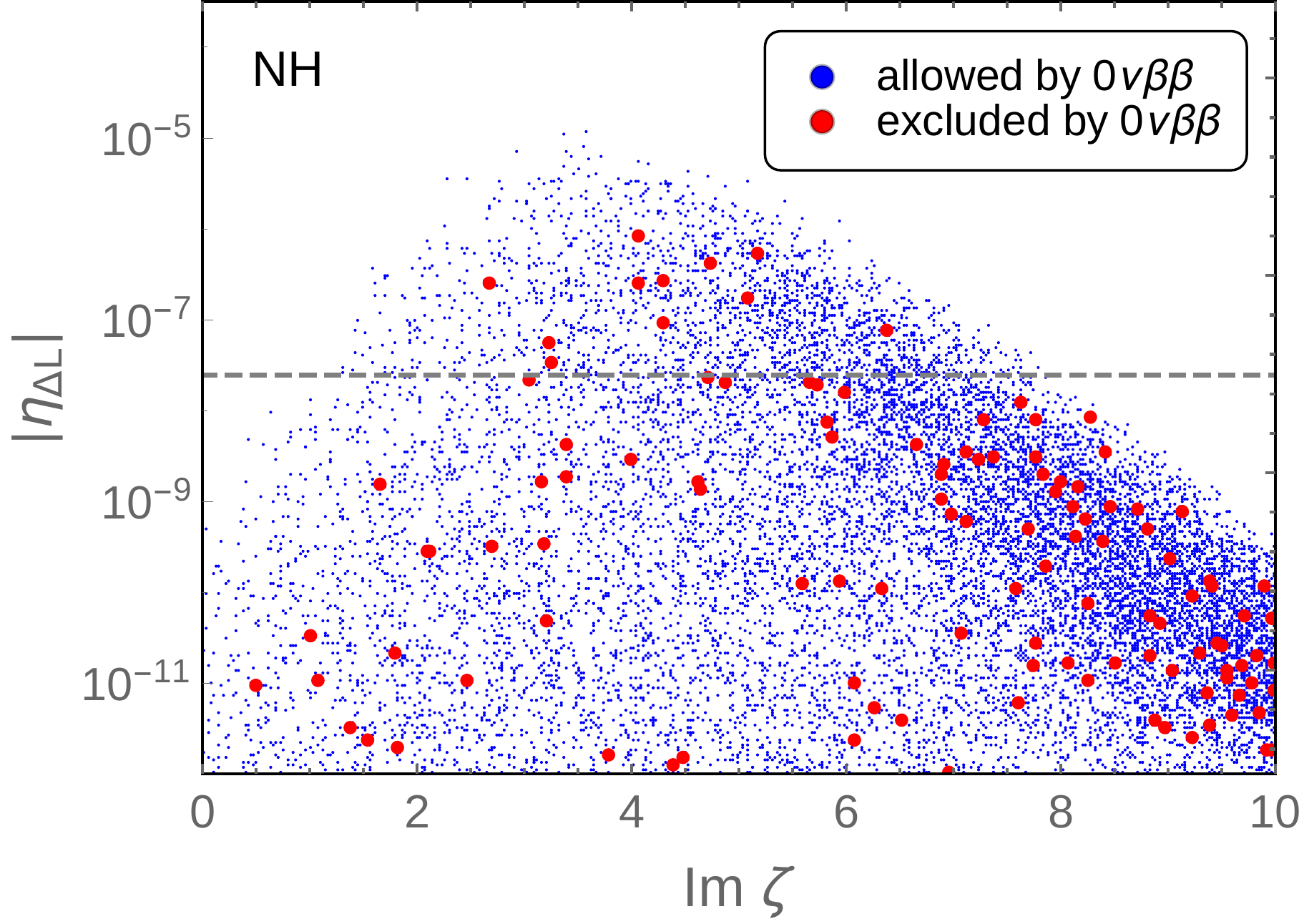}
  \includegraphics[width=0.49\textwidth]{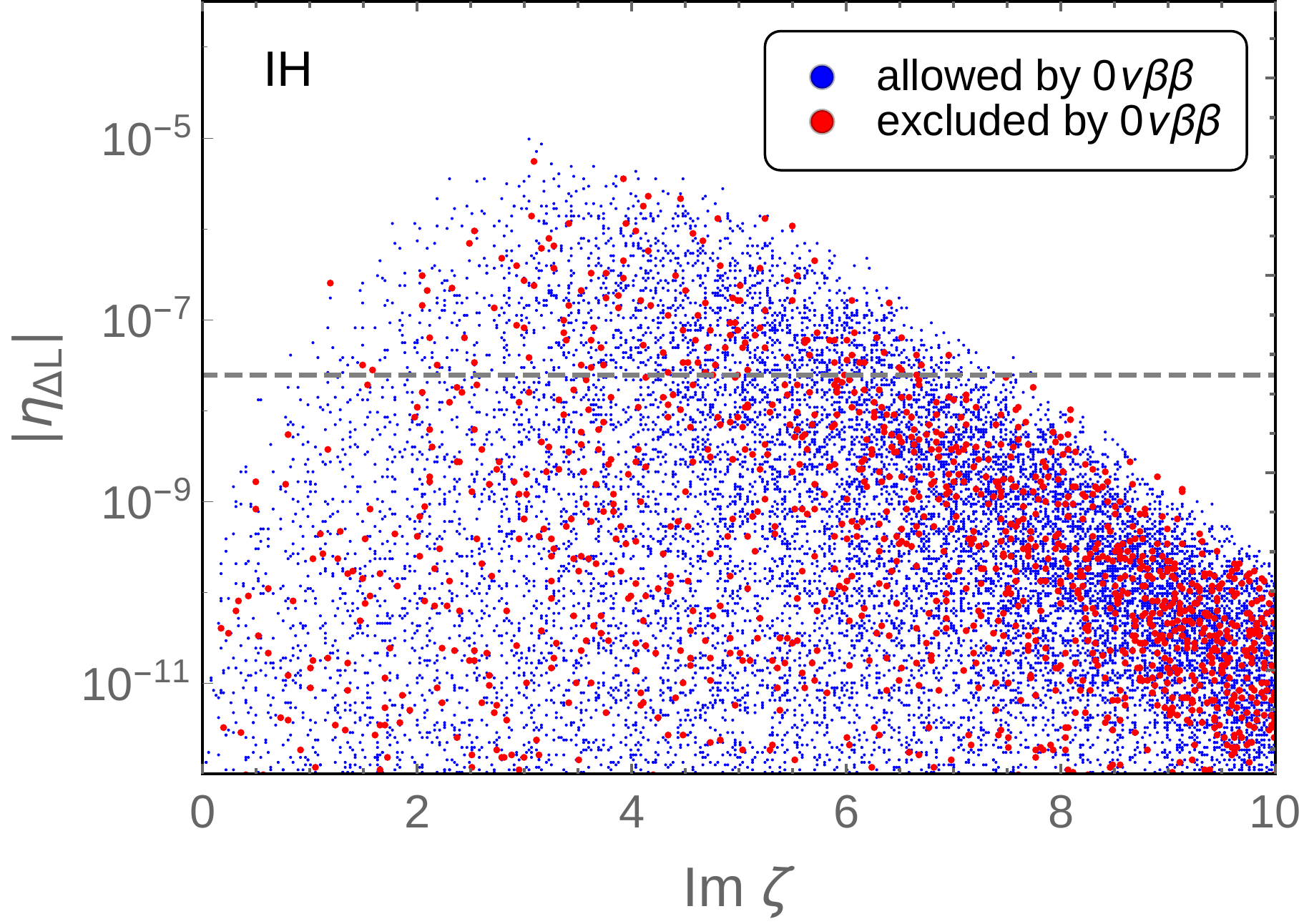}
  \caption{Scatter plot for the dependence of lepton asymmetry $|\eta_{\Delta L}|$ on the parameter $\zeta$, for the NH (left) and IH (right) cases. The blue (red) points are allowed (excluded) by the $0\nu\beta\beta$ decay constraint (see Section~\ref{sec:0nubetabeta}).}
  \label{fig:asymmetry1}
\end{figure}

To estimate the leptogenesis constraints on the $\zeta$ parameter and the resultant Yukawa couplings $|y|$ and the $W_R$ mass, we vary the neutrino oscillation parameters within their current $2\sigma$ ranges, as shown in Table~\ref{tab:neutriodata}~\cite{Tanabashi:2018oca}, and the other parameters in the following ranges:
\begin{eqnarray}
\label{eqn:ranges}
&& \delta_\nu \in [0,\,2\pi] \,, \quad
\alpha \in [0,\,2\pi] \,, \quad
\zeta \in [0,\,10]i \,, \nonumber \\
&& M_N \in [0.15,\, 10] \, {\rm TeV} \,, \quad
M_{W_R} \in [3,\,50] \, {\rm TeV} \,, \quad
\theta_R \in [0,\,2\pi] \,, \quad
\delta_R \in [0,\,2\pi] \,.
\end{eqnarray}
To be consistent with the analysis in Section~\ref{sec:SSCA}, we require that the $W_R$ boson is heavier than $N_{1,2}$, i.e. $M_{W_R} > M_N$. For the sake of simplicity, we have taken $\zeta$ to be purely imaginary, and the RHN mass splitting $\Delta M_N = \Gamma_{\rm avg}/2$ to have the maximal lepton asymmetry in Eq.~(\ref{eqn:epsilon2}). The lower bound for RHN mass is taken to be 150 GeV such that the decay $N_{1,2} \to L + \phi$ is kinematically allowed when the thermal masses of the Higgs and lepton doublets are taken into consideration~\cite{Giudice:2003jh}.\footnote{We do not consider the possibility that baryon asymmetry can also be generated from the Higgs doublet decay $\phi \to N + L$ when the RHNs are lighter than the SM Higgs doublet, see Refs.~\cite{Hambye:2016sby}.} The scatter plots for the dependence of lepton asymmetry $|\eta_{\Delta L}|$ on the parameter $|\zeta| = {\rm Im} \zeta$ are shown in Fig.~\ref{fig:asymmetry1}, with the left and right panels respectively for the NH and IH cases. In both panels the horizontal dashed gray lines correspond to the value of $\Delta \eta_L = -2.47 \times 10^{-8}$~\cite{Dev:2014laa} implied by current observations of baryon asymmetry~\cite{Aghanim:2018eyx}, and the red points are excluded by the $0\nu\beta\beta$ decays discussed in Section~\ref{sec:0nubetabeta}. It turns out that the limits from LFV decay $\mu \to e\gamma$ in Section~\ref{sec:LFV} and electron EDM in Section~\ref{sec:EDM} do not provide any limits for the parameter space shown here. As stated above, the value $|\zeta|$ can not be either too large or too small, and numerical evaluations reveal that for the parameter ranges given in Eq.~(\ref{eqn:ranges}) we have,
\begin{eqnarray}
\begin{cases}
1.3 \ \lesssim \ {\rm Im} \zeta \ \lesssim 7.8 \, \quad  & \text{for NH} \,, \\
0.8 \ \lesssim \ {\rm Im} \zeta \ \lesssim 7.7 \, \quad & \text{for NH} \,,
\end{cases}
\end{eqnarray}
and resultantly the magnitude of Yukawa couplings are required to be in the range of
\begin{equation}
\begin{cases}
1.3 \times 10^{-6} \ \lesssim \ |y|_{\rm max} \ \lesssim \ 7.2 \times 10^{-4} \, \quad   & \text{for NH} \,, \\
1.0 \times 10^{-6}  \ \lesssim \ |y|_{\rm max} \ \lesssim \ 8.6 \times 10^{-4} \, \quad  & \text{for IH} \,.
\end{cases}
\end{equation}

\begin{figure}[!t]
  \centering
  \includegraphics[width=0.55\textwidth]{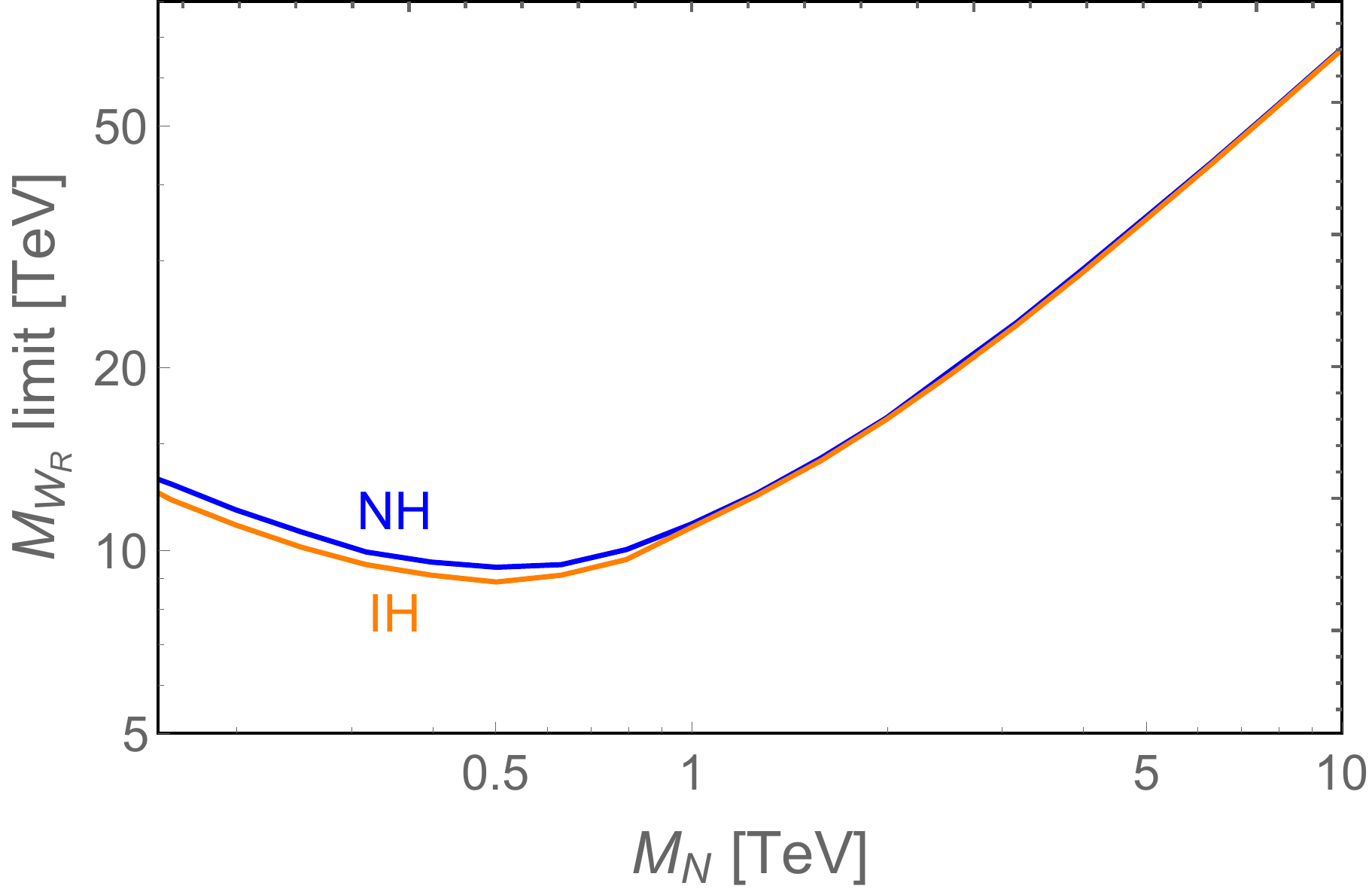}
  \caption{Leptogenesis constraint on $W_R$ mass in the case with only two quasi-degenerate RHNs, as function of the average RHN mass $M_N$. The blue and orange curves are respectively for the NH and IH cases. The regions below the curves are excluded.}
  \label{fig:WRlimit}
\end{figure}

The leptogenesis constraint on the $W_R$ mass is shown in Fig.~\ref{fig:WRlimit}, as function of the RHN mass, for the NH (blue) and IH (orange) cases. As stated above, the $W_R$ mass limit depends on the RHN mass and other parameters, which is mainly from comparing the two- and three-body decays of RHNs. In the large $M_N$ limit, the dependence is respectively
\begin{eqnarray}
\label{eqn:scalings}
\Gamma (N \to \ell q \bar{q}') \ \propto \ M_N^5/M_{W_R}^4 \,, \quad
\Gamma (N \to L\phi) \ \propto \ m_\nu M_N^{2} /M_W^2 \,.
\end{eqnarray}
For the two-body decays we have taken into account also the dependence ${\cal M}_D \propto (m_\nu M_N)^{1/2}$. When all other parameters are fixed and $M_N$ gets larger, the three-body width $\Gamma (N \to \ell q \bar{q}')$ grows faster than that for the two-body decays $N \to L\phi$, therefore the $W_R$ mass has also be larger to make sure that the two-body branching fraction ${\rm BR}_y$ and the resultant lepton asymmetry is not highly suppressed. When all other parameters are fixed and the RHN masses get smaller, the washout factor in Eq.~(\ref{eqn:Keff}) and the efficiency factor in Eq.~(\ref{eqn:epsilon2})  become larger and the dilution factor in Eq.~(\ref{eqn:dilution}) get smaller; when all these factors are combined in Eq.~(\ref{eqn:eta}), there is an
{\it absolute} lower limit on $W_R$ mass from leptogenesis for the special case with only two RHNs:
\begin{eqnarray}
\label{eqn:WRlimits}
M_{W_R} \ > \left\{\begin{array}{cc} 9.4~{\rm TeV} \;\; & \text{for NH} \,, \\
8.9~{\rm TeV} \;\; & \text{for IH} \, ,
\end{array}\right.
\end{eqnarray}
which corresponds to the RHN masses at $M_N \sim 500$ GeV in Fig.~\ref{fig:WRlimit}. This is similar to the $W_R$ mass bounds found in a different version of LRSM~\cite{DLM}, including all three RHNs. Furthermore, the scalings in Eq.~(\ref{eqn:scalings}) are not very sensitive to the neutrino oscillation parameters, thus the NH and IH limits in Fig.~\ref{fig:WRlimit} are roughly the same, in particular when the RHNs are heavy.
%When the third RHN also gets involved in leptogenesis, the lower bounds might be to some extent loosened. However, this depends largely on the third RHN mass and goes beyond the main scope of this paper.

\subsection{Testing leptogenesis at colliders} \label{sec:lepcol}

\begin{figure}[!t]
  \centering
  \includegraphics[width=0.52\textwidth]{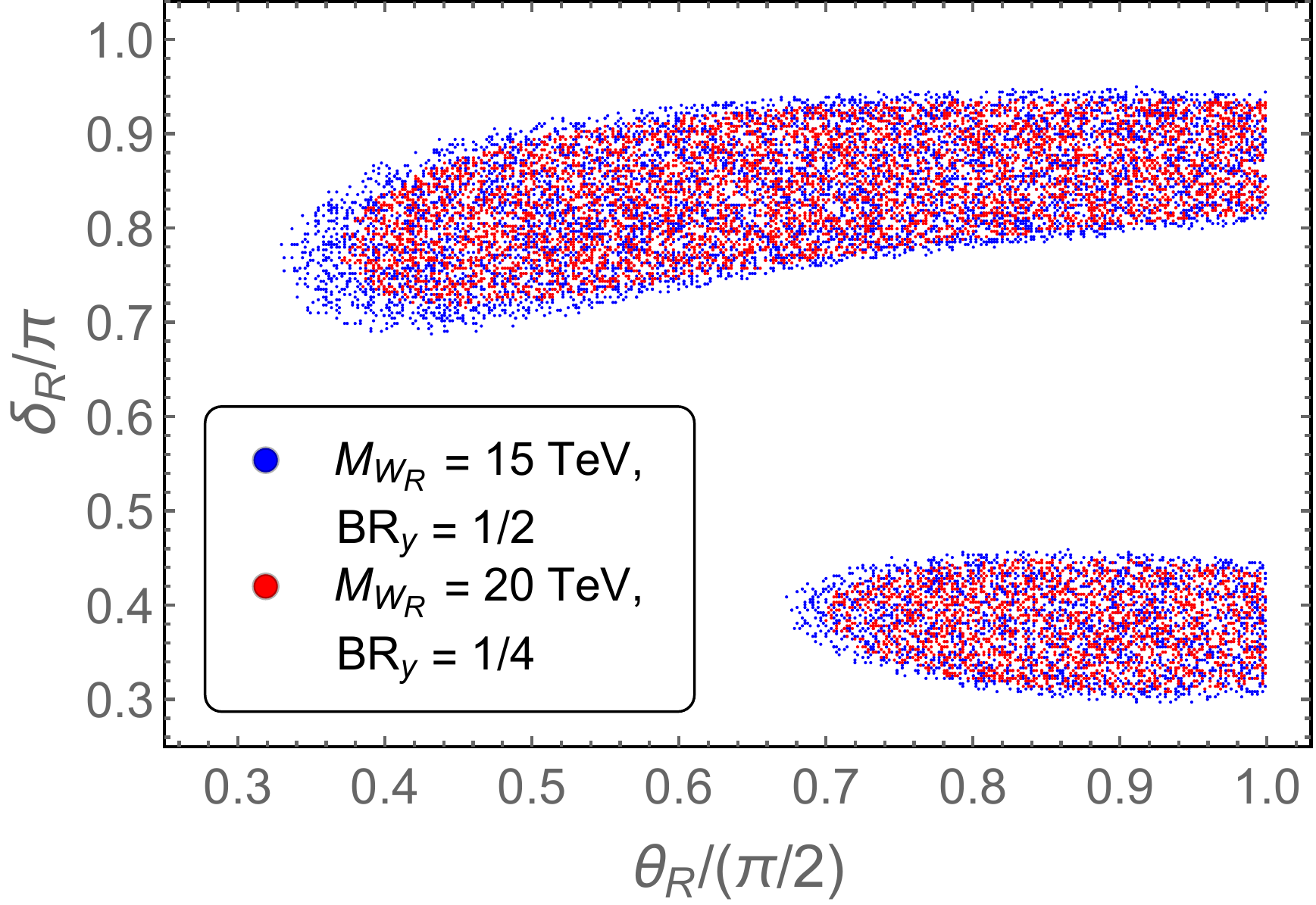}
  \caption{Scatter plot of $\theta_R$ and $\delta_R$ which can generate the observed baryon asymmetry, with two benchmark points of $M_{W_R} = 15$ TeV and ${\rm BR}_y = 1/2$ (blue) and $M_{W_R} = 20$ TeV and ${\rm BR}_y = 1/4$ (red). We have set the average RHN mass at 1 TeV in both cases.}
  \label{fig:example2}
\end{figure}

The RHN mixing angle $\theta_R$ and CP phase $\delta_R$ play an important role in Eq.~\eqref{eqn:epsilon2} for generating lepton asymmetry. Our numerical calculations reveal that only certain regions in the two-dimensional plane of $\theta_R$ and $\delta_R$ could produce the observed baryon asymmetry (in both magnitude and sign), as exemplified in Fig.~\ref{fig:example2}. We have chosen two benchmark points: (i) $M_{W_R} = 15$ TeV and ${\rm BR}_y = 1/2$ (the two- and three-body decay widths are equal), which is shown in blue; (ii) $M_{W_R} = 20$ TeV and ${\rm BR}_y = 1/4$ (the three-body decay width is three times larger than that for the two-body decays), depicted in red. In both cases, the $W_R$ mass is consistent with the absolute lower bound from leptogenesis given by Eq.~(\ref{eqn:WRlimits}). To be concrete, the average RHN mass is set at 1 TeV in both cases, and the light neutrino mass ordering is chosen to be NH. For the case of IH, the parameter space does not change too much for the two benchmark points. Some of the empty regions in Fig.~\ref{fig:example2},  for instance with $\theta_R < \pi/4$ and $\delta_R < \pi/2$, always generate a ``wrong'' sign for $\Delta \eta_L$, coming from the product of factors $(M_{N_1}^2 - M_{N_2}^2) {\rm Im} [y_{\alpha 2}^\ast y_{\alpha 2}] {\rm Re} [(y^\dagger y)_{12}]$ in Eq.~(\ref{eqn:epsilon2}).

The viable parameter space in blue and red in Fig.~\ref{fig:example2} are also shown as the brown shaded regions in Figs.~\ref{fig:contours}, \ref{fig:RCP2} and \ref{fig:RCP3}. It is clear that TeV-scale leptogenesis could be falsified at future hadron colliders by measuring the mixing angle $\theta_R$ and CP phase $\delta_R$ once the $W_R$ is discovered. For instance,  if it is found that $\theta_R < \pi/4$ and $\delta_R < \pi/2$, then leptogenesis will be excluded, at least in the case with only two quasi-degenerate RHNs. Furthermore, if the $W_R$ boson mass is found to be lower than the absolute leptogenesis limit given in Eq.~(\ref{eqn:WRlimits}), leptogenesis will also be excluded. One should note that to test leptogenesis directly at future high-energy hadron colliders, either the two-body (${\rm BR}_y$) or the three-body (${\rm BR}_g$) BRs of RHNs can not be too large or small: When the two-body decay BR is too large ($N_\alpha$ being lighter and/or $W_R$ heavier), the SSCAs signals at collider will be highly suppressed, as they originate only from the three-body decays. On the other hand, when the three-body decays dominate ($N_\alpha$ being heavier and/or $W_R$ lighter), the leptogenesis would be inefficient, as it is only from the Yukawa coupling induced two-body decays. However, we would like to stress that no matter whether leptogenesis is excluded or not, observation of mixing angle and CPV in the RHN sector at future hadron collider would be a significant step towards understanding the origin of neutrino masses.

\section{Low-energy constraints}
\label{sec:constraints}

%\begin{figure}[!t]
%  \centering
%  \includegraphics[width=0.48\textwidth]{LFV.pdf}
%  \includegraphics[width=0.48\textwidth]{DBD_1.pdf}
%  \includegraphics[width=0.48\textwidth]{DBD_2.pdf}
%  \includegraphics[width=0.48\textwidth]{EDM.pdf}
%  \caption{The dependence of LFV decay $\mu \to e\gamma$, the $0\nu\beta\beta$ decays and electron EDM on the parameter $\zeta$. }
%  \label{fig:limits}
%\end{figure}

The RHNs $N_i$ and heavy $W_R$ boson contribute to some of the low-energy and precision measurements such as $0\nu\beta\beta$ decays, LFV decay $\mu \to e\gamma$ and electron EDM, which could be used to set limits on the masses and couplings in the RH sector, in particular on the complex parameter $\zeta$ in the matrix ${\cal O}$~\cite{Dev:2014xea}. However, as we will see in this section, not all these low-energy precision measurements provide competitive limits on the parameter $\zeta$ for the parameter space of our interest.

%large imaginary angles in the matrix ${\cal O}$ are also constrained from high-precision low-energy constraints~, such as $\mu\to e\gamma$, neutrinoless double beta decay and electron electric dipole moment.

%\section{Conclusion}
%In light of the relation in Eq.~(\ref{eqn:seesaw}) with a $3\times2$ $M_D$ matrix,  there is very likely no parameter space in the minimal two RHN case to accommodate both resonant leptogenesis and SS dilepton signals at (100 TeV) colliders. We might need the third RHN to have more degrees of freedom (and fine-tuning) in the $M_D$ to have large Yukawa couplings and lighter $W_R$ boson, as in Ref.~\cite{Dev:2014iva}.

\subsection{Neutrinoless double beta decay}
\label{sec:0nubetabeta}

\begin{figure}
  \centering
  %\hspace{-30pt}
  \begin{subfigure}[b]{0.26\textwidth}
  \includegraphics[height=\textwidth]{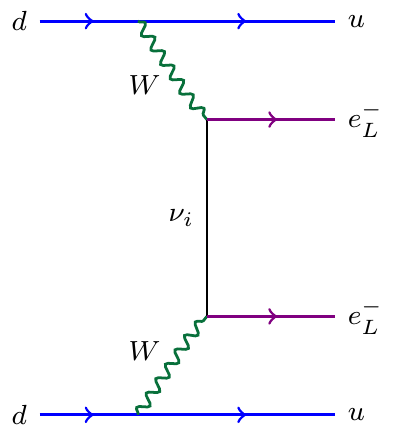}
  \caption{}
  \label{fig:a}
  \end{subfigure}
  \begin{subfigure}[b]{0.26\textwidth}
  \includegraphics[height=\textwidth]{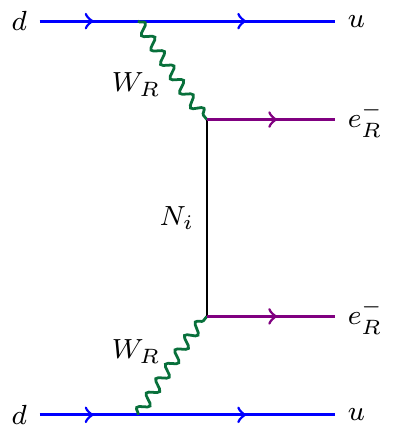}
  \caption{}
  \label{fig:b}
  \end{subfigure} \\
  \begin{subfigure}[b]{0.26\textwidth}
  \includegraphics[height=\textwidth]{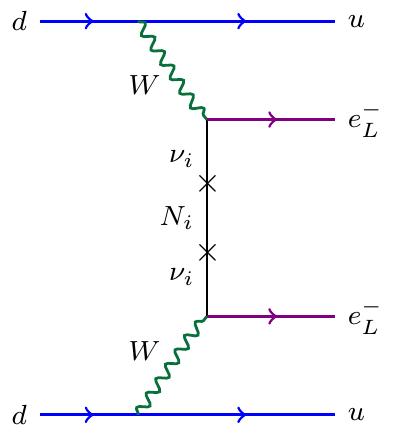}
  \caption{}
  \label{fig:c}
  \end{subfigure}
  \begin{subfigure}[b]{0.26\textwidth}
  \includegraphics[height=\textwidth]{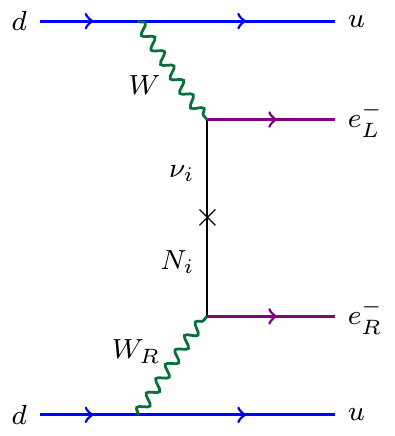}
  \caption{}
  \label{fig:d}
  \end{subfigure}
  \begin{subfigure}[b]{0.26\textwidth}
  \includegraphics[height=\textwidth]{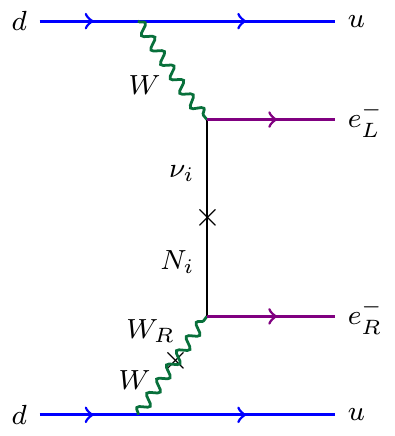}
  \caption{}
  \label{fig:e}
  \end{subfigure}
  \caption{Feynman diagrams for relevant contributions to $0\nu\beta\beta$ decays in the LRSM with type-I seesaw dominance: (a) ${\cal A}_\nu$, (b) ${\cal A}_{N}^R$, (c) ${\cal A}_{N}^L$, (d) ${\cal A}_{\lambda}$, and (e) ${\cal A}_\eta$. The $\times$ symbol on the fermion propagators denotes heavy-light neutrino mixing, and the $\times$ symbol on the gauge boson propagator in diagram (e) denotes the $W-W_R$ mixing. }
  \label{fig:diagram}
\end{figure}

If the light neutrinos in the SM are Majorana particles, they will induce $0\nu\beta\beta$ decays through the diagram illustrated in Fig.~\ref{fig:a}. In the LRSM, there are extra contributions from interactions of the heavy RHNs, heavy $W_R$ and the doubly-charged scalar $H_R^{\pm\pm}$ from the right-handed triplet $\Delta_R$~\cite{Rodejohann:2011mu}. In the LRSM we are considering, the left-handed triplet $\Delta_L$ decouples from the TeV-scale physics and the neutrino masses are type-I seesaw dominated. The $H_R^{--}$ contribution is proportional to $(f_R)_{ee} v_R/M_{H_{R}^{\pm\pm}}^{2}$ with $f_R$ being the Yukawa couplings of $H_R^{\pm\pm}$ to the RHNs and $M_{H_R^{\pm\pm}}$ the mass of $H_R^{\pm\pm}$~\cite{Bambhaniya:2015ipg, Dev:2018sel}; for simplicity we assume $H_R^{\pm\pm}$ is heavy such that its contribution is negligible. Then the dominant contributions to $0\nu\beta\beta$ decays in LRSM are all shown in Fig.~\ref{fig:diagram}, including the RHN-mediated diagrams in Fig.~\ref{fig:b}, heavy-light neutrino mixing in Figs.~\ref{fig:c} and \ref{fig:d}, and $W - W_R$ mixing in Fig.~\ref{fig:e}. For a given isotope, the lifetime of $0\nu\beta\beta$ can be factorized to be of the form~\cite{Barry:2013xxa}
\begin{equation}
\frac{1}{T_{1/2}^{0\nu}} \ = \
G_{01}^{0\nu} \Bigg[ \Big|{\cal M}_\nu^{0\nu}\eta_\nu+{\cal M}_N^{0\nu}\eta_{N_R}^L\Big|^2 + \Big|{\cal M}_N^{0\nu}\eta_{N_R}^R\Big|^2 + \Big|{\cal M}_\lambda^{0\nu}\eta_\lambda + {\cal M}_\eta^{0\nu}\eta_\eta\Big|^2 \Bigg]  \; ,
\label{half}
\end{equation}
where $G_{01}^{0\nu}$ is the phase space factor and ${\cal M}^{0\nu}_{\nu,\, N,\, \lambda,\, \eta}$ are the relevant nuclear matrix elements~\cite{Kotila:2012zza, Pantis:1996py, Meroni:2012qf}. The $\eta$'s are the dimensionless particle physics parameters obtained from the Feynman diagrams in Fig.~\ref{fig:diagram}:
\begin{eqnarray}
\label{eqn:etanu}
\eta_\nu & \ = \ & \frac{1}{m_e}\sum_i U_{ei}^2 m_i \,, \\
\label{eqn:etaRR}
\eta^R_{N} & \ = \ & m_p \left(\frac{M_{W}}{M_{W_R}}\right)^4 \sum_i  \frac{{V_{ei}^*}^2}{M_{N_i}} \,, \\
\label{eqn:etaRL}
\eta^L_{N}  & \ = \ &  m_p \sum_i \frac{S^2_{ei}}{M_{N_i}} \,, \\
\label{eqn:etalam}
\eta_{\lambda} & \ = \ &  \left(\frac{M_{W}}{M_{W_R}}\right)^2 \sum_i U_{ei} T^*_{ei} \,,  \\
\label{eqn:etaeta}
\eta_{\eta} & \ = \ &  \tan \xi \sum_i U_{ei} T^*_{ei} \,,
\end{eqnarray}
where $m_e$ and $m_p$ are respectively the masses of electron and proton, $S$ and $T$ are the heavy-light neutrino mixing matrices given by (see Appendix~\ref{sec:ST})
\begin{equation}
\label{eqn:ST}
S \ \simeq \ {\cal M}_D {\cal M}^{-1}_N V  \quad \text{and} \quad
T \ \simeq \ - ({\cal M}_D {\cal M}^{-1}_N)^{\dagger} U \,.
\end{equation}
As the Dirac mass matrix ${\cal M}_D \propto {\cal M}_N^{1/2}$, the heavy-light neutrino mixing matrices $S,\, T \propto {\cal M}_N^{-1/2}$, and therefore, the RHN and $W_R$ contributions in Eqs.~(\ref{eqn:etaRR})-(\ref{eqn:etaeta}) can always be made small by taking large RHN and/or $W_R$ masses. Furthermore, with the assumption of $\kappa' = 0$ in Section~\ref{sec:two-body-decay}, the $W - W_R$ boson mixing parameter $\xi = 0$ in Eq.~(\ref{eqn:etaeta}) (see e.g. Refs.~\cite{Deshpande:1990ip, Zhang:2007da, Dev:2016dja} for more details). With this choice, the current most stringent $0\nu\beta\beta$ decay limits from KamLAND-Zen~\cite{KamLAND-Zen:2016pfg} and GERDA~\cite{Agostini:2018tnm}
%, EXO-200~\cite{Albert:2017owj}, CUORE~\cite{Alduino:2017ehq}, and NEMO-3~\cite{Arnold:2018tmo}
could exclude some of the parameter space, as shown in Fig.~\ref{fig:asymmetry1}, but they can not provide any robust limits on the $\zeta$ parameter in the ${\cal O}$ matrix.
%\begin{eqnarray}
%\xi \ \simeq \
%- \frac{\kappa \kappa'}{v_R^2} \ \simeq \
%- \frac{\kappa'}{\kappa} \frac{v_{\rm EW}^2}{v_R^2}
%\end{eqnarray}

%with $\kappa$ and $\kappa'$ are the two VEVs of the bidoublet $\Phi$ in the LRSM (see e.g. Refs.~\cite{Deshpande:1990ip, Zhang:2007da, Dev:2016dja} for more details). To be consistent with the assumption $\kappa'=0$ we have made in Section~\ref{sec:two-body-decay}, we set the $W - W_R$ mixing to be zero, i.e. $\xi =0$. %The mixing $\xi$ is an extremely small parameter and can even be made vanishing by taking $\kappa' = 0$ ).

\subsection{LFV decay $\mu \to e\gamma$}
\label{sec:LFV}

The RHNs and $W_R$ boson might also induce new contributions to the LFV decay $\mu \to e \gamma$, which is predicted to be~\cite{Barry:2013xxa, Cirigliano:2004mv}
\begin{equation}
{\rm BR}(\mu \to e\gamma) \ = \
\frac{\alpha_w^3 s_w^2}{256\pi^2}\frac{m_\mu^4}{M_{W}^4}
\frac{m_\mu}{\Gamma_\mu}\left(|G_L^\gamma|^2+|G_R^\gamma|^2\right) \; , \label{mueg}
\end{equation}
where $m_\mu$ and $\Gamma_\mu$ are respectively the muon mass and width, $s_w \equiv \sin\theta_w$ is the weak mixing parameter, $\alpha_w \equiv g_L^2/4\pi$ is the weak coupling strength, and the form factors $G_{L,R}^\gamma$ are respectively~\cite{Barry:2013xxa}
\begin{eqnarray}
\label{eqn:mueg1}
G_L^\gamma & \ = \ &
\sum_i\left[V_{\mu i} V^*_{ei}\left( |\xi|^2G_1^\gamma(x_i)+\left(\frac{M_{W}}{M_{W_R}}\right)^2 G_1^\gamma(y_i)\right) -S^*_{\mu i}V^*_{ei} \xi \frac{M_{N_i}}{m_\mu} G_2^\gamma(x_i)\right]\,, \\
\label{eqn:mueg2}
G_R^\gamma &=& \sum_i\left[S^*_{\mu i}S_{ei} G_1^\gamma(x_i) - V_{\mu i}S_{ei} \xi \frac{M_{N_i}}{m_\mu} G_2^\gamma(x_i)\right] \,,
\end{eqnarray}
with $x_i\equiv (M_{N_i}/M_{W})^2$, $y_i\equiv (M_{N_i}/M_{W_R})^2$, and the loop functions $G^\gamma_{1,2}(x)$ are defined as
\begin{eqnarray}
G_1^\gamma(x) & \ \equiv \ &
-\frac{x(2x^2+5x-1)}{4(1-x)^3} - \frac{3x^3}{2(1-x)^4}\ln x \; ,\\
G_2^\gamma(x) & \ \equiv \ &
\frac{x^2-11x+4}{2(1-x)^2} - \frac{3x^2}{(1-x)^3}\ln x \; . \label{g2g}
\end{eqnarray}

In the case of $W - W_R$ mixing $\xi = 0$ as a result of $\kappa'=0$, the first and third terms in Eq.~(\ref{eqn:mueg1}) and the second term in Eq.~(\ref{eqn:mueg2}) are all vanishing, and the second term in Eq.~(\ref{eqn:mueg1}) is proportional to $V_{\mu 1} V_{e1}^\ast + V_{\mu 2} V_{e2}^\ast = 0$ in the limit of $M_{N_1} = M_{N_2}$; therefore, only the first term in Eq.~(\ref{eqn:mueg2}) contributes to the LFV decay $\mu \to e\gamma$. As shown in Appendix~\ref{sec:ST}, the $S_{\alpha i}$ elements are the heavy-light mixing angles to the leading order, and the $|G_R^\gamma|^2$ factor is highly suppressed by the heavy-light mixing angle to the fourth power. In addition, the prefactors in Eq.~(\ref{mueg}) are of order $10^{-3}$.
Evaluating numerically for the scatter points in Fig.~\ref{fig:asymmetry1}, it turns out that the LFV decay branching ratio ${\rm BR} (\mu \to e\gamma) \lesssim 10^{-17}$, which is orders of magnitude smaller than the current limit ${\rm BR} (\mu \to e\gamma) < 4.2 \times 10^{-13}$ from MEG experiment~\cite{Adam:2013mnn}. Therefore in the LRSM scenario we are considering, the LFV decays do not set more stringent limits on the RHN sector than the leptogenesis and $0\nu\beta\beta$ decay limits.

%The functions $G_{1,\,2}^\gamma (x)<1$ for any values of $x$, and all the terms in Eqs.~(\ref{eqn:mueg1}) and (\ref{eqn:mueg2}) can be highly suppressed in the limit of large RHN masses $M_i$, heavy $W_R$ boson and/or vanishing $W - W_R$ mixing ($\xi \to 0$). As the case for $0\nu\beta\beta$ decays, the ${\rm BR} (\mu \to e\gamma)$ limit from MEG experiment~\cite{Adam:2013mnn} can not impose any robust constraints on the $\zeta$ parameter.

\subsection{Electron EDM}
\label{sec:EDM}

The $W - W_R$ mixing might lead to a beyond SM contribution to the electron EDM at 1-loop level, which reads~\cite{Nieves:1986uk}
\begin{equation}
d_e \ \simeq \
\frac{e \alpha_w^2}{8\pi M_W^2}
{\rm Im} \left[ \sum_i \xi S_{ei} V_{ei} G_2^\gamma (x_i) M_{N_i} \right]
\label{edm}
\end{equation}
Although the RHN mixing matrix $V$ is complex in the presence of CP phase $\delta_R$, the LRSM contribution to electron EDM is vanishing in the limit of $\xi \to 0$, therefore the electron EDM limit from ACME experiment~\cite{Andreev:2018ayy} does not set any limits on the LRSM we are considering.

\section{Conclusion}
\label{sec:conclusion}

The type-I seesaw is one of the most compelling scenarios for explaining tiny neutrino masses. It uses heavy RHNs with Majorana masses as the two key ingredients. However, in absence of any evidence for new physics beyond SM, our current knowledge of RHNs is very limited. In this paper we have shown how one class of RHN models where two of the RHNs are quasi-degenerate with mixings and associated CP violation, and are part of the TeV-scale LRSM framework, can be directly probed at future high-energy hadron colliders, by measuring the charge asymmetries in the same-sign dilepton final states, e.g. number of $\ell^+ \ell^+$ versus $\ell^- \ell^-$ events (with $\ell=e,\mu$) and associated SSCA observables ${\cal A}_{\alpha\beta}$ and ${\cal R}_{\rm CP}^{(\ell)}$ defined in Eqs.~\eqref{eqn:Aab} and \eqref{eqn:Rab}, respectively.

We find that due to the large PDF uncertainties, the CP-induced SSCAs ${\cal A}_{\alpha\beta}$ can only be measured at future higher energy colliders,  such as the $\sqrt{s} = 27$ TeV HE-LHC and the 100 TeV FCC-hh or SPPC, but not at the HL-LHC, as illustrated in Fig.~\ref{fig:prospect}. The $e^\pm \mu^\pm$ channel is particularly suitable for measuring the CP phase through ${\cal A}_{e\mu}$, as it does not depend on the RHN mixing angle $\theta_R$. When combined with  the $e^\pm e^\pm$ (and/or $\mu^\pm \mu^\pm$) data, we can use the measurement of ${\cal A}_{\alpha\beta}$ to determine both the RHN mixing angle and CP phase. With the measurement of ${\cal A}_{\alpha\beta}$, the CPV in the RHN sector can be probed with $W_R$ mass up to 6.4 TeV at 27 TeV HE-LHC, which can be improved up to 26 TeV at future 100 TeV colliders.

On the other hand, the ratios ${\cal R}_{\rm CP}^{(e,\,\mu)}$ do not suffer from the PDF uncertainties and can be measured at both LHC and future higher-energy colliders.  Combining all the $e^\pm e^\pm$, $\mu^\pm \mu^\pm$ and $e^\pm \mu^\pm$ channels, measurement of ${\cal R}_{\rm CP}^{(e,\,\mu)}$ can determine both the RHN mixing angle and CP phase, as shown in Figs.~\ref{fig:RCP2} and \ref{fig:RCP3}. Furthermore, as the ratios ${\cal R}_{\rm CP}^{(e,\,\mu)}$ do not depend on the proton PDFs, they can be used to probe the RHN sector up to higher $W_R$ mass, as long as the $W_R$ boson can be produced with an observable rate in the same-sign dilepton channel. When the two-body decays of RHNs are sizable, both ${\cal A}_{ee}$ and ${\cal A}_{\mu\mu}$, as well as ${\cal R}_{\rm CP}^{(e)}$ and ${\cal R}_{\rm CP}^{(\mu)}$, are expected to be different, depending on the Dirac Yukawa coupling matrix structure; thus the measurements of ${\cal A}_{\alpha\beta}$ and ${\cal R}_{\rm CP}^{(e,\,\mu)}$ would help us get information on the seesaw mass matrix.

The RHN mixing and CP violation also play a key role in generating the observed baryon asymmetry in the framework of TeV-scale resonant leptogenesis. Thus the SSCA measurements at future high-energy colliders can be used to directly test leptogenesis as the mechanism for origin of matter, as exemplified in Figs.~\ref{fig:contours}, \ref{fig:RCP2} and \ref{fig:RCP3}. This depends largely on the branching ratios of the two- and three-body decays of RHNs, as the SSCA signatures at collider can only be induced by the gauge coupling mediated three-body decays of RHNs, while baryon asymmetry is only generated from the Yukawa coupling induced two-body decays. We find that leptogenesis requires the $\zeta$ parameter in the Casas-Ibarra parametrization to be within the range $1.3 \lesssim {\rm Im} \zeta \lesssim 7.8$ for the NH case and $0.8 \lesssim {\rm Im} \zeta \lesssim 7.7$ for the IH case (see Fig.~\ref{fig:asymmetry1}), corresponding to a maximum Dirac Yukawa coupling $1.3 \times 10^{-6} \lesssim |y|_{\rm max} \lesssim 7.2 \times 10^{-4}$ for the NH case and $1.0 \times 10^{-6} \lesssim |y|_{\rm max} \lesssim 8.6 \times 10^{-4}$ for the IH case. In the minimal LRSM leptogenesis we are considering here, there is a  lower limit on $W_R$ mass, which depends on the RHN mass, as shown in Fig.~\ref{fig:WRlimit}. The absolute lower bound $W_R$ mass was found to be 9.4 TeV for NH and 8.9 TeV for IH ordering of active neutrino masses. Regardless of whether leptogenesis can be tested at the LHC, HE-LHC or FCC-hh, observation of mixing angle and CP phase in the RHN sector at future hadron colliders would be a significant step in understanding the origin of tiny neutrino masses for this particular scenario (quasi-degenerate RHNs) for the type-I seesaw paradigm.
%Although RHN mixing and CP violation also contribute to low-energy observables such as $0\nu\beta\beta$ decays, LFV decay $\mu \to e\gamma$ and electron EDM, these precision measurements can not provide any robust limit on the RHN sector in the LRSM  considered in this paper.

\section*{Acknowledgements}

We thank Richard Ruiz for helpful discussion. The work of P.S.B.D. and Y.Z. is supported by the US Department of Energy under Grant No. DE-SC0017987. The work of R.N.M. is supported by the US National Science Foundation Grant No. PHY1620074. Y.Z. would like to thank the Institute of Theoretical Physics, Chinese Academy of Sciences, the Tsung-Dao Lee Institute, and the Institute of High Energy Physics, Chinese Academy of Sciences for generous hospitality where the paper was partially done.

\appendix

\section{Analytic formula for ${\cal M}_N^{1/2}$}
\label{sec:analytic}

\subsection{The $2\times2$ case}
\label{sec:analytic2}

The elements for the $2\times2$ matrix ${\cal M}_N^{1/2}$ are respectively
\begin{eqnarray}
\label{eqn:MNhalf11}
\left( {\cal M}_N^{1/2} \right)_{1,1} & \ = \ &
\frac{\sqrt{M_N} (z_{+} + z_{-})}{2}
- \frac{i \sqrt{M_N} (z_{+} - z_{-})}{u}
\bigg[ |\sin\theta_R| \cos\delta_R \, {\rm sign}(\sin\delta_R) \nonumber \\
&& -\frac{ir (\cos^2\theta_R - \sin^2\theta_R \cos2\delta_R)}{4|\sin\theta_R \sin\delta_R|} \bigg] \,, \\
\label{eqn:MNhalf22}
\left( {\cal M}_N^{1/2} \right)_{2,2} (r) & \ = \ & \left( {\cal M}_N^{1/2} \right)^\ast_{1,1} (-r) \,, \\
\label{eqn:MNhalf12}
\left( {\cal M}_N^{1/2} \right)_{1,2} & \ = \ &
- \frac{i \sqrt{M_N} (z_{+} - z_{-}) \cos\theta_R \, {\rm sign}(\sin\theta_R) \, {\rm sign}(\sin\delta_R)}{u}
\left( 1 + \frac{ir \cot\theta_R}{2} \right) \,, \nonumber \\ &&
\end{eqnarray}
where $r = \Delta M_N/M_N$ with $M_N$ being the averaged RHN mass and $\Delta M_N$ the mass splitting, and we have defined
\begin{eqnarray}
u & \ \equiv \ &
\frac{1}{|\sin\theta_R \sin\delta_R|}
\sqrt{\left[ \cos^2\theta_R + \sin^2\theta_R \left( \cos2\delta_R - \frac{ir}{2} \sin2\delta_R \right) \right]^2 - \left( 1 - \frac{r^2}{4} \right) } \,, \\
z_\pm & \ \equiv \ &
\sqrt{\cos^2\theta_R + \sin^2\theta_R \left( \cos2\delta_R - \frac{ir}{2} \sin2\delta_R \right) \pm u |\sin\theta_R \sin\delta_R|} \,.
\end{eqnarray}
Note that the matrix ${\cal M}_N^{1/2}$ can {\it not} be diagonalized by the $U_R$ matrix given in Eq.~(\ref{eqn:UR}), i.e.
\begin{eqnarray}
{\cal M}_N^{1/2} \ \neq \ U_R \widehat{{\cal M}}_N^{1/2} U_R^T \,.
\end{eqnarray}

In the limit of $r \to 0$, i.e. the two RHNs are mass-degenerate,
\begin{eqnarray}
u^{(0)} & \ = \ & u(r=0) \ = \
2i\sqrt{\cos^2\theta_R + \sin^2\theta_R \cos^2\delta_R} \,, \\
z_{\pm}^{(0)} & \ = \ &
z_\pm (r=0) \ = \
\sqrt{\cos^2\theta_R + \sin^2\theta_R  \cos2\delta_R \pm u |\sin\theta_R \sin\delta_R|} \,,
\end{eqnarray}
and we obtain
\begin{eqnarray}
\left( {\cal M}_N^{1/2} \right)_{1,1} & \ = \ &
\frac{\sqrt{M_N}}{2} \left(z_{+}^{(0)} + z_{-}^{(0)} \right)
- \frac{i \sqrt{M_N} |\sin\theta_R| \cos\delta_R \, {\rm sign}(\sin\delta_R)}{u^{(0)}} \left(z_{+}^{(0)} - z_{-}^{(0)} \right) \,, \\
\label{eqn:MNhalf22}
\left( {\cal M}_N^{1/2} \right)_{2,2} & \ = \ & \left( {\cal M}_N^{1/2} \right)^\ast_{1,1} \,, \\
\left( {\cal M}_N^{1/2} \right)_{1,2} & \ = \ &
- \frac{i \sqrt{M_N} \cos\theta_R \, {\rm sign}(\sin\theta_R) \, {\rm sign}(\sin\delta_R)}{u^{(0)}} \left(z_{+}^{(0)} - z_{-}^{(0)} \right) \,.
\end{eqnarray}

\subsection{Generalization to the $3\times 3$ case}
\label{sec:analytic3}

When we have the third neutrino $N_\tau$ in the TeV-scale LRSM, the analytic formulae for the elements of ${\cal M}_N^{1/2}$ obtained above can be directly generalized to the three RHN case, as long as the third RHN does not mix with the first two RHNs. If the third RHN has mass $M_{N_3}$, then the $3\times3$ matrix ${\cal M}_N^{1/2}$ can be written in the form of
\begin{eqnarray}
{\cal M}_N^{1/2} \ = \
\left( \begin{matrix}
\big( {\cal M}_N^{1/2} \Big)_{1,1} & \big( {\cal M}_N^{1/2} \Big)_{1,2} & 0 \\
\big( {\cal M}_N^{1/2} \Big)_{1,2} & \big( {\cal M}_N^{1/2} \Big)_{2,2} & 0 \\
0 & 0 & M_{N_3}^{1/2}
\end{matrix} \right) \,,
\end{eqnarray}
with the $(1,1)$, $(1,2)$ and $(2,2)$ elements given in Eqs.~(\ref{eqn:MNhalf11}) to (\ref{eqn:MNhalf12}). To be self-consistent, the arbitrary matrix in Eq.~(\ref{eqn:Ocal}) has to be altered accordingly to be of the form
\begin{eqnarray}
{\cal O} \ = \
\begin{pmatrix}
1 & 0 & 0 \\
0 & \cos\zeta & \sin\zeta  \\
0 & -\sin\zeta & \cos\zeta
\end{pmatrix} \;\; \text{ for NH} \,,  \quad \text{or} \;\;
\begin{pmatrix}
\cos\zeta & \sin\zeta & 0  \\
-\sin\zeta & \cos\zeta & 0 \\
0 & 0 & 1
\end{pmatrix} \text{ for IH} \,,
\end{eqnarray}
with the orthogonal condition ${\cal O}^{\sf T} {\cal O} = {\bf 1}_{3\times3}$. One should note that in the three RHN case, the lightest neutrino mass is not vanishing any more.

%\section{The case without CPV ($\theta_R \neq 0$ and $\delta_R = 0$)}
%
%If we have only RHN mixing but no CP violation, i.e. $\theta_R \neq 0$ and $\theta_R = 0$, then we charge asymmetry ${\cal A}_{\alpha\beta}$ in Eq.~(\ref{eqn:asymmetry}) are the same as the case without any RHN and CP violation. However, the ${\rm BR} (W_R^\pm \to \ell_\alpha^\pm \ell_\beta^\pm jj)$ will be different. We define the ratio
%\begin{eqnarray}
%R_{\alpha\beta}  =
%\frac{N (\ell_\alpha^\pm \ell_\beta^\pm)}
%{\sum_{\alpha\beta} N (\ell_\alpha^\pm \ell_\beta^\pm)} \,,
%\end{eqnarray}
%and the plots $R_{ee,\, \mu\mu}$ and $R_{e\mu}$ are shown in Fig.~\ref{fig:BR}. In particular, when there is no RHN mixing, i.e. $\theta_R = 0$, we have only the $e^\pm e^\pm$ and $\mu^\pm \mu^\pm$ events but no $e^\pm \mu^\pm$.

%\begin{figure}[!t]
%  \centering
%  \includegraphics[width=0.49\textwidth]{BR.pdf}
%  \caption{The ratio defined in Eq.~(\ref{eqn:Rab}) for $R_{ee,\,\mu\mu}$ and $R_{e\mu}$ as functions of RHN mixing angle $\theta_R$ for the case without any CPV. }
%  \label{fig:BR}
%\end{figure}

\section{Heavy-light neutrino mixing}
\label{sec:ST}

In the effective LRSM we are considering, there are three active neutrino and two heavy RHNs, and the full $5\times5$ neutrino mass matrix
\begin{equation}
{\cal M}_\nu \ = \
\left(\begin{array}{cc}
0 & {\cal M}_D \\ {\cal M}_D^{\sf T} & {\cal M}_N
\end{array}\right)
\end{equation}
can be diagonalized by a unitary matrix:
\begin{equation}
{\cal V}^{\sf T}{\cal M}_\nu {\cal V} \ = \
\left(\begin{array}{cc} \widehat{m}_\nu & {\bf 0} \\ {\bf 0} & \widehat{\cal M}_N \end{array}\right) ,
\end{equation}
where $\widehat{m}_\nu = {\rm diag}(m_1,m_2,m_3)$ and $\widehat{\cal M}_N = {\rm diag}(M_{N_1},\,M_{N_2})$. The unitary matrix ${\cal V}$ has an exact representation in terms of the matrix $\vartheta$~\cite{Korner:1992zk,Grimus:2000vj}:
\begin{equation}
{\cal V} \ = \ \left(\begin{array}{cc}
({\bf 1}+ \vartheta^*\vartheta^{\sf T})^{-1/2} &
\vartheta^*({\bf 1}+\vartheta^{\sf T}\vartheta^*)^{-1/2} \\
-\vartheta^{\sf T}({\bf 1}+\vartheta^*\vartheta^{\sf T})^{-1/2} & ({\bf 1}+\vartheta^{\sf T}\vartheta^*)^{-1/2}
\end{array}  \right)
\left(\begin{array}{cc}
U_\nu & {\bf 0} \\
{\bf 0} & V_R
\end{array}\right) \ \equiv \ \left(\begin{array}{cc}
U & S \\
T & V
\end{array}\right) \; ,
\label{V}
\end{equation}
where $\vartheta^*= {\cal M}_D {\cal M}_N^{-1}$ to leading order in a converging Taylor series expansion, $U_\nu$ and $V_R$ are respectively the unitary matrices diagonalizing the light and heavy neutrino mass matrices. To the leading order of $\vartheta$, we obtain the heavy-light neutrino mixing matrices $S$ and $T$ given in Eq.~(\ref{eqn:ST}).


\begin{thebibliography}{99}

\bibitem{seesaw1} P. Minkowski, Phys. Lett. B {\bf 67}, 421 (1977).

\bibitem{seesaw2} R. N. Mohapatra and G. Senjanovi\'{c}, Phys. Rev. Lett. {\bf 44}, 912 (1980).

\bibitem{seesaw3} T. Yanagida, Conf.  Proc.  C {\bf 7902131},  95  (1979).

\bibitem{seesaw4} M. Gell-Mann, P. Ramond and R. Slansky, Conf. Proc. C {\bf 790927}, 315 (1979) [arXiv:1306.4669 [hep-th]].

\bibitem{seesaw5} S.~L.~Glashow, NATO Sci. Ser. B {\bf 61}, 687 (1980).

\bibitem{Fukugita:1986hr}
  M.~Fukugita and T.~Yanagida,
  %``Baryogenesis Without Grand Unification,''
  Phys.\ Lett.\ B {\bf 174}, 45 (1986).
%  doi:10.1016/0370-2693(86)91126-3

\bibitem{Buchmuller:2004nz}
  W.~Buchmuller, P.~Di Bari and M.~Plumacher,
  %``Leptogenesis for pedestrians,''
  Annals Phys.\  {\bf 315}, 305 (2005)
 % doi:10.1016/j.aop.2004.02.003
  [hep-ph/0401240].

\bibitem{RL4}
  A.~Pilaftsis and T.~E.~J.~Underwood,
  %``Resonant leptogenesis,''
  Nucl.\ Phys.\ B {\bf 692}, 303 (2004)
 % doi:10.1016/j.nuclphysb.2004.05.029
  [hep-ph/0309342].

%\bibitem{Marshak:1979fm}
 % R.~E.~Marshak and R.~N.~Mohapatra,
  %``Quark - Lepton Symmetry and B-L as the U(1) Generator of the Electroweak Symmetry Group,''
  %Phys.\ Lett.\  {\bf 91B}, 222 (1980).



%\bibitem{Drewes} M.~Drewes, J.~Klari? and P.~Klose,
%  %``On Lepton Number Violation in Heavy Neutrino Decays at Colliders,''
%  arXiv:1907.13034 [hep-ph].


\bibitem{Pati:1974yy}
  J.~C.~Pati and A.~Salam,
  %``Lepton Number as the Fourth Color,''
  Phys.\ Rev.\ D {\bf 10}, 275 (1974)
  Erratum: [Phys.\ Rev.\ D {\bf 11}, 703 (1975)].
  %doi:10.1103/PhysRevD.10.275, 10.1103/PhysRevD.11.703.2

\bibitem{Mohapatra:1974gc}
  R.~N.~Mohapatra and J.~C.~Pati,
  %``A Natural Left-Right Symmetry,''
  Phys.\ Rev.\ D {\bf 11}, 2558 (1975).
  %doi:10.1103/PhysRevD.11.2558

\bibitem{Senjanovic:1975rk}
  G.~Senjanovic and R.~N.~Mohapatra,
  %``Exact Left-Right Symmetry and Spontaneous Violation of Parity,''
  Phys.\ Rev.\ D {\bf 12}, 1502 (1975).
  %doi:10.1103/PhysRevD.12.1502

%\bibitem{Jain:1994hd}
%  V.~Jain and R.~Shrock,
%  %``Models of fermion mass matrices based on a flavor dependent and generation dependent U(1) gauge symmetry,''
%  Phys.\ Lett.\ B {\bf 352}, 83 (1995)
%  doi:10.1016/0370-2693(95)00472-W
%  [hep-ph/9412367].

%\bibitem{Langacker:2000ju}
%  P.~Langacker and M.~Plumacher,
  %``Flavor changing effects in theories with a heavy $Z^\prime$ boson with family nonuniversal couplings,''
%  Phys.\ Rev.\ D {\bf 62}, 013006 (2000)
%  doi:10.1103/PhysRevD.62.013006
%  [hep-ph/0001204].

%\bibitem{Dev:2019rxh}
%  P.~S.~Bhupal Dev, R.~N.~Mohapatra and Y.~Zhang,
%  %``Probing heavy neutrino mixing and associated CP violation at future hadron colliders,''
%  arXiv:1904.04787 [hep-ph].



\bibitem{Bray:2007ru}
  S.~Bray, J.~S.~Lee and A.~Pilaftsis,
  %``Resonant CP violation due to heavy neutrinos at the LHC,''
  Nucl.\ Phys.\ B {\bf 786}, 95 (2007)
  %doi:10.1016/j.nuclphysb.2007.07.002
  [hep-ph/0702294].

  \bibitem{Blanchet:2009bu}
  S.~Blanchet, Z.~Chacko, S.~S.~Granor and R.~N.~Mohapatra,
  %``Probing Resonant Leptogenesis at the LHC,''
  Phys.\ Rev.\ D {\bf 82}, 076008 (2010)
 [arXiv:0904.2174 [hep-ph]].


\bibitem{KS}
  W.~Y.~Keung and G.~Senjanovi\'{c},
  %``Majorana Neutrinos and the Production of the Right-handed Charged Gauge Boson,''
  Phys.\ Rev.\ Lett.\  {\bf 50}, 1427 (1983).

  \bibitem{juan}  J.~C.~Vasquez,
  %``Right-handed lepton mixings at the LHC,''
  JHEP {\bf 1605}, 176 (2016).


\bibitem{GJ}
  J.~Gluza and T.~Jeli\'{n}ski,
  %``Heavy neutrinos and the pp?lljj CMS data,''
  Phys.\ Lett.\ B {\bf 748}, 125 (2015)
  %doi:10.1016/j.physletb.2015.06.077
  [arXiv:1504.05568 [hep-ph]].

\bibitem{GJ2} J.~Gluza, T.~Jelinski and R.~Szafron,
  %``Lepton number violation and ?Diracness? of massive neutrinos composed of Majorana states,''
  Phys.\ Rev.\ D {\bf 93}, no. 11, 113017 (2016)
  %doi:10.1103/PhysRevD.93.113017
  [arXiv:1604.01388 [hep-ph]].


\bibitem{DDM}
  A.~Das, P.~S.~B.~Dev and R.~N.~Mohapatra,
  %``Same Sign versus Opposite Sign Dileptons as a Probe of Low Scale Seesaw Mechanisms,''
  Phys.\ Rev.\ D {\bf 97}, no. 1, 015018 (2018)
  %doi:10.1103/PhysRevD.97.015018
  [arXiv:1709.06553 [hep-ph]].

\bibitem{Akhmedov:2007fk}
  E.~K.~Akhmedov,
  %``Do charged leptons oscillate?,''
  JHEP {\bf 0709}, 116 (2007)
  %doi:10.1088/1126-6708/2007/09/116
  [arXiv:0706.1216 [hep-ph]].

\bibitem{DM}
  P.~S.~B.~Dev and R.~N.~Mohapatra,
  %``Unified explanation of the $eejj$, diboson and dijet resonances at the LHC,''
  Phys.\ Rev.\ Lett.\  {\bf 115}, no. 18, 181803 (2015)
 % doi:10.1103/PhysRevLett.115.181803
  [arXiv:1508.02277 [hep-ph]].

\bibitem{Hirsch}
  G.~Anamiati, M.~Hirsch and E.~Nardi,
  %``Quasi-Dirac neutrinos at the LHC,''
  JHEP {\bf 1610}, 010 (2016)
 % doi:10.1007/JHEP10(2016)010
  [arXiv:1607.05641 [hep-ph]].

\bibitem{Antusch}
  S.~Antusch, E.~Cazzato and O.~Fischer,
   %``Resolvable heavy neutrino–antineutrino oscillations at colliders,''
  Mod.\ Phys.\ Lett.\ A {\bf 34}, no. 07n08, 1950061 (2019)
 % doi:10.1142/S0217732319500615
  [arXiv:1709.03797 [hep-ph]].

\bibitem{Sirunyan:2018pom}
  A.~M.~Sirunyan {\it et al.} [CMS Collaboration],
  %``Search for a heavy right-handed W boson and a heavy neutrino in events with two same-flavor leptons and two jets at $\sqrt{s}=$ 13 TeV,''
  JHEP {\bf 1805}, no. 05, 148 (2018)
%  doi:10.1007/JHEP05(2018)148
  [arXiv:1803.11116 [hep-ex]].

\bibitem{Aaboud:2019wfg}
  M.~Aaboud {\it et al.} [ATLAS Collaboration],
  %``Search for a right-handed gauge boson decaying into a high-momentum heavy neutrino and a charged lepton in $pp$ collisions with the ATLAS detector at $\sqrt{s}=13$ TeV,''
  %Submitted to: Phys.Lett.
  [arXiv:1904.12679 [hep-ex]].

\bibitem{Nir:1999mg}
  Y.~Nir,
  %``CP violation in and beyond the standard model,''
  hep-ph/9911321.


\bibitem{Zimmermann:2018wdi}
  A.~Abada {\it et al.} [FCC Collaboration],
  %``Future Circular Collider : Vol. 4 The High-Energy LHC (HE-LHC),''
  CERN-ACC-2018-0059.

\bibitem{Benedikt:2018csr}
  A.~Abada {\it et al.} [FCC Collaboration],
  %``Future Circular Collider : Vol. 3 The Hadron Collider (FCC-hh),''
  CERN-ACC-2018-0058.

\bibitem{CEPC-SPPCStudyGroup:2015csa}
  M.~Ahmad {\it et al.},
  %``CEPC-SPPC Preliminary Conceptual Design Report. 1. Physics and Detector,''
  IHEP-CEPC-DR-2015-01, IHEP-TH-2015-01, IHEP-EP-2015-01.


\bibitem{RL1}
  M.~Flanz, E.~A.~Paschos, U.~Sarkar and J.~Weiss,
  %``Baryogenesis through mixing of heavy Majorana neutrinos,''
  Phys.\ Lett.\ B {\bf 389}, 693 (1996)
  %doi:10.1016/S0370-2693(96)01337-8, 10.1016/S0370-2693(96)80011-6
  [hep-ph/9607310].

\bibitem{RL2}
  L.~Covi, E.~Roulet and F.~Vissani,
  %``CP violating decays in leptogenesis scenarios,''
  Phys.\ Lett.\ B {\bf 384}, 169 (1996)
  %doi:10.1016/0370-2693(96)00817-9
  [hep-ph/9605319].

\bibitem{RL3}
  A. Pilaftsis, Phys. Rev. {\bf D 56}, 543 (1997) [hep-ph/9707235].



\bibitem{Deppisch:2013jxa}
  F.~F.~Deppisch, J.~Harz and M.~Hirsch,
  %``Falsifying High-Scale Leptogenesis at the LHC,''
  Phys.\ Rev.\ Lett.\  {\bf 112}, 221601 (2014)
%  doi:10.1103/PhysRevLett.112.221601
  [arXiv:1312.4447 [hep-ph]].

\bibitem{Vasquez:2014mxa}
  J.~C.~Vasquez,
  %``Right-handed lepton mixings at the LHC,''
  JHEP {\bf 1605}, 176 (2016)
  %doi:10.1007/JHEP05(2016)176
  [arXiv:1411.5824 [hep-ph]].

\bibitem{Caputo:2016ojx}
  A.~Caputo, P.~Hernandez, M.~Kekic, J.~Lopez-Pavon and J.~Salvado,
  %``The seesaw path to leptonic CP violation,''
  Eur.\ Phys.\ J.\ C {\bf 77}, no. 4, 258 (2017)
  %doi:10.1140/epjc/s10052-017-4823-8
  [arXiv:1611.05000 [hep-ph]].

\bibitem{Antusch:2017pkq}
  S.~Antusch, E.~Cazzato, M.~Drewes, O.~Fischer, B.~Garbrecht, D.~Gueter and J.~Klaric,
  %``Probing Leptogenesis at Future Colliders,''
  JHEP {\bf 1809}, 124 (2018)
%  doi:10.1007/JHEP09(2018)124
  [arXiv:1710.03744 [hep-ph]].

\bibitem{Casas:2001sr}
  J.~A.~Casas and A.~Ibarra,
  %``Oscillating neutrinos and muon ---> e, gamma,''
  Nucl.\ Phys.\ B {\bf 618}, 171 (2001)
%  doi:10.1016/S0550-3213(01)00475-8
  [hep-ph/0103065].

\bibitem{Nemevsek:2012iq}
  M.~Nemevsek, G.~Senjanovic and V.~Tello,
  %``Connecting Dirac and Majorana Neutrino Mass Matrices in the Minimal Left-Right Symmetric Model,''
  Phys.\ Rev.\ Lett.\  {\bf 110}, no. 15, 151802 (2013)
  %doi:10.1103/PhysRevLett.110.151802
  [arXiv:1211.2837 [hep-ph]].

\bibitem{Frere:2008ct}
  J.~M.~Frere, T.~Hambye and G.~Vertongen,
  %``Is leptogenesis falsifiable at LHC?,''
  JHEP {\bf 0901}, 051 (2009)
  %doi:10.1088/1126-6708/2009/01/051
  [arXiv:0806.0841 [hep-ph]].

\bibitem{Dev:2014iva}
  P.~S.~Bhupal Dev, C.~H.~Lee and R.~N.~Mohapatra,
  %``Leptogenesis Constraints on the Mass of Right-handed Gauge Bosons,''
  Phys.\ Rev.\ D {\bf 90}, no. 9, 095012 (2014)
  %doi:10.1103/PhysRevD.90.095012
  [arXiv:1408.2820 [hep-ph]].

\bibitem{Dhuria:2015cfa}
  M.~Dhuria, C.~Hati, R.~Rangarajan and U.~Sarkar,
  %``Falsifying leptogenesis for a TeV scale $W^{\pm}_{R}$ at the LHC,''
  Phys.\ Rev.\ D {\bf 92}, no. 3, 031701 (2015)
  %doi:10.1103/PhysRevD.92.031701
  [arXiv:1503.07198 [hep-ph]].

\bibitem{DLM} P.~S.~B.~Dev, C.~H.~Lee and R.~N.~Mohapatra,
  %``TeV Scale Lepton Number Violation and Baryogenesis,''
  J.\ Phys.\ Conf.\ Ser.\  {\bf 631}, no. 1, 012007 (2015)
  %doi:10.1088/1742-6596/631/1/012007
  [arXiv:1503.04970 [hep-ph]].

\bibitem{zhou} W.~Chao, Z.~g.~Si, Y.~j.~Zheng and S.~Zhou,
  %``Testing the Realistic Seesaw Model with Two Heavy Majorana Neutrinos at the CERN Large Hadron Collider,''
  Phys.\ Lett.\ B {\bf 683}, 26 (2010)
  %doi:10.1016/j.physletb.2009.11.059
  [arXiv:0907.0935 [hep-ph]].

\bibitem{Kersten} J.~Kersten and A.~Y.~Smirnov,
  %``Right-Handed Neutrinos at CERN LHC and the Mechanism of Neutrino Mass Generation,''
  Phys.\ Rev.\ D {\bf 76}, 073005 (2007)
  %doi:10.1103/PhysRevD.76.073005
  [arXiv:0705.3221 [hep-ph]].


\bibitem{Ibarra:2010xw}
  A.~Ibarra, E.~Molinaro and S.~T.~Petcov,
  %``TeV Scale See-Saw Mechanisms of Neutrino Mass Generation, the Majorana Nature of the Heavy Singlet Neutrinos and $(\beta\beta)_{0\nu}$-Decay,''
  JHEP {\bf 1009}, 108 (2010)
 % doi:10.1007/JHEP09(2010)108
  [arXiv:1007.2378 [hep-ph]].

\bibitem{Mohapatra:1986pj}
  R.~N.~Mohapatra,
  %``Limits on the Mass of the Right-handed Majorana Neutrino,''
  Phys.\ Rev.\ D {\bf 34}, 909 (1986).
 % doi:10.1103/PhysRevD.34.909

\bibitem{Maiezza:2016bzp}
  A.~Maiezza, M.~Nemevšek and F.~Nesti,
  %``Perturbativity and mass scales in the minimal left-right symmetric model,''
  Phys.\ Rev.\ D {\bf 94}, no. 3, 035008 (2016)
 % doi:10.1103/PhysRevD.94.035008
  [arXiv:1603.00360 [hep-ph]].

\bibitem{Dev:2018foq}
  P.~S.~B.~Dev, R.~N.~Mohapatra, W.~Rodejohann and X.~J.~Xu,
  %``Vacuum structure of the left-right symmetric model,''
  JHEP {\bf 1902}, 154 (2019)
  % doi:10.1007/JHEP02(2019)154
  [arXiv:1811.06869 [hep-ph]].


\bibitem{Deshpande:1990ip}
  N.~G.~Deshpande, J.~F.~Gunion, B.~Kayser and F.~I.~Olness,
  %``Left-right symmetric electroweak models with triplet Higgs,''
  Phys.\ Rev.\ D {\bf 44}, 837 (1991).
  %doi:10.1103/PhysRevD.44.837

\bibitem{Zhang:2007da}
  Y.~Zhang, H.~An, X.~Ji and R.~N.~Mohapatra,
  %``General CP Violation in Minimal Left-Right Symmetric Model and Constraints on the Right-Handed Scale,''
  Nucl.\ Phys.\ B {\bf 802}, 247 (2008)
  %doi:10.1016/j.nuclphysb.2008.05.019
  [arXiv:0712.4218 [hep-ph]].

\bibitem{Dev:2016dja}
  P.~S.~B.~Dev, R.~N.~Mohapatra and Y.~Zhang,
  %``Probing the Higgs Sector of the Minimal Left-Right Symmetric Model at Future Hadron Colliders,''
  JHEP {\bf 1605}, 174 (2016)
  %doi:10.1007/JHEP05(2016)174
  [arXiv:1602.05947 [hep-ph]].

\bibitem{Dev:2014xea}
  P.~S.~B. Dev, S.~Goswami and M.~Mitra,
  %``TeV Scale Left-Right Symmetry and Large Mixing Effects in Neutrinoless Double Beta Decay,''
  Phys.\ Rev.\ D {\bf 91}, no. 11, 113004 (2015)
  %doi:10.1103/PhysRevD.91.113004
  [arXiv:1405.1399 [hep-ph]].

\bibitem{Bambhaniya:2016rbb}
  G.~Bambhaniya, P.~S.~B. Dev, S.~Goswami, S.~Khan and W.~Rodejohann,
  %``Naturalness, Vacuum Stability and Leptogenesis in the Minimal Seesaw Model,''
  Phys.\ Rev.\ D {\bf 95}, no. 9, 095016 (2017)
 % doi:10.1103/PhysRevD.95.095016
  [arXiv:1611.03827 [hep-ph]].


%\bibitem{Bray:2007ru}
%  S.~Bray, J.~S.~Lee and A.~Pilaftsis,
%  %``Resonant CP violation due to heavy neutrinos at the LHC,''
%  Nucl.\ Phys.\ B {\bf 786}, 95 (2007)
%  doi:10.1016/j.nuclphysb.2007.07.002
%  [hep-ph/0702294 [HEP-PH]].

%\bibitem{Aad:2010yt}
%  G.~Aad {\it et al.} [ATLAS Collaboration],
  %``Measurement of the $W \to \ell\nu$ and $Z/\gamma^* \to \ell\ell$ production cross sections in proton-proton collisions at $\sqrt{s} = 7$ TeV with the ATLAS detector,''
%  JHEP {\bf 1012}, 060 (2010)
  %doi:10.1007/JHEP12(2010)060
 % [arXiv:1010.2130 [hep-ex]].

%\bibitem{ATLAS:2010dda}
%  [ATLAS Collaboration],
  %``Measurement of the W->lnu production cross-section and observation of Z -> ll production in proton-proton collisions at root{s}=7 TeV with the ATLAS detector,''
%  ATLAS-CONF-2010-051.

%\bibitem{Aad:2016naf}
% G.~Aad {\it et al.} [ATLAS Collaboration],
  %``Measurement of $W^{\pm}$ and $Z$-boson production cross sections in $pp$ collisions at $\sqrt{s}=13$ TeV with the ATLAS detector,''
%  Phys.\ Lett.\ B {\bf 759}, 601 (2016)
  %doi:10.1016/j.physletb.2016.06.023
 % [arXiv:1603.09222 [hep-ex]].



\bibitem{Aaboud:2016btc}
  M.~Aaboud {\it et al.} [ATLAS Collaboration],
  %``Precision measurement and interpretation of inclusive $W^+$ , $W^-$ and $Z/\gamma ^*$ production cross sections with the ATLAS detector,''
  Eur.\ Phys.\ J.\ C {\bf 77}, no. 6, 367 (2017)
  %doi:10.1140/epjc/s10052-017-4911-9
  [arXiv:1612.03016 [hep-ex]].

%\bibitem{Chatrchyan:2011jz}
 % S.~Chatrchyan {\it et al.} [CMS Collaboration],
  %``Measurement of the lepton charge asymmetry in inclusive $W$ production in pp collisions at $\sqrt{s} = 7$ TeV,''
 % JHEP {\bf 1104}, 050 (2011)
  %doi:10.1007/JHEP04(2011)050
%  [arXiv:1103.3470 [hep-ex]].

%\bibitem{CMS:2011aa}
 % S.~Chatrchyan {\it et al.} [CMS Collaboration],
  %``Measurement of the Inclusive $W$ and $Z$ Production Cross Sections in $pp$ Collisions at $\sqrt{s}=7$ TeV,''
%  JHEP {\bf 1110}, 132 (2011)
  %doi:10.1007/JHEP10(2011)132
%  [arXiv:1107.4789 [hep-ex]].

%\bibitem{Chatrchyan:2012xt}
 % S.~Chatrchyan {\it et al.} [CMS Collaboration],
  %``Measurement of the electron charge asymmetry in inclusive $W$ production in $pp$ collisions at $\sqrt{s}=7$ TeV,''
%  Phys.\ Rev.\ Lett.\  {\bf 109}, 111806 (2012)
  %doi:10.1103/PhysRevLett.109.111806
%  [arXiv:1206.2598 [hep-ex]].

%\bibitem{Chatrchyan:2013mza}
 % S.~Chatrchyan {\it et al.} [CMS Collaboration],
  %``Measurement of the muon charge asymmetry in inclusive $pp \to W+X$ production at $\sqrt s =$ 7 TeV and an improved determination of light parton distribution functions,''
%  Phys.\ Rev.\ D {\bf 90}, no. 3, 032004 (2014)
  %doi:10.1103/PhysRevD.90.032004
%  [arXiv:1312.6283 [hep-ex]].

%\bibitem{Chatrchyan:2014mua}
%  S.~Chatrchyan {\it et al.} [CMS Collaboration],
  %``Measurement of inclusive W and Z boson production cross sections in pp collisions at $\sqrt{s}$ = 8 TeV,''
%  Phys.\ Rev.\ Lett.\  {\bf 112}, 191802 (2014)
  %doi:10.1103/PhysRevLett.112.191802
%  [arXiv:1402.0923 [hep-ex]].

\bibitem{CMS:2015ois}
  CMS Collaboration [CMS Collaboration],
  %``Measurement of inclusive W and Z boson production cross sections in pp collisions at sqrt(s)=13 TeV,''
  CMS-PAS-SMP-15-004.


\bibitem{Aad:2014bha}
  G.~Aad {\it et al.} [ATLAS Collaboration],
  %``Measurement of the production and lepton charge asymmetry of $W$ bosons in Pb+Pb collisions at $\mathbf {\sqrt{\mathbf {s}_{\mathrm {\mathbf {NN}}}}=2.76\;TeV}$ with the ATLAS detector,''
  Eur.\ Phys.\ J.\ C {\bf 75}, no. 1, 23 (2015)
  %doi:10.1140/epjc/s10052-014-3231-6
  [arXiv:1408.4674 [hep-ex]].

\bibitem{Chatrchyan:2012nt}
  S.~Chatrchyan {\it et al.} [CMS Collaboration],
  %``Study of $W$ boson production in PbPb and $pp$ collisions at $\sqrt{s_{NN}}=2.76$ TeV,''
  Phys.\ Lett.\ B {\bf 715}, 66 (2012)
  %doi:10.1016/j.physletb.2012.07.025
  [arXiv:1205.6334 [nucl-ex]].













\bibitem{Aaboud:2018spl}
  M.~Aaboud {\it et al.} [ATLAS Collaboration],
  %``Search for heavy Majorana or Dirac neutrinos and right-handed $W$ gauge bosons in final states with two charged leptons and two jets at $ \sqrt{s}=13 $ TeV with the ATLAS detector,''
  JHEP {\bf 1901}, 016 (2019)
  %doi:10.1007/JHEP01(2019)016
  [arXiv:1809.11105 [hep-ex]].

\bibitem{Mitra:2016kov}
  M.~Mitra, R.~Ruiz, D.~J.~Scott and M.~Spannowsky,
  %``Neutrino Jets from High-Mass $W_R$ Gauge Bosons in TeV-Scale Left-Right Symmetric Models,''
  Phys.\ Rev.\ D {\bf 94}, no. 9, 095016 (2016)
  %doi:10.1103/PhysRevD.94.095016
  [arXiv:1607.03504 [hep-ph]].

\bibitem{Alvarez:2016nrz}
  E.~Alvarez, D.~A.~Faroughy, J.~F.~Kamenik, R.~Morales and A.~Szynkman,
  %``Four Tops for LHC,''
  Nucl.\ Phys.\ B {\bf 915}, 19 (2017)
  %doi:10.1016/j.nuclphysb.2016.11.024
  [arXiv:1611.05032 [hep-ph]].

\bibitem{Ball:2017nwa}
  R.~D.~Ball {\it et al.} [NNPDF Collaboration],
  %``Parton distributions from high-precision collider data,''
  Eur.\ Phys.\ J.\ C {\bf 77}, 663 (2017)
  %doi:10.1140/epjc/s10052-017-5199-5
  [arXiv:1706.00428 [hep-ph]].

\bibitem{Buckley:2014ana}
  A.~Buckley, J.~Ferrando, S.~Lloyd, K.~Nordstr\"{o}m, B.~Page, M.~R\"{u}fenacht, M.~Sch\"{o}nherr and G.~Watt,
  %``LHAPDF6: parton density access in the LHC precision era,''
  Eur.\ Phys.\ J.\ C {\bf 75}, 132 (2015)
  %doi:10.1140/epjc/s10052-015-3318-8
  [arXiv:1412.7420 [hep-ph]].

\bibitem{Dulat:2015mca}
  S.~Dulat {\it et al.},
  %``New parton distribution functions from a global analysis of quantum chromodynamics,''
  Phys.\ Rev.\ D {\bf 93}, no. 3, 033006 (2016)
  %doi:10.1103/PhysRevD.93.033006
  [arXiv:1506.07443 [hep-ph]].

\bibitem{Harland-Lang:2014zoa}
  L.~A.~Harland-Lang, A.~D.~Martin, P.~Motylinski and R.~S.~Thorne,
  %``Parton distributions in the LHC era: MMHT 2014 PDFs,''
  Eur.\ Phys.\ J.\ C {\bf 75}, no. 5, 204 (2015)
  %doi:10.1140/epjc/s10052-015-3397-6
  [arXiv:1412.3989 [hep-ph]].

\bibitem{AbdulKhalek:2019bux}
  R.~Abdul Khalek {\it et al.} [NNPDF Collaboration],
  %``A First Determination of Parton Distributions with Theoretical Uncertainties,''
  arXiv:1905.04311 [hep-ph].

\bibitem{Belyaev:2012qa}
  A.~Belyaev, N.~D.~Christensen and A.~Pukhov,
  %``CalcHEP 3.4 for collider physics within and beyond the Standard Model,''
  Comput.\ Phys.\ Commun.\  {\bf 184}, 1729 (2013)
  %doi:10.1016/j.cpc.2013.01.014
  [arXiv:1207.6082 [hep-ph]].

\bibitem{Richard}
  Private communication with Richard Ruiz.

  \bibitem{Catani:2010en}
  S.~Catani, G.~Ferrera and M.~Grazzini,
  %``W Boson Production at Hadron Colliders: The Lepton Charge Asymmetry in NNLO QCD,''
  JHEP {\bf 1005}, 006 (2010)
 % doi:10.1007/JHEP05(2010)006
  [arXiv:1002.3115 [hep-ph]].

%\bibitem{Abe:2017vif}
 % K.~Abe {\it et al.} [T2K Collaboration],
  %``Measurement of neutrino and antineutrino oscillations by the T2K experiment including a new additional sample of $\nu_e$ interactions at the far detector,''
%  Phys.\ Rev.\ D {\bf 96}, no. 9, 092006 (2017)
%  Erratum: [Phys.\ Rev.\ D {\bf 98}, no. 1, 019902 (2018)]
  %doi:10.1103/PhysRevD.96.092006, 10.1103/PhysRevD.98.019902
%  [arXiv:1707.01048 [hep-ex]].

\bibitem{Abe:2018wpn}
  K.~Abe {\it et al.} [T2K Collaboration],
  %``Search for CP Violation in Neutrino and Antineutrino Oscillations by the T2K Experiment with $2.2\times10^{21}$ Protons on Target,''
  Phys.\ Rev.\ Lett.\  {\bf 121}, no. 17, 171802 (2018)
  %doi:10.1103/PhysRevLett.121.171802
  [arXiv:1807.07891 [hep-ex]].

%\bibitem{NOvA:2018gge}
%  M.~A.~Acero {\it et al.} [NOvA Collaboration],
  %``New constraints on oscillation parameters from $\nu_e$ appearance and $\nu_\mu$ disappearance in the NOvA experiment,''
%  Phys.\ Rev.\ D {\bf 98}, 032012 (2018)
  %doi:10.1103/PhysRevD.98.032012
%  [arXiv:1806.00096 [hep-ex]].

\bibitem{Acero:2019ksn}
  M.~A.~Acero {\it et al.} [NOvA Collaboration],
  %``First measurement of neutrino oscillation parameters using neutrinos and antineutrinos by NOvA,''
  arXiv:1906.04907 [hep-ex].


\bibitem{Davidson:2002qv}
  S.~Davidson and A.~Ibarra,
  %``A Lower bound on the right-handed neutrino mass from leptogenesis,''
  Phys.\ Lett.\ B {\bf 535}, 25 (2002)
 % doi:10.1016/S0370-2693(02)01735-5
  [hep-ph/0202239].

  \bibitem{Buchmuller:2003gz}
  W.~Buchmuller, P.~Di Bari and M.~Plumacher,
  %``The Neutrino mass window for baryogenesis,''
  Nucl.\ Phys.\ B {\bf 665}, 445 (2003)
%  doi:10.1016/S0550-3213(03)00449-8
  [hep-ph/0302092].

%\bibitem{Pascoli:2006ie}
%  S.~Pascoli, S.~T.~Petcov and A.~Riotto,
  %``Connecting low energy leptonic CP-violation to leptogenesis,''
%  Phys.\ Rev.\ D {\bf 75}, 083511 (2007)
  %doi:10.1103/PhysRevD.75.083511
%  [hep-ph/0609125].

%\bibitem{Pascoli:2006ci}
%  S.~Pascoli, S.~T.~Petcov and A.~Riotto,
  %``Leptogenesis and Low Energy CP Violation in Neutrino Physics,''
%  Nucl.\ Phys.\ B {\bf 774}, 1 (2007)
  %doi:10.1016/j.nuclphysb.2007.02.019
%  [hep-ph/0611338].

\bibitem{Moffat:2018smo}
  K.~Moffat, S.~Pascoli, S.~T.~Petcov and J.~Turner,
%  %``Leptogenesis from Low Energy $CP$ Violation,''
  JHEP {\bf 1903}, 034 (2019)
%  %doi:10.1007/JHEP03(2019)034
  [arXiv:1809.08251 [hep-ph]].

  \bibitem{Pilaftsis:2005rv}
  A.~Pilaftsis and T.~E.~J.~Underwood,
  %``Electroweak-scale resonant leptogenesis,''
  Phys.\ Rev.\ D {\bf 72}, 113001 (2005)
 % doi:10.1103/PhysRevD.72.113001
  [hep-ph/0506107].

  \bibitem{Dev:2014laa}
  P.~S.~B,~Dev, P.~Millington, A.~Pilaftsis and D.~Teresi,
  %``Flavour Covariant Transport Equations: an Application to Resonant Leptogenesis,''
  Nucl.\ Phys.\ B {\bf 886}, 569 (2014)
%  doi:10.1016/j.nuclphysb.2014.06.020
  [arXiv:1404.1003 [hep-ph]].


\bibitem{Dev:2017trv}
  P.~S.~B.~Dev, P.~Di Bari, B.~Garbrecht, S.~Lavignac, P.~Millington and D.~Teresi,
  %``Flavor effects in leptogenesis,''
  Int.\ J.\ Mod.\ Phys.\ A {\bf 33}, 1842001 (2018)
  %doi:10.1142/S0217751X18420010
  [arXiv:1711.02861 [hep-ph]].

\bibitem{Dev:2015kca}
  P.~S.~B.~Dev, D.~Kim and R.~N.~Mohapatra,
  %``Disambiguating Seesaw Models using Invariant Mass Variables at Hadron Colliders,''
  JHEP {\bf 1601}, 118 (2016)
  %doi:10.1007/JHEP01(2016)118
  [arXiv:1510.04328 [hep-ph]].

\bibitem{Ruiz:2017nip}
  R.~Ruiz,
  %``Lepton Number Violation at Colliders from Kinematically Inaccessible Gauge Bosons,''
  Eur.\ Phys.\ J.\ C {\bf 77}, no. 6, 375 (2017)
  %doi:10.1140/epjc/s10052-017-4950-2
  [arXiv:1703.04669 [hep-ph]].

\bibitem{CidVidal:2018eel}
  X.~Cid Vidal {\it et al.} [Working Group 3],
  %``Beyond the Standard Model Physics at the HL-LHC and HE-LHC,''
  arXiv:1812.07831 [hep-ph].



\bibitem{Tanabashi:2018oca}
  M.~Tanabashi {\it et al.} [Particle Data Group],
  %``Review of Particle Physics,''
  Phys.\ Rev.\ D {\bf 98}, no. 3, 030001 (2018).
%  doi:10.1103/PhysRevD.98.030001

\bibitem{Giudice:2003jh}
  G.~F.~Giudice, A.~Notari, M.~Raidal, A.~Riotto and A.~Strumia,
  %``Towards a complete theory of thermal leptogenesis in the SM and MSSM,''
  Nucl.\ Phys.\ B {\bf 685}, 89 (2004)
  %doi:10.1016/j.nuclphysb.2004.02.019
  [hep-ph/0310123].

\bibitem{Hambye:2016sby}
  T.~Hambye and D.~Teresi,
  %``Higgs doublet decay as the origin of the baryon asymmetry,''
  Phys.\ Rev.\ Lett.\  {\bf 117}, no. 9, 091801 (2016)
  %doi:10.1103/PhysRevLett.117.091801
  [arXiv:1606.00017 [hep-ph]].

%\bibitem{Hambye:2017elz}
%  T.~Hambye and D.~Teresi,
  %``Baryogenesis from L-violating Higgs-doublet decay in the density-matrix formalism,''
%  Phys.\ Rev.\ D {\bf 96}, no. 1, 015031 (2017)
  %doi:10.1103/PhysRevD.96.015031
%  [arXiv:1705.00016 [hep-ph]].

%\bibitem{Dev:2014tpa}
 % P.~S.~Bhupal Dev, P.~Millington, A.~Pilaftsis and D.~Teresi,
  %``Flavour Covariant Formalism for Resonant Leptogenesis,''
%  Nucl.\ Part.\ Phys.\ Proc.\  {\bf 273-275}, 268 (2016)
  %doi:10.1016/j.nuclphysbps.2015.09.037
 % [arXiv:1409.8263 [hep-ph]].

\bibitem{Aghanim:2018eyx}
  N.~Aghanim {\it et al.} [Planck Collaboration],
  %``Planck 2018 results. VI. Cosmological parameters,''
  arXiv:1807.06209 [astro-ph.CO].




%\cite{Borah:2015ufa}
%\bibitem{Borah:2015ufa}
%  D.~Borah and A.~Dasgupta,
  %``Neutrinoless Double Beta Decay in Type I+II Seesaw Models,''
 % JHEP {\bf 1511}, 208 (2015)
  %doi:10.1007/JHEP11(2015)208
%  [arXiv:1509.01800 [hep-ph]].
  %%CITATION = doi:10.1007/JHEP11(2015)208;%%
  %12 citations counted in INSPIRE as of 09 Mar 2018

%\cite{Pritimita:2016fgr}
%\bibitem{Pritimita:2016fgr}
%  P.~Pritimita, N.~Dash and S.~Patra,
  %``Neutrinoless Double Beta Decay in LRSM with Natural Type-II seesaw Dominance,''
%  JHEP {\bf 1610}, 147 (2016)
  %doi:10.1007/JHEP10(2016)147
 % [arXiv:1607.07655 [hep-ph]].
  %%CITATION = doi:10.1007/JHEP10(2016)147;%%
  %6 citations counted in INSPIRE as of 09 Mar 2018

%\cite{Borgohain:2017akh}
%\bibitem{Borgohain:2017akh}
%  H.~Borgohain and M.~K.~Das,
  %``Lepton number violation, lepton flavor violation, and baryogenesis in left-right symmetric model,''
%  Phys.\ Rev.\ D {\bf 96}, no. 7, 075021 (2017)
  %doi:10.1103/PhysRevD.96.075021
 % [arXiv:1709.09542 [hep-ph]].
  %%CITATION = doi:10.1103/PhysRevD.96.075021;%%

%\cite{Ge:2015yqa}
%\bibitem{Ge:2015yqa}
%  S.~F.~Ge, M.~Lindner and S.~Patra,
  %``New physics effects on neutrinoless double beta decay from right-handed current,''
%  JHEP {\bf 1510}, 077 (2015)
  %doi:10.1007/JHEP10(2015)077
 % [arXiv:1508.07286 [hep-ph]].
  %%CITATION = doi:10.1007/JHEP10(2015)077;%%
  %27 citations counted in INSPIRE as of 14 Mar 2018


%\bibitem{Deppisch:2014zta}
 % F.~F.~Deppisch, T.~E.~Gonzalo, S.~Patra, N.~Sahu and U.~Sarkar,
  %``Double beta decay, lepton flavor violation, and collider signatures of left-right symmetric models with spontaneous $D$-parity breaking,''
 % Phys.\ Rev.\ D {\bf 91}, no. 1, 015018 (2015)
 % doi:10.1103/PhysRevD.91.015018
 % [arXiv:1410.6427 [hep-ph]].

%\bibitem{Mohapatra:1981pm}
 % R.~N.~Mohapatra and J.~D.~Vergados,
  %``A New Contribution to Neutrinoless Double Beta Decay in Gauge Models,''
 % Phys.\ Rev.\ Lett.\  {\bf 47}, 1713 (1981).

%\bibitem{Hirsch:1996qw}
%  M.~Hirsch, H.~V.~Klapdor-Kleingrothaus and O.~Panella,
  %``Double beta decay in left-right symmetric models,''
 % Phys.\ Lett.\ B {\bf 374}, 7 (1996)
%  doi:10.1016/0370-2693(96)00185-2
%  [hep-ph/9602306].

%\bibitem{Huang:2013kma}
 % W.~C.~Huang and J.~Lopez-Pavon,
  %``On neutrinoless double beta decay in the minimal left-right symmetric model,''
 % Eur.\ Phys.\ J.\ C {\bf 74}, 2853 (2014)
%  doi:10.1140/epjc/s10052-014-2853-z
%  [arXiv:1310.0265 [hep-ph]].

%\cite{Dev:2018sel}
%\bibitem{Dev:2018sel}
 % P.~S.~B.~Dev, M.~J.~Ramsey-Musolf and Y.~Zhang,
  %``Doubly-Charged Scalars in the Type-II Seesaw Mechanism: Fundamental Symmetry Tests and High-Energy Searches,''
 % Phys.\ Rev.\ D {\bf 98}, no. 5, 055013 (2018)
  %doi:10.1103/PhysRevD.98.055013
%  [arXiv:1806.08499 [hep-ph]].

%\bibitem{Dev:2018kpa}
%  P.~S.~Bhupal Dev and Y.~Zhang,
  %``Displaced vertex signatures of doubly charged scalars in the type-II seesaw and its left-right extensions,''
%  JHEP {\bf 1810}, 199 (2018)
  %doi:10.1007/JHEP10(2018)199
%  [arXiv:1808.00943 [hep-ph]].

%\bibitem{Prezeau:2003xn}
%  G.~Prezeau, M.~Ramsey-Musolf and P.~Vogel,
  %``Neutrinoless double beta decay and effective field theory,''
%  Phys.\ Rev.\ D {\bf 68}, 034016 (2003)
  %doi:10.1103/PhysRevD.68.034016
%  [hep-ph/0303205].

\bibitem{Rodejohann:2011mu}
  W.~Rodejohann,
  %``Neutrino-less Double Beta Decay and Particle Physics,''
  Int.\ J.\ Mod.\ Phys.\ E {\bf 20}, 1833 (2011)
%  doi:10.1142/S0218301311020186
  [arXiv:1106.1334 [hep-ph]].


  \bibitem{Bambhaniya:2015ipg}
  G.~Bambhaniya, P.~S.~B.~Dev, S.~Goswami and M.~Mitra,
  %``The Scalar Triplet Contribution to Lepton Flavour Violation and Neutrinoless Double Beta Decay in Left-Right Symmetric Model,''
  JHEP {\bf 1604}, 046 (2016)
%  doi:10.1007/JHEP04(2016)046
  [arXiv:1512.00440 [hep-ph]].

  %\cite{Dev:2018sel}
\bibitem{Dev:2018sel}
  P.~S.~B.~Dev, M.~J.~Ramsey-Musolf and Y.~Zhang,
  %``Doubly-Charged Scalars in the Type-II Seesaw Mechanism: Fundamental Symmetry Tests and High-Energy Searches,''
  Phys.\ Rev.\ D {\bf 98}, no. 5, 055013 (2018)
  %doi:10.1103/PhysRevD.98.055013
 [arXiv:1806.08499 [hep-ph]].

 %\bibitem{Dev:2013vxa}
% P.~S.~B.~Dev, S.~Goswami, M.~Mitra and W.~Rodejohann,
  %``Constraining Neutrino Mass from Neutrinoless Double Beta Decay,''
%  Phys.\ Rev.\ D {\bf 88}, 091301 (2013)
%  doi:10.1103/PhysRevD.88.091301
%  [arXiv:1305.0056 [hep-ph]].


\bibitem{Barry:2013xxa}
  J.~Barry and W.~Rodejohann,
  %``Lepton number and flavour violation in TeV-scale left-right symmetric theories with large left-right mixing,''
  JHEP {\bf 1309}, 153 (2013)
  %doi:10.1007/JHEP09(2013)153
  [arXiv:1303.6324 [hep-ph]].

\bibitem{Kotila:2012zza}
  J.~Kotila and F.~Iachello,
  %``Phase space factors for double-$\beta$ decay,''
  Phys.\ Rev.\ C {\bf 85}, 034316 (2012)
  %doi:10.1103/PhysRevC.85.034316
  [arXiv:1209.5722 [nucl-th]].

\bibitem{Pantis:1996py}
  G.~Pantis, F.~Simkovic, J.~D.~Vergados and A.~Faessler,
  %``Neutrinoless double beta decay within QRPA with proton - neutron pairing,''
  Phys.\ Rev.\ C {\bf 53}, 695 (1996)
  %doi:10.1103/PhysRevC.53.695
  [nucl-th/9612036].

%\cite{Meroni:2012qf}
\bibitem{Meroni:2012qf}
  A.~Meroni, S.~T.~Petcov and F.~Simkovic,
  %``Multiple CP non-conserving mechanisms of $(\beta\beta)_{0\nu}$-decay and nuclei with largely different nuclear matrix elements,''
  JHEP {\bf 1302}, 025 (2013)
  %doi:10.1007/JHEP02(2013)025
  [arXiv:1212.1331 [hep-ph]].

\bibitem{KamLAND-Zen:2016pfg}
  A.~Gando {\it et al.} [KamLAND-Zen Collaboration],
  %``Search for Majorana Neutrinos near the Inverted Mass Hierarchy Region with KamLAND-Zen,''
  Phys.\ Rev.\ Lett.\  {\bf 117}, no. 8, 082503 (2016)
  Addendum: [Phys.\ Rev.\ Lett.\  {\bf 117}, no. 10, 109903 (2016)]
  %doi:10.1103/PhysRevLett.117.109903, 10.1103/PhysRevLett.117.082503
  [arXiv:1605.02889 [hep-ex]].

\bibitem{Agostini:2018tnm}
  M.~Agostini {\it et al.} [GERDA Collaboration],
  %``Improved Limit on Neutrinoless Double-$\beta$ Decay of $^{76}$Ge from GERDA Phase II,''
  Phys.\ Rev.\ Lett.\  {\bf 120}, no. 13, 132503 (2018)
  %doi:10.1103/PhysRevLett.120.132503
  [arXiv:1803.11100 [nucl-ex]].

%\bibitem{Albert:2017owj}
 % J.~B.~Albert {\it et al.} [EXO Collaboration],
  %``Search for Neutrinoless Double-Beta Decay with the Upgraded EXO-200 Detector,''
%  Phys.\ Rev.\ Lett.\  {\bf 120}, no. 7, 072701 (2018)
  %doi:10.1103/PhysRevLett.120.072701
%  [arXiv:1707.08707 [hep-ex]].

%\bibitem{Alduino:2017ehq}
%  C.~Alduino {\it et al.} [CUORE Collaboration],
  %``First Results from CUORE: A Search for Lepton Number Violation via $0\nu\beta\beta$ Decay of $^{130}$Te,''
 % Phys.\ Rev.\ Lett.\  {\bf 120}, no. 13, 132501 (2018)
  %doi:10.1103/PhysRevLett.120.132501
 % [arXiv:1710.07988 [nucl-ex]].

%\bibitem{Arnold:2018tmo}
%  R.~Arnold {\it et al.},
  %``Final results on $^{82}{Se}$ double beta decay to the ground state of $^{82}{Kr}$ from the NEMO-3 experiment,''
%  Eur.\ Phys.\ J.\ C {\bf 78}, no. 10, 821 (2018)
  %doi:10.1140/epjc/s10052-018-6295-x
%  [arXiv:1806.05553 [hep-ex]].

\bibitem{Cirigliano:2004mv}
  V.~Cirigliano, A.~Kurylov, M.~J.~Ramsey-Musolf and P.~Vogel,
  %``Lepton flavor violation without supersymmetry,''
  Phys.\ Rev.\ D {\bf 70}, 075007 (2004)
  %doi:10.1103/PhysRevD.70.075007
  [hep-ph/0404233].

\bibitem{Adam:2013mnn}
  J.~Adam {\it et al.} [MEG Collaboration],
  %``New constraint on the existence of the $\mu^+ \to e^+\gamma$ decay,''
  Phys.\ Rev.\ Lett.\  {\bf 110}, 201801 (2013)
  %doi:10.1103/PhysRevLett.110.201801
  [arXiv:1303.0754 [hep-ex]].

\bibitem{Nieves:1986uk}
  J.~F.~Nieves, D.~Chang and P.~B.~Pal,
  %``Electric Dipole Moment of the Electron in Left-right Symmetric Theories,''
  Phys.\ Rev.\ D {\bf 33}, 3324 (1986).
  %doi:10.1103/PhysRevD.33.3324

\bibitem{Andreev:2018ayy}
  V.~Andreev {\it et al.} [ACME Collaboration],
  %``Improved limit on the electric dipole moment of the electron,''
  Nature {\bf 562}, no. 7727, 355 (2018).
  %doi:10.1038/s41586-018-0599-8

\bibitem{Korner:1992zk}
  J.~G.~Korner, A.~Pilaftsis and K.~Schilcher,
  %``Leptonic CP asymmetries in flavor changing H0 decays,''
  Phys.\ Rev.\ D {\bf 47}, 1080 (1993)
  %doi:10.1103/PhysRevD.47.1080
  [hep-ph/9301289].

\bibitem{Grimus:2000vj}
  W.~Grimus and L.~Lavoura,
  %``The Seesaw mechanism at arbitrary order: Disentangling the small scale from the large scale,''
  JHEP {\bf 0011}, 042 (2000)
  %doi:10.1088/1126-6708/2000/11/042
  [hep-ph/0008179].

\end{thebibliography}
\end{document}